\documentclass[11pt, a4paper]{article}
\pdfoutput=1 
\usepackage[height=8.85in,width=6.75in]{geometry}
\usepackage{graphicx,rotating}     %usepackage without driver option
\usepackage[bookmarksopen,colorlinks=true,linkcolor=darkcyan,
citecolor=darkgoldenrod,urlcolor=darkgoldenrod,linktoc=all]{hyperref}

\usepackage{amsmath,amssymb}
\usepackage{slashed}
\usepackage{bm}
\usepackage{bbm}
\usepackage{cite}
\usepackage{comment}
\usepackage[utf8]{inputenc}
\usepackage{accents}
\usepackage[footnotesize]{caption}
\usepackage{cancel}

\newcommand{\vev}[1]{ \left\langle {#1} \right\rangle }

\makeatletter
\@addtoreset{equation}{section}

\makeatother

\usepackage{xcolor}
\definecolor{darkred}{rgb}{0.7, 0., 0.}
\definecolor{orangered}{rgb}{1,0.27,0.}
\definecolor{steelblue}{rgb}{0.275,0.51, 0.706}
\definecolor{forestgreen}{rgb}{0.13,0.55,0.13}
\definecolor{darkgoldenrod}{rgb}{0.72,0.53,0.04}
\definecolor{darkcyan}{rgb}{0, 0.545, 0.545}
\usepackage{tikz}
\usepackage{tikz-feynman}
\tikzfeynmanset{compat=1.0.0}
\pgfdeclarelayer{bg}    % declare background layer
\pgfsetlayers{bg,main}  % set the order of the layers (main is the standard layer)

\begin{document}
%%%%%%%%%%%%%%%%%%%%%%%%%%%%%%%%%%%%%%%

%%%%%%%%%%%%%%%%%%%%%%%%%%%%%%%%%%%%%%%
\hypersetup{pageanchor=false}
\begin{titlepage}

\begin{center}

\hfill UMN-TH-4224/23\\
\hfill FTPI-MINN-23-16\\

\vskip 0.5in

{\Huge \bfseries Cosmological Collider Signatures \vspace{5mm}\\ of Higgs-$R^2$ Inflation
} \\
\vskip .8in

{\Large Yohei Ema,$^{a,b}$}
{\Large Sarunas Verner$^{c}$}

\vskip .3in
{\begin{tabular}{ll}
$^{a}$ & \!\!\!\!\!\emph{William I. Fine Theoretical Physics Institute, School of Physics and Astronomy,}\\[-.15em]
& \!\!\!\!\!\emph{University of Minnesota, Minneapolis, MN 55455, USA}\\
$^{b}$ & \!\!\!\!\!\emph{School of Physics and Astronomy, University of Minnesota, Minneapolis, MN 55455, USA}\\
$^{c}$ & \!\!\!\!\!\emph{Institute for Fundamental Theory, Physics Department,}\\[-.15em]
& \!\!\!\!\!\emph{University of Florida, Gainesville, FL 32611, USA}
\end{tabular}}
\end{center}
\vskip .6in

\begin{abstract}
\noindent
We study the cosmological collider signatures in the Higgs-$R^2$ inflation model. We consider two distinct types of signals: one originating from the inflaton coupling to Standard Model fermions and gauge bosons, and another arising from the isocurvature mode interaction with the inflaton. In the former case, we determine that the signal magnitude is likely too small for detection by upcoming probes, primarily due to suppression by both the Planck scale and slow-roll parameters. 
However, we provide a detailed computation of the signal which could be potentially applicable to various Higgs inflation variants. For the isocurvature mode signals, we observe that the associated couplings remain unsuppressed when the isocurvature mode is relatively light or comparable to the inflationary scale. In this case, we study the Higgs-$R^2$ inflation parameter space that corresponds to the quasi-single-field inflation regime and find that the signal strength could be as large as $|f_{\rm NL}| > 1$, making Higgs-$R^2$ inflation a viable candidate for observation by future 21-cm surveys.

\end{abstract}

\end{titlepage}
%%%%%%%%%%%%%%%%%%%%%%%%%%%%%%%%%%%%%%%

%%%%%%%%%%%%%%%%%%%%%%%%%%%%%%%%%%%%%%%
\tableofcontents
\renewcommand{\thepage}{\arabic{page}}
\renewcommand{\thefootnote}{$\natural$\arabic{footnote}}
\setcounter{footnote}{0}
%\newpage
\hypersetup{pageanchor=true}
%%%%%%%%%%%%%%%%%%%%%%%%%%%%%%%%%%%%%%%

%%%%%%%%%%%%%%%%%%%%%%%%%%%%%%%%%%%%%%%
\section{Introduction}
\label{sec:introduction}
%%%%%%%%%%%%%%%%%%%%%%%%%%%%%%%%%%%%%%%

Cosmic inflation plays a central role in modern cosmology~\cite{Guth:1980zm,Sato:1980yn,Linde:1981mu,Albrecht:1982wi}.\footnote{For reviews on inflation, see~\cite{Linde:1990flp, Lyth:1998xn, Baumann:2009ds,Baumann:2018muz}.}
In its simplest form, inflation is driven by the potential energy of a slowly rolling scalar field, called the inflaton. Despite its success in explaining the cosmic microwave background (CMB) anisotropy and the seeds of the large-scale structure (LSS), the particle physics origin of the inflaton remains unknown, and unraveling its nature is one of the most important goals in modern cosmology.

Since the Higgs boson is the only elementary scalar field within the Standard Model (SM), it naturally invites speculation about its relationship to inflaton, another scalar field. The original Higgs model of inflation~\cite{Futamase:1987ua,Fakir:1990eg,Cervantes-Cota:1995ehs,Bezrukov:2007ep}, where the SM Higgs is associated with the inflaton,
introduces a large non-minimal coupling of order $\xi \sim \mathcal{O}(10^4)$ 
between the Higgs boson and the Ricci curvature scalar.\footnote{
	In models of critical Higgs inflation, this requirement can be relaxed if the Higgs quartic coupling $\lambda$ is small at the inflationary scale~\cite{Hamada:2014iga,Bezrukov:2014bra}.
} 
As a result, the cut-off scale is lowered to $M_P/\xi$, where $M_P$ is the reduced Planck scale~\cite{Burgess:2009ea,Barbon:2009ya,Burgess:2010zq,Hertzberg:2010dc}. 
% at the vacuum
This scale is lower than the inflationary energy scale and hence it casts doubts on the validity of the model. In particular, while unitarity may be preserved during inflation 
due to the large Higgs field value~\cite{Ferrara:2010in,Bezrukov:2010jz},
it is violated by the production of longitudinal gauge bosons right after inflation 
during preheating~\cite{Ema:2016dny}. This violation occurs due to the mass term arising from the target space curvature in the Einstein frame~\cite{Sfakianakis:2018lzf,Ema:2021xhq}.\footnote{
	This critically depends on the fact that the SM Higgs doublet contains four scalar degrees of freedom. While only the radial direction is important during inflation, the Goldstone modes, or equivalently, the longitudinal gauge bosons, are efficiently produced
	after inflation, leading to a violation of unitarity. For the case involving only a single 
	scalar degree of freedom, see~\cite{Lebedev:2023zgw}.
} 
Therefore, a UV-completion of Higgs inflation is necessary for understanding the inflationary dynamics until the end of preheating and reheating. Inflationary observables, such as the spectral index and the tensor-to-scalar ratio, depend on the reheating temperature via the number of e-folds of inflation, and this presents a critical issue in Higgs inflation.

Higgs-$R^2$ inflation introduces a squared Ricci scalar term, $R^2$, in the action. With an additional scalar degree of freedom---the scalaron---arising from the $R^2$ contribution~\cite{Starobinsky:1980te,Barrow:1983rx,
Whitt:1984pd,Barrow:1988xh},
the model remains perturbative up to $M_P$, as long as $\xi^2/\alpha \lesssim 1$, where $\alpha$ is the coefficient of $R^2$.
This ensures that Higgs-$R^2$ inflation is a UV completion of Higgs inflation~\cite{Ema:2017rqn,Gorbunov:2018llf}. 
A large value of $\alpha$ naturally arises from 
the renormalization group (RG) running~\cite{Salvio:2015kka,Calmet:2016fsr,Ema:2019fdd,Ema:2020evi} when a non-minimal coupling $\xi$ is large. This UV-completion is best understood through the nonlinear sigma model~\cite{Ema:2020zvg,He:2023vlj,He:2023fko}.
In this framework, the target space encompasses the conformal mode of the metric and remains invariant under the Weyl transformation,
with the scalaron identified with the sigma meson that flattens the target space of Higgs inflation.

In Higgs-$R^2$ inflation, the inflaton is a mixture of the Higgs and the scalaron. Consequently, the inflaton naturally couples
to SM particles through both its Higgs component and the conformal factor, $\Omega$. A distinguishing feature of Higgs-$R^2$ inflation is that these couplings are all explicitly given, allowing for unambiguous study of their effects. Specifically, the preheating and reheating \emph{after} inflation, associated with these couplings, have been thoroughly investigated in~\cite{He:2018mgb,Bezrukov:2019ylq,He:2020ivk,Bezrukov:2020txg,He:2020qcb,Aoki:2022dzd}.

In this paper, we study the cosmological signatures arising from the couplings between the inflationary sector and the SM particles \emph{during} inflation. It is well known that a particle with a mass comparable to the Hubble parameter leaves a unique imprint in the squeezed limit of the non-gaussianity of the curvature perturbation when it couples to the inflatonary sector. These signals are referred to as cosmological collider signatures~\cite{Chen:2009we,Chen:2009zp,Baumann:2011nk,Assassi:2012zq,Chen:2012ge,Pi:2012gf,Noumi:2012vr,Arkani-Hamed:2015bza}.

In this context, SM particles are particularly interesting. In Higgs-$R^2$ inflation, even though the Higgs field value is typically much larger than the Hubble parameter during inflation, there still exist particles with masses as light as the Hubble parameter
due to the large hierarchy of the Yukawa couplings~\cite{Chen:2016nrs,Chen:2016uwp,Chen:2016hrz}. Furthermore, the inflaton sector now contains multiple scalar fields, namely the Higgs and the scalaron, and the isocurvature mode could potentially give rise to additional cosmological collider signatures. Therefore, the main aim of this paper is to investigate these cosmological collider signatures arising from the SM particles. While we show that these signatures from the SM fermions and gauge bosons are too small to be observable in the near future, the isocurvature mode may produce a substantial effect, detectable by future 21 cm observations~\cite{Chen:2016zuu,Meerburg:2016zdz}.

The remainder of this paper is organized as follows: In Section~\ref{sec:preliminary}, we review Higgs-$R^2$ inflation, with a focus on the coupling between the SM particles and the inflaton sector, as well as the isocurvature mode. We then compute the cosmological collider signatures of the SM fermions and gauge bosons in Section~\ref{sec:CC_psiA}, and those of the isocurvature mode in Section~\ref{sec:CC_isocurvature}. Finally, we summarize our findings in Section~\ref{sec:conclusions}.
We aim to keep our discussion free of technical details, focusing solely on key results, as the computations are quite involved and may obscure the main findings. Instead, all technical details are provided in the appendices. In Appendix~\ref{app:conventions}, we summarize the conventions and notation used throughout this paper. Subsequently, we review the covariant formalism of multi-field inflation in Appendix~\ref{app:covariant_formalism}. The results of this appendix are extensively used in the computation of the cosmological collider signatures of the isocurvature mode. Finally, we derive the propagators for scalars, fermions, and gauge bosons in de~Sitter spacetime in Appendix~\ref{app:propagators}. 
%%%%%%%%%%%%%%%%%%%%%%%%%%%%%%%%%%%%%%%
\section{Preliminaries}
\label{sec:preliminary}
%%%%%%%%%%%%%%%%%%%%%%%%%%%%%%%%%%%%%%%

In this section, we briefly review the inflationary dynamics of 
the Higgs-$R^2$ model~\cite{Ema:2017rqn,Wang:2017fuy,He:2018gyf,Gundhi:2018wyz,Enckell:2018uic}.
We pay particular attention to the couplings between the inflaton field and other SM particles, as these are essential for understanding cosmological collider signatures.

%%%%%%%%%%%%%%%%%%%%%%%%%%%%%%%%%%%%%%%
\subsection{Higgs-$R^2$ model and Weyl transformation}
\label{subsec:Weyl}
%%%%%%%%%%%%%%%%%%%%%%%%%%%%%%%%%%%%%%%

The action of Higgs-$R^2$ inflation in the Jordan frame is given by
\begin{align}
    \label{eq:actionjord}
	S \; = \; \int d^4x \sqrt{-g}\left[
	\frac{M_P^2}{2}R\left(1+\frac{2\xi \vert \Phi\vert^2}{M_P^2}\right) + \alpha R^2
	+ g^{\mu\nu}(D_\mu \Phi)^\dagger (D_\nu \Phi) - \lambda \vert \Phi\vert^4
	\right]
	+ S_{\psi A} \, ,
\end{align}
where $g_{\mu\nu}$ is the spacetime metric and $g$ is its determinant, 
$R$ is the Ricci curvature scalar, $\Phi$ represents the Higgs doublet, and $\lambda$ is the Higgs quartic coupling. The covariant derivative is defined as
\begin{align}
	D_\mu \Phi \; = \; \left(\partial_\mu - i\frac{g}{2}\tau^a W^a_\mu + i\frac{g'}{2}B_\mu\right)\Phi \, ,
\end{align}
where $W^a_\mu$ and $B_\mu$ are the SU(2) and U(1) gauge bosons, with $g$ and $g'$ their gauge couplings, respectively,
and $\tau^a$ is the Pauli matrix.
Although the $R^2$ term induces the Higgs mass term and the cosmological constant due to the RG running,
its effect is suppressed compared to the terms given above during inflation~\cite{Ema:2020evi}. Therefore, we omit these terms in the following discussion. The matter action is given by
\begin{align}
    \label{eq:matsecjord}
	S_{\psi A}
	 = \int d^4x \sqrt{-g}&\left[i\bar{\psi}_i\slashed{\nabla}\psi_i - 
	Y_{l}^{ij} \bar{L}_i \Phi l_{Rj} - Y_d^{ij} \bar{Q}_i \Phi d_{Rj} - Y_{u}^{ij}\bar{Q}_i \tilde{\Phi} u_{Rj}
	%\right. \nonumber \\ &\left.
	-\frac{1}{4}W_{\mu\nu}^a W^{a\mu\nu}
	-\frac{1}{4}B_{\mu\nu}B^{\mu\nu}
	\right]
	+ \cdots,
\end{align}
where $\psi_i$ includes all the SM fermions, $Q_i$ is the left-handed quark doublet, $L_i$ is the left-handed lepton doublet,
$u_{Ri}$ and $d_{Ri}$ are the right-handed up-type and down-type quarks, $l_{Ri}$ are the right-handed leptons,
and $\tilde{\Phi} \equiv i\tau^2 \Phi^*$.
The Yukawa coupling matrices are given by $Y_{l}^{ij}$, $Y_d^{ij}$, and $Y_u^{ij}$, with the flavor indices $i$ and $j$.
We omit the hermitian conjugate of the Yukawa interaction, 
as well as the gluon kinetic term and the interaction between the fermions and gauge bosons that are irrelevant to our study.
The covariant derivative for fermions is given by
\begin{align}
	\slashed{\nabla}\psi = e^{\mu}_a \gamma^a \left(\partial_\mu + \frac{1}{4}\omega_{\mu}^{bc}\gamma_{bc}\right)\psi,
\end{align}
where $e_{\mu}^{a}$ is the tetrad defined by $g_{\mu \nu} = \eta_{ab} e_{\mu}^a e_{\nu}^b$, $\omega_{\mu}^{ab}$ is the spin connection, and $\gamma^{ab} \equiv \frac{1}{2} \left[\gamma^a, \gamma^b \right]$ (see App.~\ref{app:conventions} for more details). We use the Latin characters for both the gauge indices and local Lorentz indices; however, their distinction is clear from the context.

The action~(\ref{eq:actionjord}) is defined in the Jordan frame. We now transform to the Einstein frame by performing a Weyl transformation,
as it is more convenient for the analysis of inflation. We first introduce an auxiliary field $\tilde{\sigma}$ to extract the scalar degree of freedom from the $R^2$ term. In this case, the action becomes~\cite{Whitt:1984pd, Maeda:1988ab}
\begin{align}
	S \; = \; \int d^4 x \sqrt{-g}\left[
	\frac{M_P^2}{2}R\left(1+\frac{2\xi \vert \Phi\vert^2 + 4\tilde{\sigma}}{M_P^2}\right) -\frac{\tilde{\sigma}^2}{\alpha}
	+ g^{\mu\nu}(D_\mu \Phi)^\dagger (D_\nu \Phi) - \lambda \vert \Phi\vert^4
	\right]
	+ S_{\psi A} \, .
\end{align}
We perform the Weyl transformation by redefining the metric as
\begin{align}
	g_{\mu\nu} \to \Omega^{-2}g_{\mu\nu},
	\quad~\mathrm{where}~\quad
	\Omega^2 \; = \; 1+\frac{2\xi \vert \Phi\vert^2 + 4\tilde{\sigma}}{M_P^2} \, .
\end{align}
The Ricci scalar transforms under the Weyl transformation according to Eq.~(\ref{eq:riccitrans})\footnote{Further details related to the Weyl transformation are provided in App.~\ref{app:conventions}}, and the action in the Einstein frame becomes
\begin{align}
	S \; = \; \int d^4 x \sqrt{-g}\left[
	\frac{M_P^2}{2}R
	+ \frac{1}{2}g^{\mu\nu}\partial_\mu \sigma \partial_\nu \sigma
	+ e^{-\sqrt{\frac{2}{3}}\frac{\sigma}{M_P}}g^{\mu\nu}(D_\mu \Phi)^\dagger (D_\nu \Phi) - V(\sigma, \Phi)
	\right]
	+ S_{\psi A} \, ,
\end{align}
where the scalaron field $\sigma$ is defined as 
\begin{equation}
    \label{eq:omegasigma}
    \frac{\sigma}{M_P} \; \equiv \; \sqrt{6} \ln{\Omega} (\tilde{\sigma}, \Phi) \, ,
\end{equation}
and the scalar potential is given by
\begin{align}
	V(\sigma, \Phi) \; = \; e^{-\sqrt{\frac{8}{3}}\frac{\sigma}{M_P}}
	\left[\lambda \vert \Phi\vert^4
	+ \frac{M_P^4}{16\alpha}
	\left(e^{\sqrt{\frac{2}{3}}\frac{\sigma}{M_P}}-1-\frac{2\xi \vert \Phi\vert^2}{M_P^2}\right)^2\right] \, .
\end{align}
By redefining the fermions as $\psi_i \to \Omega^{3/2}\psi_i$, the matter sector action~(\ref{eq:matsecjord}) in the Einstein frame becomes
\begin{align}
	S_{\psi A}
	\; = \; \int d^4x \sqrt{-g}&\left[i\bar{\psi}_i\slashed{\nabla}\psi_i - 
	\frac{1}{\Omega}\left(Y_{l}^{ij} \bar{L}_i \Phi l_{Rj} + Y_d^{ij} \bar{Q}_i \Phi d_{Rj} + Y_{u}^{ij}\bar{Q}_i \tilde{\Phi} u_{Rj}
	\right)
	%\right. \nonumber \\ &\left.
	-\frac{1}{4}W_{\mu\nu}^a W^{a\mu\nu}
	-\frac{1}{4}B_{\mu\nu}B^{\mu\nu}
	\right] \, ,
\end{align}
where we only keep the terms that are relevant to our study. We work in the unitary gauge and take the Higgs doublet as
$\Phi = (0, h/\sqrt{2})^T$. The electroweak (EW) gauge bosons are defined as
\begin{align}
	W_\mu^\pm = \frac{1}{\sqrt{2}}(W_\mu^1 \mp i W_\mu^2),
	\quad
	Z_\mu = \cos\theta_W W_\mu^3 + \sin\theta_W B_\mu,
    \quad
	\sin\theta_W = \frac{g'}{\sqrt{g^2 + g'^2}}.
\end{align}
The final action in the Einstein frame that we use for our analysis is given by
\begin{align}
    \label{eq:acteinstein}
	S \; = \; \int d^4x \sqrt{-g}
	&\left[
	\frac{M_P^2}{2}R + \frac{1}{2}(\partial \sigma)^2 + \frac{1}{2\Omega^2}(\partial h)^2 - V(\sigma, h)
	+\bar{l}_i\left(i\slashed{\nabla}-m_{l_i}\right)l_i
	+\bar{u}_i\left(i\slashed{\nabla}-m_{u_i}\right)u_i
	\right. \nonumber \\ &\left.
	+\bar{d}_i\left(i\slashed{\nabla}-m_{d_i}\right)d_i
	-\frac{1}{2}W^+_{\mu\nu}W^{-\mu\nu} + {m_W^2}W^+_{\mu}W^{-\mu}
	-\frac{1}{4}Z_{\mu\nu}Z^{\mu\nu} + \frac{m_Z^2}{2}Z_\mu Z^\mu
	\right] \, .
\end{align}
Here the mass terms are the functions of $\sigma$ and $h$, and can be expressed as
\begin{align}
	m_{l_i} \; = \; \frac{y_{l_i} h}{\sqrt{2}\Omega} \, ,
	\quad
	m_{u_i} \; = \; \frac{y_{u_i} h}{\sqrt{2}\Omega}\, ,
	\quad
	m_{d_i} \; = \; \frac{y_{d_i} h}{\sqrt{2}\Omega}\, ,
	\quad
	m_W^2 \; = \; \frac{g^2 h^2}{4\Omega^2}\,,
	\quad
	m_Z^2 \; = \; \frac{m_W^2}{\cos^2\theta_W}\,,
\end{align}
and the scalar potential is given by
\begin{align}
	V(\sigma, h) \; = \; e^{-\sqrt{\frac{8}{3}}\frac{\sigma}{M_P}}
	\left[\frac{\lambda}{4}h^4
	+ \frac{M_P^4}{16\alpha}
	\left(e^{\sqrt{\frac{2}{3}}\frac{\sigma}{M_P}}-1-\frac{\xi h^2}{M_P^2}\right)^2\right] \, .
\end{align}
Here, we rotate the fermions to the mass eigenbasis, making their mass matrices diagonal with eigenvalues $m_{\psi_i}$. As before, we keep only the terms relevant to our study. The Einstein frame action~(\ref{eq:acteinstein}) is our starting point of the analysis.
Note that the inflaton sector couples to the SM fermions and gauge bosons solely through the mass terms, that appear in the combination of $h/\Omega$ terms.

\begin{figure}[t]
	\centering
 	\includegraphics[width=0.45\linewidth]{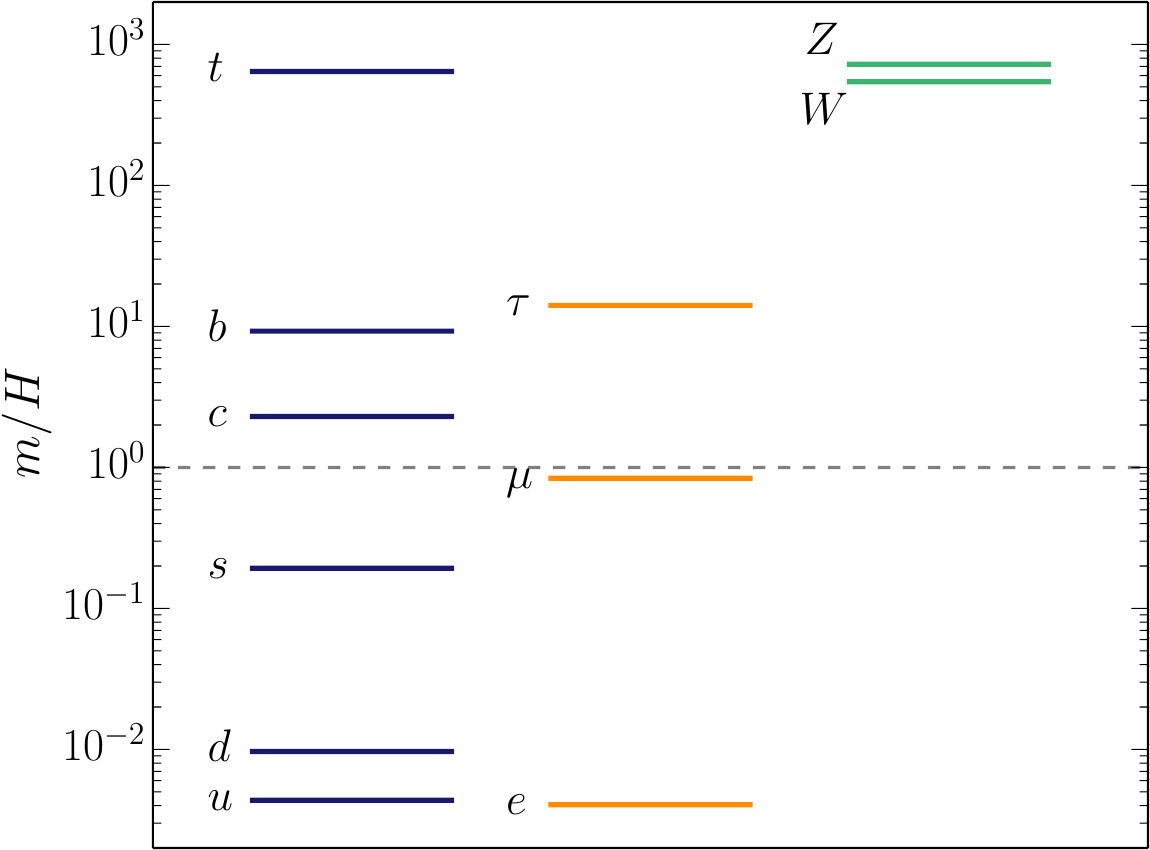}
	\caption{\small An example of the SM mass spectrum during inflation.}
	\label{fig:SM_spectrum}
\end{figure}

In Fig.~\ref{fig:SM_spectrum}, we illustrate the SM mass spectrum during inflation by solving the RG equations
with \texttt{SARAH}~\cite{Staub:2013tta}, where the coupling values at the electroweak (EW) scale are taken from Ref.~\cite{Buttazzo:2013uya}\footnote{
	Famously, with the current central values of the Higgs mass and the top quark mass,
	the Higgs quartic coupling becomes negative at an intermediate scale, 
	$\mu \sim 10^{10}\,\mathrm{GeV}$~\cite{Degrassi:2012ry,Buttazzo:2013uya}.
	To avoid this, we take the top quark mass to be light, $m_t = 170.5\,\mathrm{GeV}$, 
	ensuring that  $\lambda > 0$ up to the inflationary scale.
	This choice is only for illustration; any new physics below the inflationary scale could potentially stabilize the EW vacuum for a larger value of the top quark mass.
}
(see Sec.~\ref{subsec:inflation} for the discussion of the size of the Higgs field value and the Hubble parameter during inflation).
As an example, we choose the parameter $\alpha = 3\times 10^{8}$ and set the RG scale to $\mu = H \simeq 1.4 \times 10^{13} \, \rm{GeV}$ for this plot. We ignore the scalaron's contribution to the runnings, as it becomes important only above the scalaron mass scale.
The figure shows that even though the Higgs field value is large compared to the Hubble parameter during inflation, with $h/\Omega H \sim 2\times 10^{3}$ for our chosen parameters, the SM fermions can be as light as the Hubble parameter due to the large hierarchy between the Yukawa couplings. Consequently, these particles could produce significant cosmological collider signatures.

%%%%%%%%%%%%%%%%%%%%%%%%%%%%%%%%%%%%%%%
\subsection{Inflationary predictions}
\label{subsec:inflation}
%%%%%%%%%%%%%%%%%%%%%%%%%%%%%%%%%%%%%%%

%%
\begin{figure}[t]
	\centering
 	\includegraphics[width=0.45\linewidth]{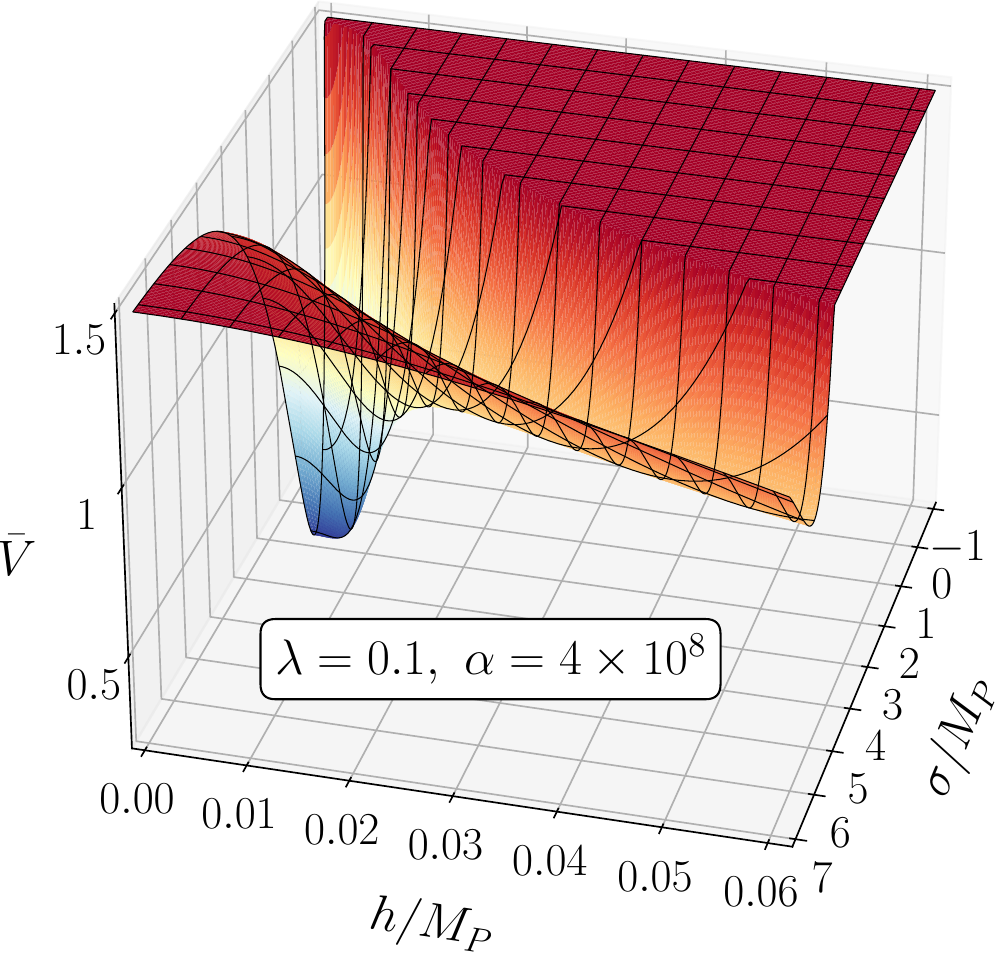}
	\hspace{2.5mm}
	\includegraphics[width=0.45\linewidth]{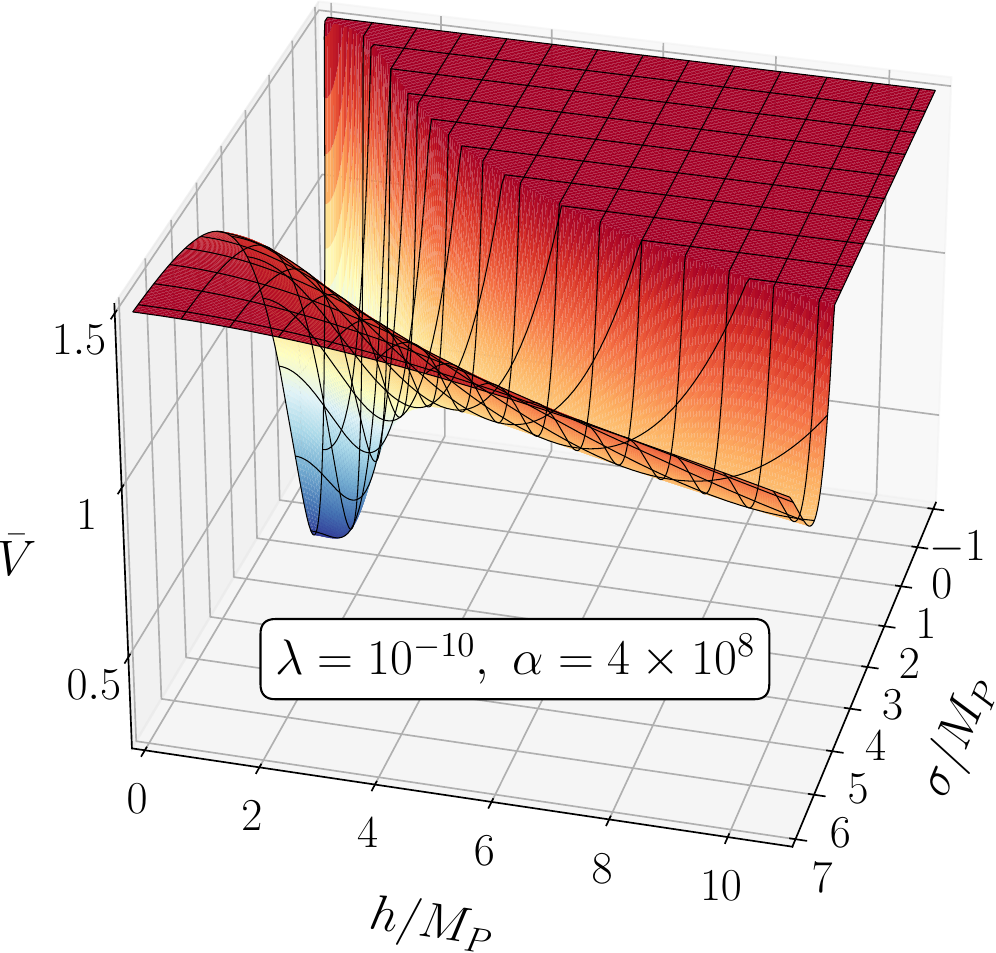}
	\caption{\small The scalar potential in the Einstein frame. The $z$-axis represents the normalized potential $\bar{V}$, where $V = \lambda M_P^4 \bar{V}/4(\xi^2 + 4\lambda \alpha)$, with a cutoff at $\bar{V} = 1.5$.
	We have set the parameters as $(\lambda, \alpha) = (0.1, 4\times 10^{8})$, yielding $\xi \simeq 9\times 10^{3}$ 
	due to the CMB normalization with $N_e = 60$ in the left panel, and $(\lambda, \alpha) = (10^{-10}, 4\times 10^{8})$ 
	corresponding to $\xi \simeq 0.3$ in the right panel, respectively. A valley-like structure is observable, as defined by Eq.~\eqref{eq:valle_approx}, along which inflation occurs. The typical value of the Higgs field is $h \sim M_P/\sqrt{\xi}$, and this value depends on the magnitude of $\lambda$, which in turn affects the curvature scale of the isocurvature direction and the turning rate.
 }
	\label{fig:pot}
\end{figure}
Here, we review the inflationary dynamics of the Higgs-$R^2$ model~\cite{Ema:2017rqn,Wang:2017fuy,He:2018gyf,Gundhi:2018wyz,Enckell:2018uic}. In this model, the inflaton is a combination of the Higgs field and the scalaron, with inflation occurring in the region where $\sigma \gg M_P$.
We use the approximation that in the Einstein frame, the inflationary potential takes a valley-like form~(see Fig.~\ref{fig:pot}), allowing us to use a single-field slow-roll approximation, given by~\cite{Wang:2017fuy, He:2018gyf}
\begin{align}
	\frac{\partial V}{\partial h} \; = \; 0
	~~\Longrightarrow~~
	h_0^2 \; = \; \frac{\xi M_P^2}{\xi^2 + 4\lambda \alpha}\left(e^{\sqrt{\frac{2}{3}}\frac{\sigma_0}{M_P}}-1\right)\,,
	\label{eq:valle_approx}
\end{align}
where the subscript ``$0$" denotes background values.\footnote{
	If the turning rate of the inflationary trajectory is large, as in the case of the small $\lambda$ scenario that we consider later, the inflationary dynamics can deviate from this original minimum.
	Since the potential derivatives in the isocurvature direction are suppressed by a large value of $\alpha$,
	the effect of this shift is minor and we ignore it in the following discussion.
}
Ignoring the effects of the isocurvature mode, the background action is given by
\begin{align}
	S_0 \; = \; \int d^4x \sqrt{-g} \left[
	\frac{M_P^2}{2}R + \frac{1}{2}\left(1+c^2\right)\dot{\sigma}_0^2
	-\frac{\lambda M_P^4}{4(\xi^2 + 4\lambda \alpha)}\left(1-e^{-\sqrt{\frac{2}{3}}\frac{\sigma_0}{M_P}}\right)^2
	\right],
	\quad
	c^2 \; = \; \frac{\xi}{6(\xi^2+4\lambda \alpha)} \, ,
\end{align}
where terms suppressed by $e^{-\sqrt{\frac{2}{3}}\frac{\sigma_0}{M_P}}$ in the kinetic term were neglected. We focus on the parameter region where $c \lesssim 1$, necessary for a single-field slow-roll approximation as we see later.
Using the principal CMB observables from the \textit{Planck} analysis~\cite{Planck:2018vyg}, we find that the CMB normalization fixes the parameters according to
\begin{align}
	\left(4\alpha + \frac{\xi^2}{\lambda}\right)\left(1+c^2\right) \; \simeq \; 2.4\times 10^{9}\times \left(\frac{N_e}{60}\right)^2 \, ,
	\label{eq:COBE_norm}
\end{align}
where $N_e$ represents the number of e-folds of inflation, given by
\begin{align}   
    \label{eq:numofefolds}
	N_e \; = \; \frac{3(1+c^2)}{4}e^{\sqrt{\frac{2}{3}}\frac{\sigma_0}{M_P}} \, ,
\end{align}
and the spectral index and tensor-to-scalar ratio are given by
\begin{align}
	n_s \; \simeq \; 1 - \frac{2}{N_e},
	\qquad
	r \; \simeq \; \frac{12}{N_e^2}(1+c^2) \, .
\end{align}
The spectral index is consistent with \textit{Planck} observations~\cite{Planck:2018vyg},
and the tensor-to-scalar ratio is approximately $\mathcal{O}(10^{-3})$, and within reach of future experiments like CMB-S4~\cite{CMB-S4:2016ple} and LiteBIRD~\cite{LiteBIRD:2022cnt}.
Furthermore, the inflationary dynamics is stable against possible Planck suppressed operators~\cite{Lee:2023wdm}.

The CMB normalization~\eqref{eq:COBE_norm} 
requires a large value of $\alpha$ and/or $\xi^2/\lambda$.
If $\alpha \gg \xi^2/\lambda$, the inflaton is predominantly driven by the scalaron, whereas in the other case, it is primarily driven by the SM Higgs field. In the remainder of this paper, we focus on scenarios where there is no large hierarchy between $\alpha$ and $\xi^2/\lambda$,
allowing both the Higgs and the scalaron to contribute to the inflationary sector. This requires a large value of $\xi \sim 10^{4}$, provided that the coupling $\lambda$ is not too small. While this represents the  ``standard" parameter region of the Higgs-$R^2$ model most often discussed in the literature,
we also explore cases where $\xi \gg 1$ and $\xi \lesssim \mathcal{O}(0.1)$, with a small value of $\lambda$ at the inflationary scale.
The latter scenario is especially important when we examine the cosmological collider signatures of the isocurvature mode.

%%%%%%%%%%%%%%%%%%%%%%%%%%%%%%%%%%%%%%%
\subsection{Covariant formalism and isocurvature mode}
\label{subsec:isocurvature}
%%%%%%%%%%%%%%%%%%%%%%%%%%%%%%%%%%%%%%%

Higgs-$R^2$ inflation has two scalar degrees of freedom, adiabatic and isocurvature modes.
To study the mode dynamics, we use the covariant formalism of multi-field inflation~\cite{Sasaki:1995aw,
GrootNibbelink:2000vx,GrootNibbelink:2001qt,Langlois:2008mn,Peterson:2010np,Gong:2011uw,
Kaiser:2012ak}.
We write down the action in the Einstein frame as
\begin{align}
	S \; = \; \int d^4x \sqrt{-g}\left[\frac{M_P^2}{2}R + \frac{1}{2}h_{ab}\partial_\mu \phi^a \partial^\mu \phi^b - V
	\right] \, ,
\end{align}
where $\phi^1 = \sigma$ and $\phi^2 = h$. The target space metric in our case is given by
\begin{align}
	h_{ab} \; = \; \mathrm{diag}\left(1, e^{-\sqrt{\frac{2}{3}}\frac{\sigma}{M_P}}\right) \, .
\end{align}
The quadratic action of the curvature mode $\zeta$ and the isocurvature mode $\chi$, both of which are mixtures of the Higgs field and the scalaron, is given by
\begin{align}
    \label{eq:quadraticact1}
	S_\mathrm{quad} \; = \; \int d^4x a^3\left[\frac{1}{2}\frac{\dot{\phi}_0^2}{H^2} 
	\left(\dot{\zeta}^2-\frac{1}{a^2}\left(\partial_i \zeta\right)^2\right)
	+ \frac{1}{2}\left(\dot{\chi}^2 - \frac{1}{a^2}\left(\partial_i \chi\right)^2
	-m_\chi^2 \chi^2\right)
	-\frac{2\dot{\theta} \dot{\phi}_0}{H} \dot{\zeta}\chi\right] \, .
\end{align}
Here, the velocity of the background field is defined as $\dot{\phi}_0 = \sqrt{h_{ab}\dot{\phi}_0^a\dot{\phi}_0^b}$, and $\dot{\theta}$ is the turning rate that parametrizes the curvature of the inflationary trajectory, or the mixing between adiabatic and isocurvature modes. The mass of the isocurvature mode is defined as
\begin{align}
    \label{eq:massisocurv}
	m_\chi^2 \; = \; N^a N^b \nabla_b V_a + \frac{\dot{\phi}_0^2}{2}\mathcal{R} - \dot{\theta}^2,
	\qquad
	\mathcal{R} \; = \; \mathcal{R}_{ab}h^{ab} \; = \; \mathcal{R}_{acbd}h^{ab}h^{cd} \, ,
\end{align}
with $\nabla_a$ and $\mathcal{R}_{acbd}$ constructed from the target space metric $h_{ab}$. Further details of the derivation are provided in Appendix~\ref{app:covariant_formalism}. We assume that the inflaton field value and its velocity vary along the inflationary valley, while during inflation the Higgs field evolves according to~\eqref{eq:valle_approx}. In this case, the adiabatic and isocurvature directions are defined as
\begin{align}
    \label{eq:adandisodir}
	T^a \; = \; \frac{1}{\sqrt{1+c^2}}\begin{pmatrix} 1 \\ c\,e^{\sigma_0/\sqrt{6}M_P} \end{pmatrix},
	\quad
	N^a = \frac{1}{\sqrt{1+c^2}}\begin{pmatrix} -c \\ e^{\sigma_0/\sqrt{6}M_P} \end{pmatrix} \, .
\end{align}
We ignore terms suppressed by $1/N_e$ here and in subsequent discussion.
By substituting the explicit forms of the target space metric and potential, we find
\begin{align}
	N^a N^b \nabla_b V_a \; \simeq \; \frac{\xi (24\lambda \alpha +\xi(1+6\xi))}{\lambda \alpha}H^2,
	\quad
	\mathcal{R} = -\frac{1}{3M_P^2} \, ,
	\quad
	\frac{\dot{\theta}}{H} \; \simeq \; 3c
	\; = \; \sqrt{\frac{3\xi}{2(4\lambda\alpha +\xi^2)}} \, .
\end{align}
In the standard parameter region of the Higgs-$R^2$ model, $\xi \gg 1$ and $\lambda \sim \mathcal{O}(0.1)$, the isocurvature mode is heavy and the turning rate is small. Consequently, the isocurvature mode can be ignored for both cosmological perturbation and cosmological collider signature studies. On the other hand, if we take the parameters as
\begin{align}
	\xi \sim \lambda \alpha \lesssim \mathcal{O}(0.1),
\end{align}
we have $m_\chi^2/H^2 \sim \mathcal{O}(1)$ and $c \sim \mathcal{O}(0.1)$ 
making the isocurvature mode light and the turning rate significant. This corresponds to the so-called quasi-single field inflation regime~\cite{Chen:2009we,Chen:2009zp}.\footnote{
	If the isocurvature mode is light enough, with $m_\chi \lesssim H$,
	the model becomes a true multi-field type. The inflationary dynamics for this case has been studied in~\cite{Gundhi:2018wyz},
	and we do not consider this regime in this paper.
}
The observational effects of this regime are the main focus of Sec.~\ref{sec:CC_isocurvature}.
%%%%%%%%%%%%%%%%%%%%%%%%%%%%%%%%%%%%%%%
\subsection{Order of magnitude estimation}
\label{subsec:estimation}
%%%%%%%%%%%%%%%%%%%%%%%%%%%%%%%%%%%%%%%

Before proceeding to the actual computation, we estimate the order of magnitude of cosmological collider signals arising from the SM particles. Future experiments on large-scale structure~\cite{Amendola:2016saw} and 21 cm surveys~\cite{Meerburg:2016zdz, Furlanetto:2006jb} are expected to constrain the primordial bispectrum down to values of $f_{NL} \sim 1$, 
potentially probing non-Gaussianities even below this value. As previously demonstrated, the inflaton couples to SM fermions and gauge bosons via a coupling of the form $h/\Omega$. By expanding this function, we see that the coupling of the inflaton to the SM fermions and gauge bosons is of the form
\begin{align}
    \label{eq:fermgaugebosonact1}
	{S}_\mathrm{int} \; = \; \int d^dx \sqrt{-g}\,\mathcal{O}_\mathrm{SM}
	\left(1+c_1\frac{\varphi}{N_e M_P} + c_2 \frac{\varphi^2}{2M_P^2} + \cdots\right) \,,
\end{align}
where
\begin{equation}
    \mathcal{O}_\mathrm{SM} \; = \; 
    \begin{cases}
         -m \bar{\psi}\psi\,, & \textrm{Fermions}\, , \\ 
        \displaystyle \frac{m^2}{2}g^{\mu\nu}A_\mu A_\nu\,, & \textrm{Gauge bosons} \, ,
    \end{cases}
\end{equation}
and $c_1, c_2$ are constants of order unity. To estimate the size of non-Gaussianity and, consequently, the signal, we first combine the linear and quadratic couplings to form the three-point function, which leads to an overall factor of $1/N_e M_P^3$. The non-Gaussianity scales as $f_{NL} \propto P_{\zeta}^{-2} \times H^3/\dot{\phi}_0^3$,\footnote{This approximation is valid only when the couplings between the inflaton and the SM particles do not have any derivative couplings.} where $P_\zeta = H^4/4\pi^2 \dot{\phi}_0^2 \simeq 2.1\times 10^{-9}$ is the amplitude of the power spectrum fixed by the CMB normalization. Lastly, since the coupling depends on the mass, and the cosmological collider signatures are maximized when the mass is comparable to the Hubble parameter, with $m \sim H$, we obtain a simple estimate for non-Gaussianity using dimensional analysis
\begin{align}
	\vert f_\mathrm{NL}\vert \; \lesssim \; \frac{\dot{\phi}_0H}{N_e M_P^3}
	\; \sim \;  3.5\times 10^{3}\frac{H^3}{N_e M_P^3} \, .
	\label{eq:estimate_psiA}
\end{align}
Due to the suppression by the Planck scale, this value is too small to be observed by upcoming observations. Nevertheless, we rigorously compute this signal in Sec.~\ref{sec:CC_psiA} to verify the accuracy of our estimation.

As discussed in Sec.~\ref{subsec:isocurvature}, the isocurvature mode is too heavy to leave a signal when $\xi \gg 1$. Therefore, when computing the cosmological signatures, we focus on the case $\xi \sim \lambda \alpha \lesssim 1$, which corresponds to a light isocurvature mode and the quasi-single field regime. The isocurvature mode mixes with the inflaton through the turning rate $\dot{\theta}$ and also couples to the inflaton via the potential. In particular, the derivative with respect to the isocurvature direction $N^a$ is not necessarily suppressed by the slow-roll parameters, and we find that the potential induces couplings of the form
\begin{align}   
    \label{eq:actioncubic}
	S_\mathrm{cubic} \; = \; \int d\tau d^3x\,a^4 \left[
	-\frac{1}{6}V_{N^3}\chi^3 -\frac{1}{2}V_{T^2N}\varphi^2 \chi 
	\right] \, ,
\end{align}
where $d \tau = d t/a$ is the conformal time, and 
\begin{align}
	\frac{V_{N^3}}{H} \; \sim \; \frac{V_{T^2N}}{H} \; \sim \; \alpha^{-1/2} \, ,
\end{align}
where we assumed $\xi \sim \lambda \alpha \sim 1$
(see Sec.~\ref{sec:CC_isocurvature} for the precise forms of the couplings). 
Since Higgs-$R^2$ inflation requires $\alpha \sim 10^{9}$, the couplings are relatively suppressed compared to the Hubble parameter. However, this suppression is compensated by the smallness of the power spectrum. The non-Gaussianity resulting from these couplings can be estimated as 
\begin{equation}
\label{eq:estimate_isocurvature}
    f_\mathrm{NL} \; \sim \; 
    \begin{cases}
         \displaystyle P_\zeta^{-1/2} \frac{V_{N^3}}{H} \left(\frac{\dot{\theta}}{H}\right)^3\,, & \textrm{for}~\chi^3\, , \vspace{1.5mm}\\ 
        \displaystyle  P_\zeta^{-1/2}\frac{V_{T^2N}}{H} \frac{\dot{\theta}}{H}\,, & \textrm{for}~\varphi^2 \chi \, ,
    \end{cases}
\end{equation}
where the factor $P_\zeta^{-1/2}$ arises from the definition of the non-Gaussianity. Since $P_\zeta\times \alpha \sim \mathcal{O}(1)$ and the turning rate $\dot{\theta}$ can be of the order of the Hubble parameter, 
we expect that non-Gaussianity could potentially be as large as order unity, thus yielding an observable signature.
In Sec.~\ref{sec:CC_isocurvature}, we compute the cosmological collider signatures of the isocurvature mode in detail to confirm this expectation.
%%%%%%%%%%%%%%%%%%%%%%%%%%%%%%%%%%%%%%%
\section{Cosmological collider signatures of fermions and gauge bosons}
\label{sec:CC_psiA}
%%%%%%%%%%%%%%%%%%%%%%%%%%%%%%%%%%%%%%%

In Higgs-$R^2$ inflation, the inflaton naturally couples to the SM fermions and gauge bosons, potentially giving rise to cosmological collider signatures. The inflaton couples to the fermions and gauge bosons through the coupling of the form $h/\Omega$, as demonstrated in Sec.~\ref{subsec:Weyl}, and this coupling is expanded as
\begin{align}
	\frac{h}{\Omega} \; = \; \frac{h_0}{\Omega_0} + \varphi^a \nabla_a \left(\frac{h}{\Omega}\right)_0
	+ \frac{1}{2}\varphi^a \varphi^b \nabla_b \nabla_a \left(\frac{h}{\Omega}\right)_0 + \cdots \, ,
\end{align}
where the subscript ``0" indicates that the quantities are evaluated using background field values 
and $\varphi^a = \phi^a - \phi^a_0$. 
We provide technical details related to this expansion in target space in App.~\ref{app:covariant_formalism}. Focusing on the adiabatic direction $T^a$, from Eqs.~(\ref{eq:omegasigma}) and (\ref{eq:adandisodir}), we obtain
\begin{align}
	\frac{h}{\Omega} \; \simeq \; \frac{h_0}{\Omega_0}\left(1 + \sqrt{\frac{4\alpha\lambda + \xi^2}{24\alpha \lambda + \xi(1+6\xi)}}
	e^{-\sqrt{\frac{2}{3}}\frac{\sigma_0}{M_P}}\frac{\varphi}{2M_P}
	+ \frac{\varphi^2}{12M_P^2}\right) \, ,
\end{align}
where we only kept the leading order terms in the slow-roll expansion. In this section, we focus on the ``standard" parameter region of the Higgs-$R^2$ model, with $\lambda\alpha \sim \xi^2 \gg \xi$. In this limit, the above expression simplifies to
\begin{align}
	\frac{h}{\Omega} \; \simeq \; \frac{h_0}{\Omega_0}
	\left(1 + \frac{\sqrt{6}}{16 N_e}\frac{\varphi}{M_P} + \frac{\varphi^2}{12M_P^2}\right) \, ,
\end{align}
where we used the number of e-folds~$(\ref{eq:numofefolds})$. We note that the linear term is suppressed by the slow-roll parameter, or equivalently $1/N_e$, while the quadratic term is not suppressed.
The coupling between the inflaton and SM fermions/gauge bosons is represented by the action~(\ref{eq:fermgaugebosonact1}), with $(c_1, c_2) = (\sqrt{6}/16, 1/6)$ in the case of fermion and $(\sqrt{6}/8, 1/3)$ 
for the gauge bosons, respectively.
Here the mass $m$ is evaluated using the background field values of $\sigma$ and $h$.

In the following analysis, we evaluate the contributions of SM fermion and gauge bosons to the three-point functions. 
The computation is analogous to Ref.~\cite{Chen:2016hrz}.
We have two contributions: one with two $\mathcal{O}_\mathrm{SM}$ insertions and another one with three $\mathcal{O}_\mathrm{SM}$ insertions. Given that the linear coupling is suppressed by an additional factor of $1/N_e$, our focus is on the former case. We are primarily interested in the squeezed limit of the bispectrum, with $k_3 \ll k_1, k_2$, 
and thus only consider the diagram containing $\hat{\varphi}_{\vec{k}_3}$ on the linear side and $\hat{\varphi}_{\vec{k}_1}\hat{\varphi}_{\vec{k}_2}$ on the quadratic side of the vertices.
Lastly, since the cosmological collider signatures are free of UV divergences, there is no need for a regularization scheme to evaluate the diagrams. Bearing these points in mind, the diagram of interest can be expressed as follows:
\begin{align}
	\langle \hat{\varphi}_{\vec{k}_1}\hat{\varphi}_{\vec{k}_2}\hat{\varphi}_{\vec{k}_3} \rangle
	&= 
	\begin{tikzpicture}[baseline=(v1)]
	\begin{feynman}[inline = (base.v1)]
		\vertex (f1);
		\vertex [right = 0.5 of f1,label=270:\({\scriptstyle k_1~}\)] (f2);
		\vertex [right = 0.25 of f2] (v1s);
		\vertex [below = 0.75 of v1s] (v1);
		\vertex [right = 0.25 of v1s,label=270:\({~~\scriptstyle k_2}\)] (f3);
		\vertex [right = 1.5 of f3, label=270:\({~~~\,\scriptstyle k_3}\)] (f4);
		\vertex [right = 0.5 of f4] (f5);
		\vertex [below = of f4] (v2);
		\node [left = -0.125 of v1, circle, fill = gray, draw = black, inner sep = 2.5pt] (b1);
		\node [left = -0.125 of v2, circle, fill = gray, draw = black, inner sep = 2.5pt] (b2);
		\begin{pgfonlayer}{bg}
		\diagram*{
		(f1) -- [very thick] (f5),
		(f2) -- (v1) -- (f3),
		(f4) -- (v2),
		(v1) -- [scalar, quarter left] (v2) -- [scalar, quarter left] (v1),
		};
		\end{pgfonlayer}
	\end{feynman}
	\end{tikzpicture}
	\nonumber \\
	&= -\frac{c_1c_2}{N_e M_P^3}\sum_{\lambda_1,\lambda_2 = \pm}
	\lambda_1 \lambda_2 \int_{-\infty}^0 \frac{d\tau_1d\tau_2}{(H^2\tau_1\tau_2)^4}
	\Delta_{+\lambda_1}(k_1; 0,\tau_1) \Delta_{+\lambda_1}(k_2; 0, \tau_1)
	\Delta_{+\lambda_2}(k_3;0,\tau_2)
	\nonumber \\
	&\times
	\int d^{3}x_1\,e^{-i(\vec{k}_1 + \vec{k}_2)\cdot\vec{x}_1}
	\int d^{3}x_2\,e^{-i\vec{k}_3\cdot\vec{x}_2}
	\langle \mathcal{O}_{SM}(\tau_1,\vec{x}_1)\mathcal{O}_{SM}(\tau_2,\vec{x}_2) \rangle_{\lambda_1 \lambda_2} \,,
\end{align}
where the thick line indicates the asymptotic future time slice, 
the gray blob indicates a vertex from either the time-ordered or anti-time-ordered contours in the Schwinger-Keldysh formalism,
with $\lambda_1$ and $\lambda_2$ as their corresponding labels,
and the dashed lines denote the SM fermions or gauge bosons. The Schwinger-Keldysh formalism is reviewed in App.~\ref{app:propagators}. As the correlator depends only on $\vec{x}_{12} = \vec{x}_1 - \vec{x}_2$, 
we can factor out the overall space integral as the delta function corresponding to momentum conservation. We define the dimensionless non-Gaussianity function $S_\mathrm{NG}$ as
\begin{equation}
	\langle \zeta_{\vec{k}_1}\zeta_{\vec{k}_2}\zeta_{\vec{k}_3} \rangle'
	\; = \; (2\pi)^4 S_\mathrm{NG}(\vec{k}_1,\vec{k}_2,\vec{k}_3) \frac{P_\zeta^2}{k_1^2 k_2^2 k_3^2} \, ,
\end{equation}
where $\zeta_{\vec{k}} = -(H/\dot{\phi}_0)\hat{\varphi}_{\vec{k}}$\footnote{
	This relation receives a correction at the next-to-leading order, which contributes to the non-Gaussianity~\cite{Maldacena:2002vr}.
	As this contribution is local and does not generate a cosmological collider signal, we do not consider it here.
}
and the prime indicates that we have removed the factor $(2\pi)^3 \delta^{(3)}(\vec{k}_1 + \vec{k}_2 + \vec{k}_3)$.
The SM fermion and gauge boson contribution to $S_\mathrm{NG}$ is expressed as
\begin{align}
	S_\mathrm{NG}
	&= \frac{c_1c_2 H \dot{\phi}_0}{8N_e M_P^3}\frac{1}{k_1k_2k_3}
	\sum_{\lambda_1,\lambda_2 = \pm}
	\lambda_1 \lambda_2 \int_{-\infty}^0 \frac{d\tau_1d\tau_2}{(H^2\tau_1\tau_2)^4}
	e^{i\lambda_1 (k_1 + k_2)\tau_1 + i\lambda_2 k_3 \tau_2}
	\nonumber \\
	&\times
	\left(1-i\lambda_1 k_1 \tau_1\right)(1-i\lambda_1 k_2 \tau_1)(1-i\lambda_2 k_3 \tau_2)
	%\nonumber \\
	%&\times
	\int d^{3}x_{12}\,e^{i \vec{k}_3\cdot\vec{x}_{12}}
	\langle \mathcal{O}_\mathrm{SM}(\tau_1,\vec{x}_1)
	\mathcal{O}_\mathrm{SM}(\tau_2,\vec{x}_2) \rangle_{\lambda_1 \lambda_2} \, ,
\end{align}
where we used the explicit form of the massless scalar propagator $\Delta$ derived in App.~\ref{app:scalar}.

The remaining task is to evaluate the two-point function of $\mathcal{O}_\mathrm{SM}$.
For this, we use the late-time expansion, as described in Ref.~\cite{Chen:2016hrz}, to estimate the order of magnitude of the signal.
In the late-time expansion, we fix $\vec{x}_{12}$ and take the limit
$\tau_1, \tau_2 \to 0$, and different propagators become equivalent (see App.~\ref{app:propagators}). Therefore, we can omit the subscripts
$\lambda_1$ and $\lambda_2$ from the two-point function of $\mathcal{O}_\mathrm{SM}$.
%%%%%%%%%%%%%%%%%%%%%%%%%%%%%%%%%%%%%%%
\subsection{Fermion contribution}
%%%%%%%%%%%%%%%%%%%%%%%%%%%%%%%%%%%%%%%
We first discuss the fermion contribution.
As we noted above, we can drop the contour subscripts as long as we focus on the late-time behavior.
Therefore, we may evaluate the two-point function as
\begin{align}
	\langle \mathcal{O}_\mathrm{SM}(\tau_1,\vec{x}_1)\mathcal{O}_\mathrm{SM}(\tau_2,\vec{x}_2) \rangle
	\; = \; -m^2 \mathrm{Tr}\left[S(\vec{x}_{12};\tau_1, \tau_2) S(\vec{x}_{21};\tau_2, \tau_1)\right] \, ,
\end{align}
where $\vec{x}_{21} = \vec{x}_2 - \vec{x}_1$. The fermion propagator in de~Sitter spacetime is given by
\begin{align}
	S(\vec{x}_{12};\tau_1, \tau_2)
	\; = \; \frac{H^{2}}{(4\pi)^{2}}\left[a\left(i\slashed{\nabla}+m\right)\right]_{1}
	\frac{1}{\sqrt{a_1 a_2}}\left[P_+ I_{\nu_-}(Z_{12}) + P_- I_{\nu_+}(Z_{12})\right] \,,
\end{align}
where $m$ represents the fermion mass, $P_\pm = (1\pm\gamma^0)/2$ is the projection operator, $a_i \equiv a(\tau_i)$, $\nu_{\pm} = 1/2 \mp im/H$, $Z_{12}$ is the embedding distance, and the definition of $I_\nu(Z_{12})$ and its derivation is given in App.~\ref{app:fermion}. To take the trace, it is convenient to keep track of the projection operators $P_\pm$ and to write the propagators as
\begin{align}
	S(\vec{x}_{12};\tau_1,\tau_2)
	= \frac{H^2}{(4\pi)^2}\frac{1}{\sqrt{a_1a_2}}&\left\{
	P_+ \left[i\left(\partial_{\tau_1}+\frac{\nu_--3/2}{\tau_1}\right)I_{\nu_-}(Z_{12})
	+ i\gamma^i \partial_{1i}I_{\nu_+}(Z_{12})\right]
	\right. \nonumber \\
	&\left.+
	P_-\left[-i\left(\partial_{\tau_1}+\frac{\nu_+-3/2}{\tau_1}\right)I_{\nu_+}(Z_{12})
	+ i\gamma^i \partial_{1i}I_{\nu_-}(Z_{12})\right]
	\right\},
\end{align}
and
\begin{align}
	S(\vec{x}_{21};\tau_2,\tau_1)
	= \frac{H^2}{(4\pi)^{2}}\frac{1}{\sqrt{a_1a_2}}
	&\left\{
	\left[i\left(\partial_{\tau_2}+\frac{\nu_- - 3/2}{\tau_2}\right)I_{\nu_-}(Z_{21}) + i\gamma^i \partial_{2i}I_{\nu_-}(Z_{21})
	\right]P_+
	\right. \nonumber \\
	&\left.+
	\left[-i\left(\partial_{\tau_2}+\frac{\nu_+-3/2}{\tau_2}\right)I_{\nu_+}(Z_{21})
	+ i\gamma^i\partial_{2i}I_{\nu_+}(Z_{21})\right]P_-
	\right\}.
\end{align}
Using these definitions, we obtain
\begin{align}
	&\mathrm{Tr}\left[S(\vec{x}_{12};\tau_1, \tau_2) S(\vec{x}_{21};\tau_2, \tau_1)\right]
	\nonumber \\
	&= \frac{2H^4}{(4\pi)^4}\frac{1}{a_1a_2}
	\left[-\left(\partial_{\tau_1}+\frac{\nu_--3/2}{\tau_1}\right)I_{\nu_-}
	\left(\partial_{\tau_2}+\frac{\nu_--3/2}{\tau_2}\right)I_{\nu_-} + (\partial_{1i}I_{\nu_+}) (\partial_{2i}I_{\nu_-})
	+ (\nu_- \leftrightarrow \nu_+)\right],
\end{align}
where we do not distinguish between the arguments $Z_{12}$ and $Z_{21}$ 
as it is irrelevant for the late-time expansion. We note that the time derivative, together with the factor $\nu_\pm -3/2$, eliminates the leading order term in the late-time expansion of $I_{\nu_\pm}$.
Using Eqs.~\eqref{eq:Zderiv1} and~\eqref{eq:Zderiv2}, we evaluate the late-time expansion $Z \to \infty$ as
\begin{align}
	&\mathrm{Tr}\left[S(\vec{x}_{12};\tau_1, \tau_2) S(\vec{x}_{21};\tau_2, \tau_1)\right]
	\nonumber \\
	&=
	-\frac{H^6}{\pi^4}\frac{\frac{2\pi m}{H} (1+\frac{m^2}{H^2})}{\sinh(2\pi m/H)}(-2Z_{12})^{-3}
	-\frac{3H^6}{2\pi^5}
	\left[
	\Gamma\left(\nu_-\right)
	\Gamma\left(\nu_- - 1\right)
	\Gamma^2\left(\frac{5}{2}-\nu_-\right)
	(-2Z_{12})^{-5+2\nu_-}
	+ (\nu_- \to \nu_+)
	\right],
\end{align}
where we ignore the terms of $\mathcal{O}(Z_{12}^{-5})$ and keep both the local and non-local contributions. 
We verify the consistency of our findings by considering the massless limit. If we take the limit $m \rightarrow 0$, our result simplifies to
\begin{align}
	\mathrm{Tr}\left[S(\vec{x}_{12};\tau_1, \tau_2) S(\vec{x}_{21};\tau_2, \tau_1)\right]
	&\simeq -\frac{1}{\pi^4 a_1^3 a_2^3}\left[\frac{1}{x_{12}^6}+\frac{3(\tau_1-\tau_2)^2}{x_{12}^8} + \cdots\right] \, ,
\end{align}
where we expanded the $Z_{12}$ term. On the other hand, if $m = 0$, the fermion field becomes conformal, allowing the scale factor to be factored out by redefining $\psi = a^{-3/2}\psi_\mathrm{flat}$. The massless fermion propagator in flat spacetime
is well-known and given in coordinate space by (see e.g.~\cite{Novikov:1983gd})
\begin{align}
	S_\mathrm{flat}(x) \; = \; \frac{i\slashed{x}}{2\pi^2 x^4} \, ,
\end{align}
where $x$ here refers to a four-dimensional space-time coordinate. Therefore, if $m = 0$, we find
\begin{align}
	\mathrm{Tr}\left[S(\vec{x}_{12};\tau_1, \tau_2) S(\vec{x}_{21};\tau_2, \tau_1)\right]
	&= \frac{1}{a_1^3a_2^3}\times \mathrm{Tr}\left[S_\mathrm{flat}(x_{12}) S_\mathrm{flat}(x_{21})\right]
	= -\frac{1}{\pi^4 a_1^3 a_2^3}\left[\frac{1}{x_{12}^6} + \frac{3(\tau_1-\tau_2)^2}{x_{12}^8} + \cdots\right],
\end{align}
which coincides with our result.
Focusing on the non-local part and keeping only the leading-order term, we obtain
\begin{align}
	\langle \mathcal{O}_{SM}(\tau_1,\vec{x}_1)\mathcal{O}_{SM}(\tau_2,\vec{x}_2) \rangle
	&\simeq
	\frac{3H^6m^2}{2\pi^5}
	\left[
	\Gamma\left(\nu_-\right)
	\Gamma\left(\nu_- - 1\right)
	\Gamma^2\left(\frac{5}{2}-\nu_-\right)
	\left(\frac{\tau_1\tau_2}{x_{12}^2}\right)^{5-2\nu_-}
	+ (\nu_- \to \nu_+)
	\right] \, .
\end{align}
This agrees with the result in Ref.~\cite{Lu:2019tjj} (which corrected an error in Ref.~\cite{Chen:2016hrz}).
After performing the Fourier transformation and conformal time integrals, the non-Gaussianity function in the squeezed limit $k_3 \ll k_1, k_2$ can be expressed as
\begin{align}
	S_\mathrm{NG} 
	\; = \; -\frac{3c_1c_2 H\dot{\phi}_0}{8\pi^4 N_eM_P^3}\frac{m^2}{H^2}C_{1/2}(\nu_-)\left(\frac{k_3}{k_1}\right)^{4-2\nu_-}
	+ (\nu_- \to \nu_+) \, ,
\end{align}
where
\begin{align}
	C_{1/2}(\nu) \; = \; 2^{2\nu}(3-\nu)(3-2\nu)^2\sin^2(\nu\pi)\sin(2\nu\pi)\Gamma(\nu)\Gamma(\nu-1)\Gamma(-8+4\nu_-)
	\Gamma^2\left(\frac{5}{2}-\nu\right)\Gamma^2(2-2\nu) \, .
\end{align}
The coefficient $(m^2/H^2)C_{1/2}(\nu_\pm)$ peaks at $m \sim 0.1H$ and is generally of order unity. This expression aligns with our earlier estimation~\eqref{eq:estimate_psiA}, albeit with an additional suppression factor $\sim (2\pi)^{-4}$ arising from the loops.
Unfortunately, mainly due to Planck scale suppression, the signal is too small to be observable in the foreseeable future.
%%%%%%%%%%%%%%%%%%%%%%%%%%%%%%%%%%%%%%%
\subsection{Gauge boson contribution}
%%%%%%%%%%%%%%%%%%%%%%%%%%%%%%%%%%%%%%%
We next compute the contribution from the gauge bosons.
The two-point function of $\mathcal{O}_{\mathrm{SM}}$ is given by
\begin{align}
	\langle \mathcal{O}_\mathrm{SM}(\tau_1,\vec{x}_1)\mathcal{O}_\mathrm{SM}(\tau_2,\vec{x}_2) \rangle
	\; = \; \frac{m^4}{2a_1^2 a_2^2}
	\eta^{\alpha\beta}\eta^{\gamma\delta}G_{\alpha\gamma}(\vec{x}_{12};\tau_1,\tau_2)
	G_{\beta\delta}(\vec{x}_{12};\tau_1,\tau_2) \, ,
\end{align}
where $m$ represents the gauge boson mass, and we again drop the contour indices.
The gauge boson propagator in de~Sitter spacetime is given by 
\begin{align}
	G_{\alpha\beta}(\vec{x}_1-\vec{x}_2;\tau_1, \tau_2)
	&= \frac{H^2}{m^2(4\pi)^{2}}K_{\alpha\beta}(Z_{12}),
\end{align}
where $Z_{12}$ is the embedding distance and
\begin{align}
	K_{\alpha\beta} (Z_{12}) 
	&= \left[Z_{12}I_\nu''+ 3I_\nu'\right] (\partial_{1\alpha}Z_{12})(\partial_{2\beta}Z_{12})
	%\nonumber \\
	+ \left[(1-Z_{12}^2)I_\nu'' - 3Z_{12}I_\nu'\right] \partial_{1\alpha}\partial_{2\beta}Z_{12} \, ,
\end{align}
with $\nu = \sqrt{1/4-m^2/H^2}$ and the primes denoting the derivatives with respect to $Z_{12}$. The full details of the derivation are given in App.~\ref{app:gauge_boson}. The late-time behavior of the two-point function then becomes
\begin{align}
	\langle \mathcal{O}_{SM}(\tau_1,\vec{x}_1)\mathcal{O}_{SM}(\tau_2,\vec{x}_2) \rangle
	\; = \; \frac{3H^8}{128\pi^5}
	(1+2\nu)^2 \Gamma^2\left(\frac{5}{2}-\nu\right)\Gamma^2(\nu)\left(\frac{\tau_1\tau_2}{x_{12}^2}\right)^{3-2\nu}
	+ (\nu \to -\nu) \, ,
\end{align}
where we focus on the non-local contribution.\footnote{
	This agrees with~\cite{Lu:2019tjj} up to an overall factor two.
}
After performing the Fourier transformation and computing the conformal time integrals, the non-Gaussianity function $S_\mathrm{NG}$ becomes
\begin{align}
	S_\mathrm{NG} &= -\frac{3c_1c_2}{8\pi^4}\frac{H\dot{\phi}_0}{N_eM_P^3}\frac{m^4}{H^4}C_1(\nu)
	\left(\frac{k_3}{k_1}\right)^{2-2\nu} + (\nu \to -\nu),
\end{align}
where
\begin{align}
	C_1(\nu) = 2^{2\nu}(2-\nu)\sin^2(\nu\pi)\sin(2\nu\pi)
	\Gamma(-4+4\nu)
	\Gamma^2\left(\frac{5}{2}-\nu\right)\Gamma^2(\nu)\Gamma^2(-2\nu).
\end{align}
This expression correctly reproduces our estimation~\eqref{eq:estimate_psiA}, with an additional suppression factor $\sim(2\pi)^{-4}$ arising from the loops.
Similar to the fermion case, the signal is too small for near-future observation due to suppression by the Planck scale, $M_P$.
Moreover, the gauge boson contribution is likely further suppressed exponentially 
by $m_W/H$ or $m_Z/H$ (see Fig.~\ref{fig:SM_spectrum}).
Therefore, we turn our attention to the isocurvature mode signatures.
%%%%%%%%%%%%%%%%%%%%%%%%%%%%%%%%%%%%%%%
\section{Cosmological collider signatures of isocurvature mode}
\label{sec:CC_isocurvature}
%%%%%%%%%%%%%%%%%%%%%%%%%%%%%%%%%%%%%%%
As discussed in the previous section, the inflationary sector in Higgs-$R^2$ inflation naturally couples to SM fermions and gauge bosons, but the resulting cosmological collider signatures are too small for near-future detection. Therefore, in this section, we focus on the isocurvature mode. In the ``standard" parameter region of Higgs-$R^2$ inflation, $\xi^2 \gg 1$, the isocurvature mode is too heavy to be excited during inflation, leaving no observable signal. However, if $\lambda \ll 1$ such that $\xi \sim \lambda \alpha \lesssim \mathcal{O}(0.1)$, the isocurvature mode can be as light as the Hubble parameter to be excited during inflation. In this case, the isocurvature mode can produce a substantial cosmological collider signature, potentially detectable by future 21 cm observations~\cite{Chen:2016zuu,Meerburg:2016zdz}. This serves as a concrete example of the clock signals discussed in Refs.~\cite{Chen:2009we,Chen:2009zp,Chen:2015lza}, originating from a UV-complete model (up to the Planck scale) motivated by particle physics.

%%%%%%%%%%%%%%%%%%%%%%%%%%%%%%%%%%%%%%%
\subsection{Mixed propagator}
%%%%%%%%%%%%%%%%%%%%%%%%%%%%%%%%%%%%%%%
We first define a mixed propagator of the adiabatic and isocurvature modes, following Ref.~\cite{Chen:2017ryl}. The relevant part of the quadratic action is given by Eq.~(\ref{eq:quadraticact1}), where all details of the derivation can be found in App.~\ref{app:covariant_quadratic}.
Since the curvature mode $\zeta$ is related to $\varphi$ as $\zeta = -H \varphi/\dot{\phi}_0$,\footnote{
	 As before, we ignore the next-to-leading order correction that generates only local-type non-Gaussianity.
}
within the leading order of the slow-roll approximation, the action is equivalently expressed as
\begin{align}
	S_\mathrm{quad} \; \simeq \; \int d\tau d^3x 
	\left[\frac{a^2}{2} 
	\left({\varphi'}^2-\left(\partial_i \varphi\right)^2\right)
	+ \frac{a^2}{2}\left({\chi'}^2 - \left(\partial_i \chi\right)^2
	-a^2 m_\chi^2 \chi^2\right)
	+2\dot{\theta}a^3 \varphi'\chi\right] \, .
\end{align}
Ultimately, we evaluate the three-point function of the curvature perturbation at the asymptotic future where $\tau = 0$. Therefore, we define the mixed propagator as 
\begin{align}
	\vev{\hat{\varphi}(0,\vec{x}_1) \hat{\chi}_\pm(\tau,\vec{x}_2)}
	&= \int\frac{d^3k}{(2\pi)^3}e^{i\vec{k}\cdot(\vec{x}_1 - \vec{x}_2)}\mathcal{G}_\pm(k;\tau) \, ,
\end{align}
where the subscript $\pm$ denotes the contour on which $\chi$ resides.
Using the Feynman rules of the Schwinger-Keldysh formalism, it is computed as
(see App.~\ref{app:scalar} for the discussion of scalar propagators)
\begin{align}
	\mathcal{G}_\pm(k;\tau) &\equiv
	\begin{tikzpicture}[baseline=(v1)]
	\begin{feynman}[inline = (base.v1)]
		\vertex (f1);
		\vertex [right = of f1] (f2);
		\vertex [right = of f2] (f3);
		\vertex [below = 0.75 of f2] (v1);
		\vertex [below = 0.75 of v1] (c1);
		\diagram*{
		(f1) -- [very thick] (f3),
		(f2) -- [insertion={1.0}] (v1) -- [scalar] (c1),
		};
	\end{feynman}
	\end{tikzpicture}
	= 
	\begin{tikzpicture}[baseline=(v1)]
	\begin{feynman}[inline = (base.v1)]
		\vertex (f1);
		\vertex [right = of f1] (f2);
		\vertex [right = of f2] (f3);
		\vertex [below = 0.75 of f2] (v1);
		\vertex [below = 0.75 of v1] (c1);
		\node [left = -0.125 of v1, circle, fill = black, draw = black, inner sep = 2.5pt] (b1);
		\begin{pgfonlayer}{bg}
		\diagram*{
		(f1) -- [very thick] (f3),
		(f2) -- (v1) -- [scalar] (c1),
		};
		\end{pgfonlayer}
	\end{feynman}
	\end{tikzpicture}
	+
	\begin{tikzpicture}[baseline=(v1)]
	\begin{feynman}[inline = (base.v1)]
		\vertex (f1);
		\vertex [right = of f1] (f2);
		\vertex [right = of f2] (f3);
		\vertex [below = 0.75 of f2] (v1);
		\vertex [below = 0.75 of v1] (c1);
		\node [left = -0.125 of v1, circle, fill = white, draw = black, inner sep = 2.5pt] (b1);
		\begin{pgfonlayer}{bg}
		\diagram*{
		(f1) -- [very thick] (f3),
		(f2) -- (v1) -- [scalar] (c1),
		};
		\end{pgfonlayer}
	\end{feynman}
	\end{tikzpicture}
	\nonumber \\
	&= 2i\dot{\theta}\int_{-\infty}^{0}\frac{d\tau'}{(-H\tau')^3} \left[\partial_{\tau'}\Delta(k;0,\tau') \times 
	G_{\pm +}(k;\tau,\tau') - \partial_{\tau'} \Delta(k;\tau',0) \times G_{\pm -}(k;\tau,\tau')
	\right] \, .
\end{align}
Here, the black (white) circle represents the vertex from the time-ordered (anti-time-ordered) contour, the solid line represents $\varphi$,
and the dashed line indicates $\chi$. Additionally, we assume that the turning rate $\dot{\theta}$ is constant. We find that
\begin{align}
	\mathcal{G}_\pm(k;\tau)
	&= \frac{\pi \dot{\theta} H}{4k^3}\mathcal{I}_\pm(-k\tau; \nu) \, ,
\end{align}
where
\begin{equation}
    \nu \; = \; \sqrt{\frac{9}{4} - \frac{m_{\chi}^2}{H^2}} \, ,
\end{equation}
and
\begin{align}
	\mathcal{I}_\pm(z;\nu) = z^{3/2}
	&\left[-iH_\nu^{(1)}(z)\int_0^{\infty(1-i0)}\frac{dz'}{\sqrt{z'}}
	H_\nu^{(2)}(z')e^{-iz'}
	+iH_\nu^{(2)}(z)\int_0^{\infty(1+i0)}\frac{dz'}{\sqrt{z'}}
	H_\nu^{(1)}(z')e^{iz'}
	\right. \nonumber \\ &~\left.
	+ iH_\nu^{(1)}(z)\int_0^{z(1\mp i0)} \frac{dz'}{\sqrt{z'}} {H_\nu^{(2)}}(z')e^{\mp iz'}
	- i {H_\nu^{(2)}}(z) \int_0^{z(1\mp i0)} \frac{dz'}{\sqrt{z'}} {H_\nu^{(1)}}(z')e^{\mp iz'}
	\right] \,,
	\label{eq:Ipm}
\end{align}
where we explicitly show the $i\epsilon$ prescription. We note that we correctly reproduce the result from~\cite{Chen:2017ryl} by setting their 
$\lambda_2$ equal to our $2\dot{\theta}$. The integral can be evaluated analytically and is expressed in terms of the generalized hypergeometric functions ${}_2F_2$. We refer interested readers to Ref.~\cite{Chen:2017ryl} for its explicit form. 
We may focus on the parameter region $m_\chi/H > 1$ hereafter since
otherwise the isocurvature mode would not be sufficiently heavy, 
and the model is eventually turning into a multi-field inflation regime.\footnote{
	The expression~\eqref{eq:Ipm} is IR finite as long as $\mathrm{Re}[\nu] < 3/2$~\cite{Chen:2009zp},
	or the mass is finite,
	and in this sense $m/H > 1$ is not a strict bound.
	Nevertheless we focus on this parameter region for simplicity.
} Finally, we introduce the following asymptotic limit which is useful when considering the squeezed limit:
\begin{align}
	\mathcal{I}_\pm(z;\nu) \xrightarrow[z\to0]{} -\frac{2}{\sqrt{\pi}}
	\left[\frac{2^{\nu}z^{3/2-\nu}\Gamma(\nu)}{\cos(\pi\nu/2)+\sin(\pi\nu/2)}
	+ (\nu \to -\nu)\right] \, ,
	\label{eq:Ipm_squeezed}
\end{align}
which does not depend on the choice of $\mathcal{I}_{+}$ or $\mathcal{I}_{-}$, and here we kept only the non-local contribution.
%%%%%%%%%%%%%%%%%%%%%%%%%%%%%%%%%%%%%%%
\subsection{Bispectrum}
%%%%%%%%%%%%%%%%%%%%%%%%%%%%%%%%%%%%%%%
Equipped with the mixed propagator, we now proceed to calculate the bispectrum arising from the isocurvature mode.
The full cubic action for both the adiabatic and isocurvature modes 
is derived in App.~\ref{app:covariant_cubic}, but most terms are suppressed by slow-roll parameters. Therefore, we focus on the dominant cubic interactions, given by Eq.~(\ref{eq:actioncubic}),
where
\begin{align}
	V_{N^3} &= N^a N^b N^c \nabla_c \nabla_b V_a 
	\simeq \frac{(12\lambda \alpha + \xi(1+3\xi))\sqrt{\xi(24\lambda \alpha + \xi(1+6\xi)}}
	{\alpha\sqrt{2\lambda(4\alpha\lambda + \xi^2)}}H \,,
	\\
	V_{T^2N} &= \frac{1}{3}\left(T^a T^b N^c + T^a N^b T^c + N^a T^b T^c\right)\nabla_c \nabla_b V_a
	\simeq \frac{\xi \sqrt{\xi(24\lambda \alpha + \xi(1+6\xi))}}{6\alpha \sqrt{2\lambda (4\alpha \lambda +\xi^2)}}H \, .
\end{align}
In this expression, we only keep the terms of the leading order in $1/N_e$.
Using the first coupling $V_{N^3}$, we obtain the expression
\begin{align}
	\left.\vev{\hat{\varphi}_{\vec{k}_1} \hat{\varphi}_{\vec{k}_2}\hat{\varphi}_{\vec{k}_3}}'\right\vert_{N^3}
	&= 
	\begin{tikzpicture}[baseline=(c2)]
	\begin{feynman}[inline = (base.c2)]
		\vertex (f1);
		\vertex [right = 0.5 of f1] (z1);
		\vertex [right = 0.75 of z1] (z2);
		\vertex [right = 0.75 of z2] (z3);
		\vertex [right = 0.5 of z3] (f2);
		\vertex [below = 1.5 of z2] (v1);
		\vertex [below = 0.75 of z2] (c2);
		\vertex [left = 0.375 of c2] (c1);
		\vertex [right = 0.375 of c2] (c3);
		\node [left = -0.125 of v1, circle, fill = black, draw = black, inner sep = 2.5pt] (b1);
		\begin{pgfonlayer}{bg}
		\diagram*{
		(f1) -- [very thick] (f2),
		(z1) -- [insertion={1.0}] (c1) -- [scalar] (v1),
		(z2) -- [insertion={1.0}] (c2) -- [scalar] (v1),
		(z3) -- [insertion={1.0}] (c3) -- [scalar] (v1),
		};
		\end{pgfonlayer}
	\end{feynman}
	\end{tikzpicture}
	+
	\begin{tikzpicture}[baseline=(c2)]
	\begin{feynman}[inline = (base.c2)]
		\vertex (f1);
		\vertex [right = 0.5 of f1] (z1);
		\vertex [right = 0.75 of z1] (z2);
		\vertex [right = 0.75 of z2] (z3);
		\vertex [right = 0.5 of z3] (f2);
		\vertex [below = 1.5 of z2] (v1);
		\vertex [below = 0.75 of z2] (c2);
		\vertex [left = 0.375 of c2] (c1);
		\vertex [right = 0.375 of c2] (c3);
		\node [left = -0.125 of v1, circle, fill = white, draw = black, inner sep = 2.5pt] (b1);
		\begin{pgfonlayer}{bg}
		\diagram*{
		(f1) -- [very thick] (f2),
		(z1) -- [insertion={1.0}] (c1) -- [scalar] (v1),
		(z2) -- [insertion={1.0}] (c2) -- [scalar] (v1),
		(z3) -- [insertion={1.0}] (c3) -- [scalar] (v1),
		};
		\end{pgfonlayer}
	\end{feynman}
	\end{tikzpicture}
	\nonumber \\
	&= -\frac{i\pi^3 H^3}{64 k_2^3k_3^3}\frac{V_{N^3}}{H}\left(\frac{\dot{\theta}}{H}\right)^3
	\int_0^\infty \frac{dz}{z^4}\left[\mathcal{I}_+(z;\nu) \mathcal{I}_+
	\left(\frac{k_2}{k_1}z;\nu\right) \mathcal{I}_+ \left(\frac{k_3}{k_1}z;\nu\right)
	-\left(\mathcal{I}_+\to \mathcal{I}_-\right)\right] \, ,
\end{align}
where the prime indicates that we have removed the factor of $(2\pi)^3\delta^{(3)}(\vec{k}_1+\vec{k}_2+\vec{k}_3)$.
From the second coupling, we find
\begin{align}
	\left.\vev{\hat{\varphi}_{\vec{k}_1} \hat{\varphi}_{\vec{k}_2}\hat{\varphi}_{\vec{k}_3}}'\right\vert_{T^2N}
	&= 
	\begin{tikzpicture}[baseline=(c2)]
	\begin{feynman}[inline = (base.c2)]
		\vertex (f1);
		\vertex [right = 0.5 of f1] (z1);
		\vertex [right = 0.75 of z1] (z2);
		\vertex [right = 0.75 of z2] (z3);
		\vertex [right = 0.5 of z3] (f2);
		\vertex [below = 1.5 of z2] (v1);
		\vertex [below = 0.75 of z2] (c2);
		\vertex [left = 0.375 of c2] (c1);
		\vertex [right = 0.375 of c2] (c3);
		\node [left = -0.125 of v1, circle, fill = black, draw = black, inner sep = 2.5pt] (b1);
		\begin{pgfonlayer}{bg}
		\diagram*{
		(f1) -- [very thick] (f2),
		(z3) -- [insertion={1.0}] (c3) -- [scalar] (v1),
		(z2) -- (v1),
		(z1) -- (v1),
		};
		\end{pgfonlayer}
	\end{feynman}
	\end{tikzpicture}
	+
	\begin{tikzpicture}[baseline=(c2)]
	\begin{feynman}[inline = (base.c2)]
		\vertex (f1);
		\vertex [right = 0.5 of f1] (z1);
		\vertex [right = 0.75 of z1] (z2);
		\vertex [right = 0.75 of z2] (z3);
		\vertex [right = 0.5 of z3] (f2);
		\vertex [below = 1.5 of z2] (v1);
		\vertex [below = 0.75 of z2] (c2);
		\vertex [left = 0.375 of c2] (c1);
		\vertex [right = 0.375 of c2] (c3);
		\node [left = -0.125 of v1, circle, fill = white, draw = black, inner sep = 2.5pt] (b1);
		\begin{pgfonlayer}{bg}
		\diagram*{
		(f1) -- [very thick] (f2),
		(z3) -- [insertion={1.0}] (c3) -- [scalar] (v1),
		(z2) -- (v1),
		(z1) -- (v1),
		};
		\end{pgfonlayer}
	\end{feynman}
	\end{tikzpicture}
	+ \mathrm{perm.}
	\nonumber \\
	&= -\frac{i\pi H^3}{16 k_2^3k_3^3}\frac{V_{T^2N}}{H}\frac{\dot{\theta}}{H}
	\int_0^\infty \frac{dz}{z^4}\,\left[
	(1+iz)\left(1+i\frac{k_2}{k_1}z\right)\mathcal{I}_+\left(\frac{k_3}{k_1}z;\nu\right)e^{-i\left(1+\frac{k_2}{k_1}\right)z}
	\right. \nonumber \\ &\left.~~~~~~~~~~~~~~~~~~~~~~~~~~~~~~~~~~~~~~
	-(1-iz)\left(1-i\frac{k_2}{k_1}z\right)\mathcal{I}_-\left(\frac{k_3}{k_1}z;\nu\right)e^{i\left(1+\frac{k_2}{k_1}\right)z}
	+ \mathrm{perm.}
	\right],
\end{align}
where ``perm." indicates the permutations $k_3 \leftrightarrow k_1$ and $k_3 \leftrightarrow k_2$, which become negligible in the squeezed limit $k_3 \to 0$. The first expression serves as an explicit example of contributions discussed in~\cite{Chen:2009we,Chen:2009zp},
while the second corresponds to those discussed in~\cite{Chen:2015lza}.
In the squeezed limit, $k_3 \ll k_1, k_2$, these bispectra simplify to
\begin{align}
	\left.\vev{\hat{\varphi}_{\vec{k}_1} \hat{\varphi}_{\vec{k}_2}\hat{\varphi}_{\vec{k}_3}}'\right\vert_{N^3}
	&= \frac{2\pi H^3}{k_1^3k_3^3}\frac{V_{N^3}}{H}\left(\frac{\dot{\theta}}{H}\right)^3
	C_{N^3}(\nu) \left(\frac{k_3}{k_1}\right)^{3/2-\nu} + (\nu \to -\nu) \, ,
\end{align}
where
\begin{align}
	C_{N^3}(\nu) \; = \; \frac{i\pi^{3/2}2^{\nu-6} \Gamma(\nu)}{\cos(\pi\nu/2)+\sin(\pi\nu/2)}
	\int_0^{\infty}dz\,z^{-(5/2+\nu)}\left[\mathcal{I}_+^2(z;\nu) - \mathcal{I}_-^2(z;\nu)\right] \,,
\end{align}
for the former coupling, and
\begin{align}
	\left.\vev{\hat{\varphi}_{\vec{k}_1} \hat{\varphi}_{\vec{k}_2}\hat{\varphi}_{\vec{k}_3}}'\right\vert_{T^2N}
	&= \frac{2\pi H^3}{k_1^3k_3^3}\frac{V_{T^2N}}{H}\frac{\dot{\theta}}{H}
	C_{T^2N}(\nu)\left(\frac{k_3}{k_1}\right)^{3/2-\nu} + (\nu \to -\nu) \, ,
\end{align}
where
\begin{align}
	C_{T^2N}(\nu) &= \frac{2^{\nu-4}(5-2\nu)}{\sqrt{\pi}(3+2\nu)}\Gamma(\nu)\Gamma\left(\frac{1}{2}-\nu\right)
	\tan\left(\frac{\pi(1-2\nu)}{4}\right) \, ,
\end{align}
for the latter coupling. In particular, the conformal time integral can be analytically computed in the latter case.
The corresponding non-Gaussianity function is given by
\begin{align}
    \label{eq:nongaussfunc1}
	S_\mathrm{NG}
	&= P_\zeta^{-1/2}
	\left[\frac{V_{N^3}}{H}\left(\frac{\dot{\theta}}{H}\right)^{3}C_{N^3}(\nu) 
	+ \frac{V_{T^2N}}{H}\frac{\dot{\theta}}{H}C_{T^2N}(\nu)\right]
	\left(\frac{k_3}{k_1}\right)^{1/2-\nu} + (\nu \to -\nu) \,,
\end{align}
where we assumed $\dot{\phi}_0 < 0$ so that $P_\zeta^{-1/2} = -2\pi \dot{\phi}_0/H^2$. This accurately reproduces our estimate, given by Eq.~\eqref{eq:estimate_isocurvature},
up to the numerical coefficients $C_{N^3}$ and $C_{T^2N}$. We represent the coefficient of the non-Gaussianity function as
\begin{align}
	f_\mathrm{NL} \; = \; \frac{10}{9}P_\zeta^{-1/2}
	\left[\frac{V_{N^3}}{H}\left(\frac{\dot{\theta}}{H}\right)^{3}C_{N^3}(\nu) 
	+ \frac{V_{T^2N}}{H}\frac{\dot{\theta}}{H}C_{T^2N}(\nu)\right] \, ,
\end{align}
which characterizes the size of the non-Gaussianity.\footnote{
	This coefficient is motivated by the normalization given by Ref.~\cite{Baumann:2018muz}. However, this is just a convention since we evaluate non-Gaussianity only in the squeezed limit and not in the equilateral configuration. 
} However, when $m/H > 3/2$, the coefficient $\nu$ becomes imaginary, and this quantity cannot be evaluated at the equilateral configuration.
\begin{figure}[t]
	\centering
 	\includegraphics[width=0.495\linewidth]{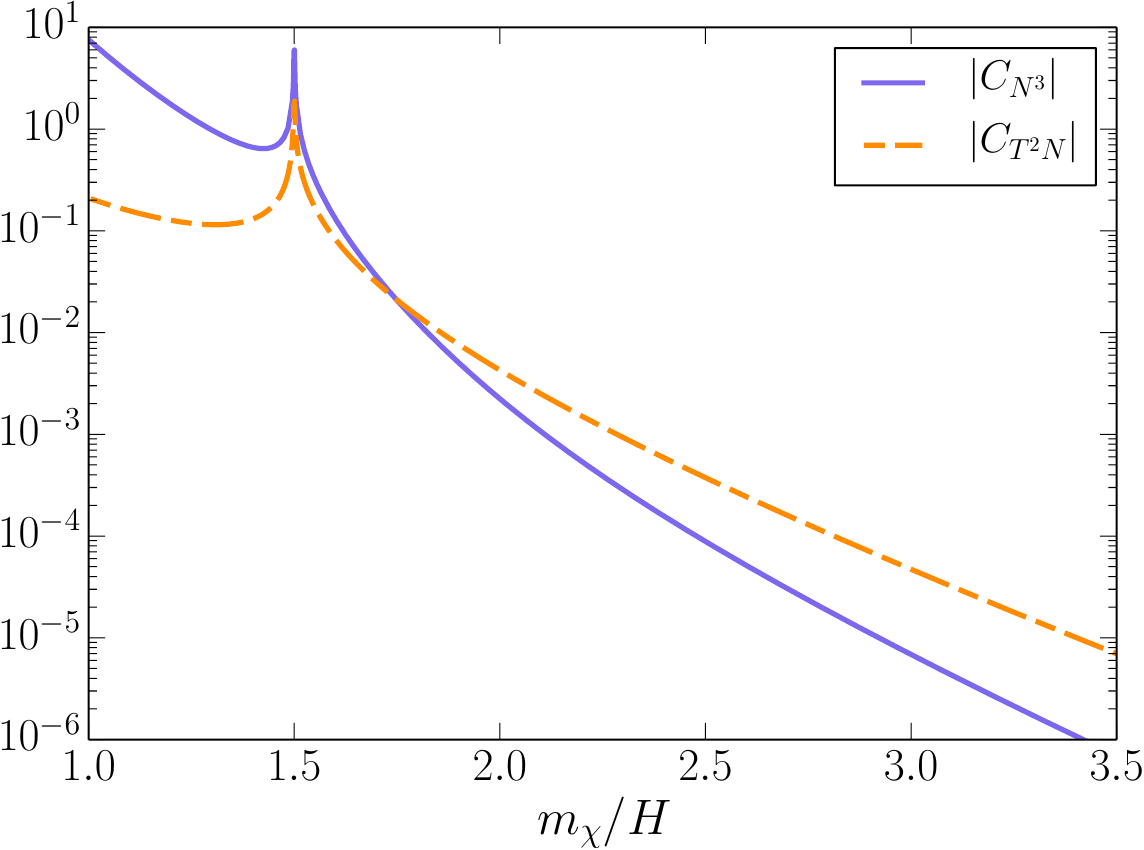}
	\includegraphics[width=0.495\linewidth]{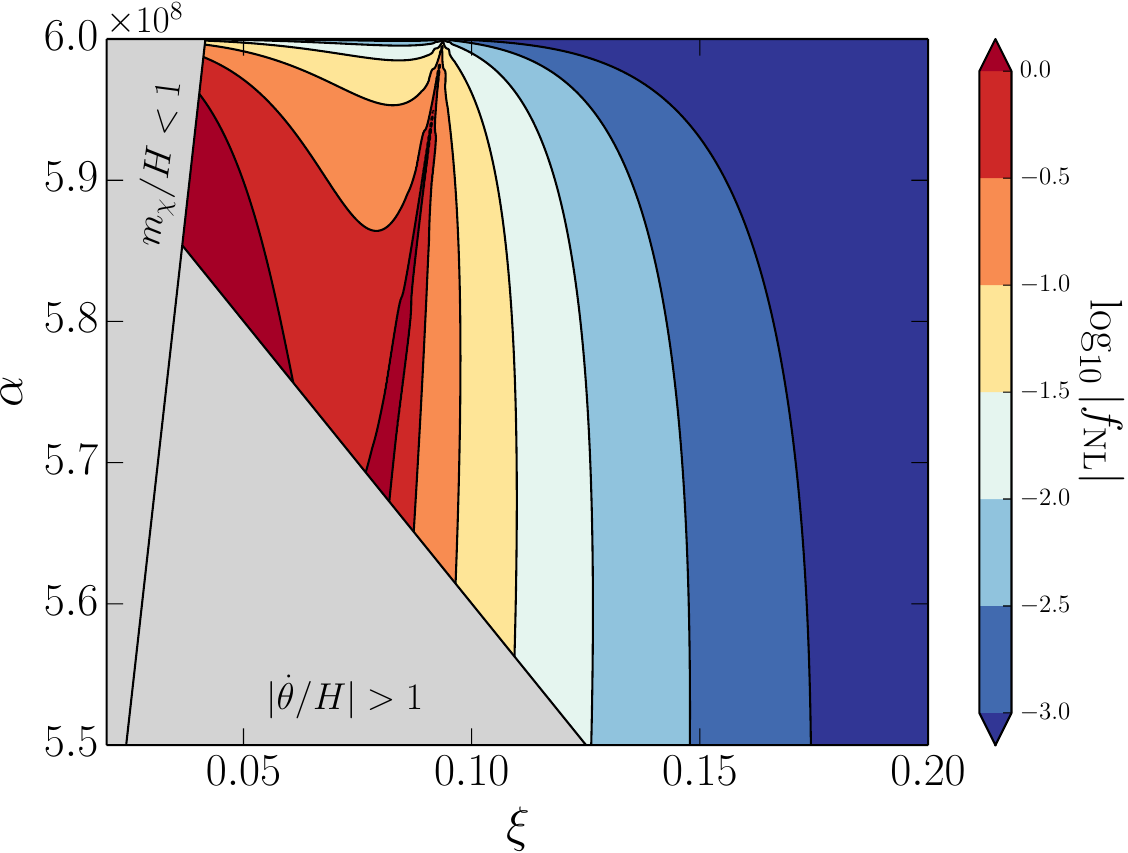}
	\caption{\small \emph{Left panel:} $C_{N^3}(\nu)$ and $C_{T^2N}(\nu)$ as functions of $m_{\chi}/H$. Note that $\nu$ is real for $m_\chi/H < 3/2$ and it is purely imaginary for $m_\chi/H > 3/2$.
	We show only the cases where $\nu > 0$ for $m_\chi/H < 3/2$ because this contribution is less suppressed by $k_3/k_1$ in the squeezed limit.
	On the other hand, for $m_\chi/H > 3/2$, the coefficients $C_{N^3}(-\nu)$ and $C_{T^2N}(-\nu)$ can be related to the complex conjugated functions $C_{N^3}^*(\nu)$ and $C_{T^2N}^*(\nu)$.
	We note that the peak at $m_\chi/H = 3/2$ is spurious since this originates from the late time expansion~\eqref{eq:Ipm_squeezed} 
	that is invalid for $\nu$ close to $0$, as discussed in~\cite{Chen:2009zp}.
	\emph{Right panel:} The coefficients of the non-Gaussianity function for different values of $\xi$ and $\alpha$,
	with $\lambda$ fixed using the CMB normalization~\eqref{eq:COBE_norm} with $N_e = 60$ e-folds (ignoring the $c^2$ contribution).
	The small region, where $\vert f_\mathrm{NL}\vert > 1$ with $\xi \sim 0.08$ (which is spurious; see above), corresponds to $m_\chi/H \simeq 3/2$, and for larger $\xi$, $m_\chi/H > 3/2$, and the signal exhibits oscillatory behavior with respect to $k_3/k_1$. We focus on the region $m_\chi/H > 1$ and $\vert \dot{\theta}/H \vert < 1$ 
	so that the model is in the quasi-single field regime.
	Notably, these two conditions imply that $\vert \dot{\theta}/m_\chi \vert < 1$.
    }
	\label{fig:CN3_CT2N}
\end{figure}

In the left panel of Fig.~\ref{fig:CN3_CT2N}, we display the coefficients $C_{N^3}(\nu)$ and $C_{T^2N}(\nu)$ as functions of $m_\chi/H$. We note that the parameter $\nu$ is real for $m_\chi/H < 3/2$ and purely imaginary for $m_\chi/H > 3/2$. These function are exponentially suppressed for large values of $m_{\chi}/H$, aligning with the expectation that cosmological collider signatures arise from the non-local propagation of particles. In the large mass limit $m_{\chi} \gg H$, correlation functions usually include two contributions: one power-suppressed and the other exponentially suppressed by the mass term.
The former corresponds to higher-dimensional operators and gives rise only to local effects (see, for example, Refs.~\cite{Banyeres:2018aax,Domcke:2019qmm}), while the latter is associated with particle production and, consequently, cosmological collider signatures.

In the right panel of Fig.~\ref{fig:CN3_CT2N}, we show the non-Gaussianity coefficient $\vert f_\mathrm{NL}\vert$ 
as a function of $\xi$ and $\alpha$. In this figure, we set $\lambda$ according to the CMB normalization~\eqref{eq:COBE_norm} (with $c^2$ ignored). As is evident from the figure, the coefficient can attain values as large as order unity while satisfying the conditions $m_\chi/H > 1$ and $\vert \dot{\theta}/H \vert < 1$,\footnote{
	For a larger value of the turning rate $\dot{\theta}$, the isocurvature mode can be classically excited, and the 
	inflationary trajectory can be oscillatory, giving rise to unique features in the curvature perturbation~\cite{Chen:2011zf,Shiu:2011qw,Gao:2012uq,Cheong:2022gfc}.
	This is distinct from the cosmological collider signatures arising from quantum particle production,
	and we do not consider it here.
}
so that the model is in the quasi-single field inflation regime. Therefore, this signal is potentially detectable by future 21 cm observations~\cite{Chen:2016zuu,Meerburg:2016zdz} within this specific parameter space region.

\begin{figure}[t]
	\centering
 	\includegraphics[width=0.98\linewidth]{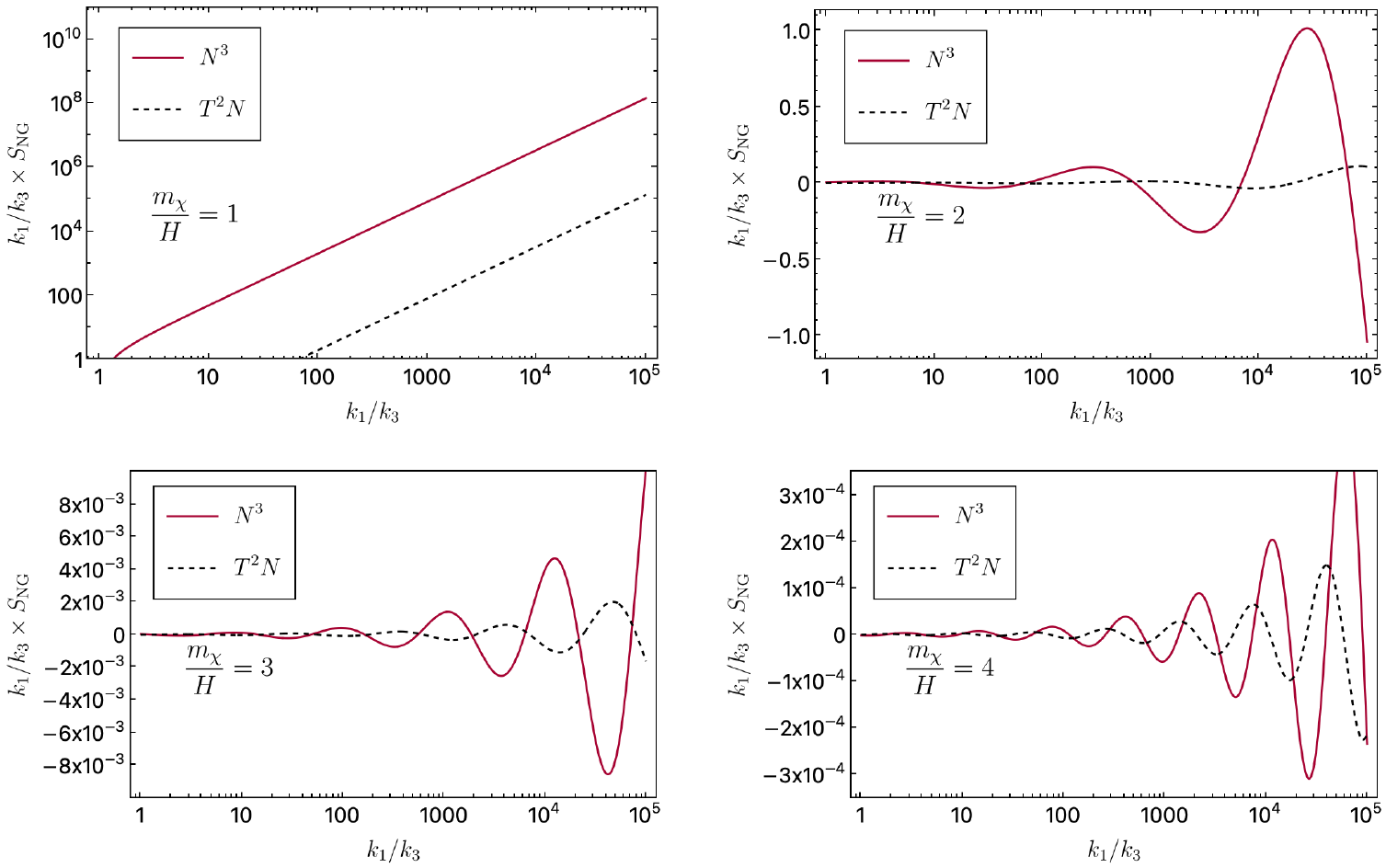}
	\caption{\small The non-Gaussianity function in the squeezed limit~(\ref{eq:nongaussfunc1}) as a function of $k_1/k_3$ for the mass range $m_{\chi}/H = 1,2,3,$ and $4$ . To enhance the oscillatory components of the clock signals, we scale $S_{\rm NG}$ by a factor of $k_1/k_3$. We consider a nominal choice of the Higgs-$R^2$ inflation parameters in the range $0.04 \lesssim \xi \lesssim 0.4$, $4 \times 10^{8} \lesssim \alpha \lesssim 6 \times 10^{8}$, where $\lambda$ is fixed by the CMB normalization~\eqref{eq:COBE_norm} (where we ignore the $c^2$ contribution) with $N_{e} = 60$ e-folds, and the mass of the isocurvature mode is determined from Eq.~(\ref{eq:massisocurv}). Although the signal strength peaks when $m_{\chi} \sim H$ and $\nu$ is real, which maximizes the value of the coupling $|C_{N^3}(\nu)|$, the oscillatory behavior is only evident for the imaginary values of $\nu$, with $m_{\chi}/H = 2,3,$ and $4$. The panels also illustrate that the $N^3$ coupling is always more significant than the $T^2N$ coupling.
    }
	\label{fig:panelsosc}
\end{figure}

In Fig.~\ref{fig:panelsosc}, we plot the full non-Gaussianity function in the squeezed limit~(\ref{eq:nongaussfunc1}) for different values of $m_{\chi}/H$. Here we consider a choice of Higgs-$R^2$ inflation parameters in the range $0.04 \lesssim \xi \lesssim 0.4$, $4 \times 10^{8} \lesssim \alpha \lesssim 6 \times 10^{8}$, and using the CMB normalization~\eqref{eq:COBE_norm} (where we ignore the $c^2$ contribution) with $N_{e} = 60$ e-folds, we determine the value of $\lambda$ and the mass of the isocurvature mode~(\ref{eq:massisocurv}). The left panel of Fig.~\ref{fig:CN3_CT2N} illustrates that the signal peaks when $|C_{N^3}| \sim \mathcal{O}(1)$, with $m_{\chi}/H \sim 1$. However, we show in Fig.~\ref{fig:panelsosc}, that when $m_{\chi}/H > 3/2$, the parameter $\nu$ becomes imaginary and the squeezed limit displays an oscillatory feature, leading to a distinct \textit{clock signal}. However, this signal has a reduced non-Gaussianity amplitude compared to the $m_{\chi} \simeq H$ case. We highlight these important oscillatory features in non-Gaussianities across three panels, where $\nu$ is imaginary for the values $m_{\chi}/H =  2, 3,$ and $4$. For a more detailed discussion of clock signals, see Ref.~\cite{Chen:2015lza}.

Finally, we comment on the trispectrum. In Higgs-$R^2$ inflation, we find that the quartic couplings of the adiabatic and isocurvature modes are given by
\begin{align}
	S_\mathrm{quartic} \; = \; \int d\tau d^3x\,a^4\left[\frac{1}{24}V_{N^4}\chi^4 + \frac{1}{4}V_{T^2N^2}\chi^2 \varphi^2\right] \, ,
\end{align}
where
\begin{align}
	V_{N^4} &\simeq  \frac{7+3\xi(14 + 9\xi)}{18\alpha} + \lambda\left(6-\frac{14}{9(\xi^2 + 4\lambda\alpha)}\right),
	\quad
	V_{T^2 N^2} \simeq \frac{\xi(132\lambda \alpha + \xi(7+33\xi)}{108\alpha (\xi^2 + 4\lambda \alpha))},
\end{align}
with the other combinations suppressed by the slow-roll parameters.
These couplings both scale as $V_{N^4} \sim V_{T^2N^2} \sim 1/\alpha$ for $\xi \sim \lambda \alpha \sim \mathcal{O}(1)$.
Therefore, an estimation similar to Sec.~\ref{subsec:estimation} provides us with
\begin{align}
	g_\mathrm{NL},~\tau_\mathrm{NL} \sim P_\zeta^{-1} \frac{V_{N^4}}{H}\left(\frac{\dot{\theta}}{H}\right)^4,
	~~P_\zeta^{-1} \frac{V_{T^2N^2}}{H}\left(\frac{\dot{\theta}}{H}\right)^2.
\end{align}
We expect a similar size of the signal from the combination of the cubic interactions.
Since $P_\zeta \times \alpha \sim 1$, this can be of order unity for $\xi \sim \lambda \alpha \sim 1$,
and hence we expect a sizable trispectrum in the parameter region of our interest in this section.
We leave a detailed study on the trispectrum, including its precise size and spectral feature in the squeezed limit,
as a possible future work.

%%%%%%%%%%%%%%%%%%%%%%%%%%%%%%%%%%%%%%%
\section{Conclusions and Discussion}
\label{sec:conclusions}
%%%%%%%%%%%%%%%%%%%%%%%%%%%%%%%%%%%%%%%
In this work, we examined and computed the cosmological collider signals for Higgs-$R^2$ inflation. We considered two distinct types of potential signals. The first originates from the inflaton coupling to the SM fermions and gauge bosons through the coupling $h/\Omega$. The second is an isocurvature mode coupling that couples to the inflaton through the turning rate 
$\dot{\theta}$. 

We found that the cosmological collider signatures from the SM fermions and gauge bosons are relatively weak due to their Planck suppression and further suppression by slow-roll parameters, or the number of e-folds $N_e$. Consequently, they are unlikely to be detected even by forthcoming 21 cm probes. However, a considerably stronger signal might emerge from the isocurvature mode and inflaton couplings. In the parameter space, where $\xi \sim \lambda \alpha \sim \mathcal{O}(0.1)$, the isocurvature mode remains light, and the turning rate $\dot{\theta}$, and subsequently coupling to the inflaton, is large. This parameter region aligns with the quasi-single inflation regime of Higgs-$R^2$ inflation, with parameters spanning from $0.5\lesssim \xi\lesssim 2$ and $4 \times 10^{8} \lesssim \alpha \lesssim 6 \times 10^{8}$. Importantly, in this scenario, the non-Gaussianity could be significant, with $\vert f_{\rm NL}\vert \sim \mathcal{O}(1)$, for $m_{\chi}/H \lesssim 3/2$. Although the isocurvature mode contribution to $f_{\rm NL}$ is large in this parameter space, it might be difficult for future 21 cm observations to isolate it from the background. However, when $m_{\chi}/H \gtrsim 3/2$, the cubic interaction of the isocurvature mode, and consequently $f_{\rm NL}$, still remain sizable. This oscillatory clock signal might be distinguishable from the 21 cm background, offering hope for detection by future experiments.

We note that detecting a signal originating from the isocurvature mode couplings could provide strong evidence for multi-field models of inflation. The Higgs-$R^2$ model remains a highly appealing scenario due to several of its features, including a UV-completion. It would be interesting to explore the full features and the parameter space of a multi-field Higgs-$R^2$ model and its associated cosmological collider signatures. However, such a study is quite involved, and we hope to investigate it in future work.

In this paper, we only focused on the SM particles.
However, once we introduce new particles, the inflaton typically couples to them through the conformal factor, $\Omega$, unless these new particles are conformal. For example, the inflaton can couple to right-handed neutrinos via the Yukawa interaction and Majorana mass terms. Our estimates from~Sec.~\ref{subsec:estimation} likely remain valid in this scenario, with $c_1, c_2 \sim \mathcal{O}(1)$.
Therefore, we do not expect that these new particles would significantly enhance the cosmological collider signatures in Higgs-$R^2$ inflation. The situation might change if a stronger coupling between the inflaton and the new particles is introduced,
although this might spoil the UV-completeness of the model by introducing an additional scale.

Another intriguing question is if there are variants of Higgs inflation where the SM fermions and gauge bosons
leave observable cosmological collider signatures.
One such variant of Higgs inflation scenario, known as Palatini Higgs inflation~\cite{Bauer:2008zj,Bauer:2010jg,Shaposhnikov:2020gts},
has recently attracted attention. This model has the same action as the original Higgs inflation model but relies on the Palatini formalism of gravity, where the spin connection and vierbein are treated independently. This model avoids the unitarity issues both during~\cite{Bauer:2010jg} 
and after inflation~\cite{Rubio:2019ypq,Ema:2021xhq,Dux:2022kuk}, in contrast to the metric formalism. In general, using the same arguments as in Sec.~\ref{subsec:estimation}, we expect that the non-Gaussianity arising from the SM particles takes the form
\begin{align}
	f_\mathrm{NL} \; \sim \; \frac{\dot{\phi}_0 H}{\Lambda^3}
	\sim 3.5\times 10^{3}\frac{H^3}{\Lambda^3} \, ,
\end{align}
where $\Lambda$ is the scale of the inflaton coupling to the SM particles. 
In Palatini Higgs inflation, the cut-off scale is given by $\Lambda \sim M_P/\sqrt{\xi}$,
which is significantly smaller than the Planck scale
since from the CMB normalization, we find $\xi \sim 5\times 10^{10}\lambda$. However, in this model the Hubble parameter is also relatively small, with $H\sim \sqrt{\lambda}M_P/\xi$. Therefore, the size of non-Gaussianity can be estimated as
\begin{align}
	f_\mathrm{NL} \; \sim \; 3.5\times 10^{3} \times \left(\frac{\lambda}{\xi}\right)^{3/2} \sim 10^{-13} \, ,
\end{align}
which is too small to be observed in the near future. Nonetheless, it is still interesting to explore if there are other variants that predict a larger cosmological collider signal.

%%%%%%%%%%%%%
\paragraph{Acknowledgements ---}
%%%%%%%%%%%%%
The work of Y.E. and S.V. was performed in part at the Aspen Center for Physics, which is supported by National Science Foundation grant PHY-2210452. We would also like to thank Xingang Chen, 
Soubhik Kumar and Yiming Zhong for useful discussions. Y.E. is supported in part by DOE grant DE-SC0011842.
The work of S.V. was supported in part by DOE grant DE-SC0022148. The Feynman diagrams in this paper are drawn with \texttt{TikZ-Feynman}~\cite{Ellis:2016jkw}.

%%%%%%%%%%%%%%%%%%%%%%%%%%%%%%%%%%%%%%%
%%%%%%%%%%%%%%%%%%%%%%%%%%%%%%%%%%%%%%%
\appendix
%%%%%%%%%%%%%%%%%%%%%%%%%%%%%%%%%%%%%%%
%%%%%%%%%%%%%%%%%%%%%%%%%%%%%%%%%%%%%%%

%%%%%%%%%%%%%%%%%%%%%%%%%%%%%%%%%%%%%%%
\section{Conventions and notation}
\label{app:conventions}
%%%%%%%%%%%%%%%%%%%%%%%%%%%%%%%%%%%%%%%

%%%%%%%%%%%%%%%%%%%%%%%%%%%%%%%%%%%%%%%
\subsection{Conventions}
%%%%%%%%%%%%%%%%%%%%%%%%%%%%%%%%%%%%%%%

Here we summarize the conventions used in this paper.
%%%%%%%%%%
\subsubsection*{Metric sector}
%%%%%%%%%%
We take almost the plus convention,
\begin{align}
	\eta_{\mu\nu} \; = \; \mathrm{diag}(+1,-1,-1,-1) \, .
\end{align}
The geometrical quantities are defined as
\begin{align}
	\Gamma^{\mu}_{\nu\rho} &= \frac{1}{2}g^{\mu\alpha}\left(\partial_\nu g_{\rho\alpha} + \partial_\rho g_{\nu\alpha}
	-\partial_\alpha g_{\nu\rho}\right), \quad
	{R_{\mu\nu\rho}}^\sigma = \partial_\mu \Gamma^{\sigma}_{\nu\rho} - \partial_\nu \Gamma^{\sigma}_{\mu\rho}
	+ \Gamma^\alpha_{\nu\rho}\Gamma^\sigma_{\alpha\mu} - \Gamma^\alpha_{\mu\rho}\Gamma^\sigma_{\alpha\nu},
	\\
	R_{\mu\nu} &=  {R_{\mu\alpha\nu}}^{\alpha}
	= \partial_\mu \Gamma^\alpha_{\alpha \nu} - \partial_\alpha \Gamma^\alpha_{\mu\nu} + \Gamma^\alpha_{\beta\nu}
	\Gamma^\beta_{\alpha\mu} - \Gamma^\alpha_{\mu\nu}\Gamma^\beta_{\alpha\beta},
	\quad
	R = g^{\mu\nu}R_{\mu\nu} \,.
\end{align}
With this definition, the sign in front of the Einstein-Hilbert action is positive, 
and the conformal coupling corresponds to $\xi = -1/6$.
The Ricci scalar transforms under the Weyl transformation as
\begin{align}
    \label{eq:riccitrans}
	{g}_{\mu\nu} \to \Omega^{-2}g_{\mu\nu},
	\quad
	{R} \to \Omega^{2}\left[R 
	+ \frac{3}{2}(\partial_\mu \ln \Omega^{2})(\partial^\mu \ln \Omega^{2})
	- 3\Box \ln \Omega^{2}\right] \, .
\end{align}
Finally the commutator acting on a vector field $V^\alpha$ satisfies
\begin{align}
	[\nabla_\mu, \nabla_\nu] V^\alpha = {R_{\mu\nu\beta}}^{\alpha}V^\beta.
\end{align}
%%

%%%%%%%%%%
\subsubsection*{Target space sector}
%%%%%%%%%%
Since the target space metric is positive definite, we may change the sign convention from the metric sector.
The Christoffel symbol is defined as
\begin{align}
	\Gamma^{a}_{bc} \; = \; \frac{1}{2}h^{ad}\left(h_{db,c} + h_{dc,b} - h_{bc,d}\right) \, ,
\end{align}
and the Riemann tensor is defined as
\begin{align}
	{\mathcal{R}^{a}}_{bcd} \; = \; \partial_c \Gamma^{a}_{db} - \partial_d \Gamma^{a}_{cb}
	+ \Gamma^{a}_{ce}\Gamma^{e}_{db} - \Gamma^{a}_{de}\Gamma^{e}_{cb} \, ,
\end{align}
which has the opposite sign convention from the metric sector.
The Ricci tensor and Ricci scalar are defined as
\begin{align}
	\mathcal{R}_{ab} \; = \; h^{cd}\mathcal{R}_{acbd} \, ,
	\quad
	\mathcal{R} \; = \; h^{ab}\mathcal{R}_{ab} \,.
\end{align}
%%

%%%%%%%%%%
\subsubsection*{Fermion sector}
%%%%%%%%%%

The covariant derivative in the fermion sector is given by
\begin{align}
	\nabla_\mu \psi &= \left(\partial_\mu + \frac{1}{4}\omega_{\mu}^{ab}\gamma_{ab}\right)\psi.
\end{align}
We define the gamma matrices as
\begin{align}
	\left\{\gamma^a, \gamma^b\right\} &= 2\eta^{ab},
	\quad
	\gamma_{ab} = \frac{1}{2}\left(\gamma_a \gamma_b - \gamma_b \gamma_a\right),
	\quad
	\gamma_5 = i\gamma^0 \gamma^1 \gamma^2 \gamma^3
	= - \frac{i}{4!}\epsilon^{abcd}\gamma_a \gamma_b \gamma_c\gamma_d,
\end{align}
with $\epsilon^{0123} = +1$.
We use $a, b, \cdots$ for the local Lorentz indices and $\mu, \nu, \cdots$ for the spacetime indices.
The spin connection and the vierbein are defined as
\begin{align}
	\omega_{\mu}^{ab} &= e^a_\nu \left[\partial_\mu e^{\nu b} + {\Gamma^\nu}_{\sigma\mu} e^{\sigma b}\right],
	\quad
	g_{\mu\nu} = e_{\mu}^a \eta_{ab} e_{\nu}^b.
\end{align}
Under the Weyl transformation, the fermion kinetic term transforms as
\begin{align}
	{\slashed{\nabla}}{\psi} \to \Omega^{5/2}\slashed{\nabla}
	\left(\Omega^{-3/2}\psi \right) \,,
\end{align}
in four spacetime dimensions.

%%%%%%%%%%%%%%%%%%%%%%%%%%%%%%%%%%%%%%%
\subsection{ADM decomposition}
\label{app:ADM}
%%%%%%%%%%%%%%%%%%%%%%%%%%%%%%%%%%%%%%%

Here we may explain the $3+1$ decomposition of the Ricci scalar 
that we use in the computation of the perturbation. We may use the ADM decomposition~\cite{Arnowitt:1959ah}
\begin{align}
	ds^2 \; = \; N^2 dt^2 - \gamma^{ij}(dx^i + \beta^i dt)(dx^j + \beta^j dt) \, ,
\end{align}
or equivalently
\begin{align}
	g_{\mu\nu} \; = \; \begin{pmatrix} N^2 - \beta^i \beta_i & -\beta_i \\ -\beta_j & -\gamma_{ij}\end{pmatrix} \,,
	\quad
	g^{\mu\nu} \; = \; \begin{pmatrix} N^{-2} & -N^{-2}\beta^i \\ -N^{-2}\beta^j & -\gamma^{ij} + N^{-2}\beta^i \beta^j
	\end{pmatrix}\,,
\end{align}
where the spatial indices $i, j$ are raised and lowered by $\gamma_{ij}$.
We may define the temporal part of the vierbein $n_\mu$ and the extrinsic curvature $K_{\mu\nu}$ as
\begin{align}
	n_{\mu} \; \equiv \; (N, \vec{0}),
	\quad
	K_{\mu\nu} \; = \; \nabla_\nu n_{\mu} \, .
\end{align}
Our goal is to express the Ricci scalar in terms of $K_{\mu\nu}$ and the three-dimensional curvature tensors.
For this purpose, we first take the Gauss normal coordinate, $N = 1$ and $\beta^i = 0$,
and compute everything in these coordinates.
We then express the result by the covariant quantities so that we can go back to the general coordinate.
In the Gauss normal coordinates, we can easily show that
\begin{align}
	&\Gamma^{0}_{00} \; = \; \Gamma^{0}_{0i} = \Gamma^{i}_{00} = 0\,,
	\quad
	\Gamma^{0}_{ij} \; = \; \frac{1}{2}\dot{\gamma}_{ij} = -K_{ij} \, ,
	\quad
	\Gamma^{i}_{0j} \; = \; \frac{1}{2}\gamma^{ik}\dot{\gamma}_{jk} \; = \; - {K^i}_j \,.
\end{align}
The Ricci scalar is decomposed as
\begin{align}
	R \; = \; 2{R_{0\alpha0}}^{\alpha} - \gamma^{ij}{R_{ikj}}^{k} \, .
\end{align}
The first term is equally expressed as
\begin{align}
	{R_{0\alpha0}}^{\alpha} \; = \; n^\mu n^\nu {R_{\mu\alpha\nu}}^\alpha = n^\mu [\nabla_\mu, \nabla_\nu] n^\nu \, , 
\end{align}
which is now in covariant form. The second term is also easily computed as
\begin{align}
	-\gamma^{ij}{R_{ikj}}^k = R^{(3)} - K_{ij}K^{ij} + K^2,
\end{align}
where $K = \gamma^{ij} K_{ij}$.
We further note that
\begin{align}
	n^\mu K_{\mu\rho} \; = \; n^\mu \nabla_\rho n_\mu = 0 \, .
\end{align}
Therefore we obtain after integration by parts
\begin{align}
	R \; = \; R^{(3)} + K_{ij}K^{ij} - K^2 + (\mathrm{total}~\mathrm{deriv}) \, ,
\end{align}
which is the desired $3+1$ decomposition of the Ricci scalar.
Following~\cite{Maldacena:2002vr}, we define
\begin{align}
	K_{ij} \; = \; -\frac{1}{N}E_{ij} \,. 
\end{align}
In the general ADM coordinate, it is given by
\begin{align}
	K_{ij} \; = \; -\frac{1}{2N}\left(\dot{\gamma}_{ij} - \nabla_j^{(3)}\beta_i - \nabla_i^{(3)}\beta_j\right)\,.
\end{align}
%%

%%%%%%%%%%%%%%%%%%%%%%%%%%%%%%%%%%%%%%%
\section{Covariant formalism of multi-field inflation}
\label{app:covariant_formalism}
%%%%%%%%%%%%%%%%%%%%%%%%%%%%%%%%%%%%%%%

In this appendix, we review the covariant formalism of multi-field inflation~\cite{Sasaki:1995aw,
GrootNibbelink:2000vx,GrootNibbelink:2001qt,Langlois:2008mn,Peterson:2010np,Gong:2011uw,
Kaiser:2012ak}.

%%%%%%%%%%%%%%%%%%%%%%%%%%%%%%%%%%%%%%%
\subsection{Background dynamics}
\label{app:covariant_background}
%%%%%%%%%%%%%%%%%%%%%%%%%%%%%%%%%%%%%%%

We first consider the background dynamics. We take
\begin{align}
	ds^2 \; = \; N^2(t) dt^2 - a^2(t) dx^i dx^i \,,
	\quad
	\phi^a \; = \; \phi^a_0(t) \,.
\end{align}
The Ricci scalar is given by
\begin{align}
	R \; = \; -\frac{6H^2}{N^2} + (\mathrm{total}~\mathrm{deriv}),
	\quad
	H = \frac{\dot{a}}{a} \,.
\end{align}
By taking the derivative of the action with respect to  $\phi^a$, $N$, and $a$, and then setting $N = 1$,
we obtain
\begin{align}
	&\mathcal{D}_t \dot{\phi}^a_0 + 3H\dot{\phi}_0^a + \partial^a V \; = \; 0 \, ,
	\quad
	H^2 \; = \; \frac{1}{3M_P^2}\left(\frac{1}{2}h_{ab}\dot{\phi}^a_0 \dot{\phi}^b_0 + V\right) \, ,
	\quad
	\dot{H} \; = \; -\frac{1}{2M_P^2}h_{ab}\dot{\phi}_0^a \dot{\phi}_0^b \, ,
\end{align}
where the covariant derivative is defined as
\begin{align}
	\mathcal{D}_t X^a \; = \; \dot{X}^a + \Gamma^{a}_{bc}\dot{\phi}_0^b X^c \, ,
\end{align}
with $\Gamma^{a}_{bc}$ being the Christoffel symbol constructed from $h_{ab}$.
We define
\begin{align}
	T^a \; \equiv \; \frac{\dot{\phi}_0^a}{\dot{\phi}_0}\,,
	\quad
	\mathcal{D}_t T^a \; \equiv \; -\dot{\theta}N^a \,,	
	\quad
	\dot{\phi}_0 \; = \; \sqrt{h_{ab}\dot{\phi}_0^a\dot{\phi}_0^b} \, ,
\end{align}
where $T^a$ indicates the direction of the inflationary trajectory while $N^a$ the orthogonal direction
with the normalization $N^a N_a = 1$.
The turning rate $\dot{\theta}$ parametrizes the curvature of the inflationary trajectory, or the mixing between the adiabatic and isocurvature modes as we will see.
By using the equation of motion of $\phi_0^a$, we see that
\begin{align}
	\mathcal{D}_t T^a \; = \; -\frac{T^a}{\dot{\phi}_0}\left(\ddot{\phi}_0 + 3H\dot{\phi}_0 + V_\phi\right)
	- \frac{1}{\dot{\phi}_0}\left(V^a - T^a V_\phi\right)\,,
	\quad
	V_\phi \; = \; T^a V_a \,,
\end{align}
where $V^a = \partial^a V$ and so on.
Since $T^a$ is normalized, we have $T_a (DT^a/dt) = 0$, from which we obtain
\begin{align}
	\ddot{\phi}_0 + 3H\dot{\phi}_0 + V_\phi &= 0,
	\quad
	\dot{\theta}N^a = \frac{1}{\dot{\phi}_0}\left(V^a - T^a V_\phi\right) \,.
\end{align}
The latter tells us that
\begin{align}
	V_a \; = \; N_a V_N + T_a V_\phi \,,
	\quad
	V_N \; = \; N^a V_a \,,
	\quad
	\dot{\theta} \; = \; \frac{V_N}{\dot{\phi}_0} \,,
	\quad
	N^a \; = \; \frac{V^a - T^a V_\phi}{\sqrt{h_{ab}(V^a - T^a V_\phi)(V^b - T^b V_\phi)}} \, .
\end{align}
Note that $\dot{\theta}$ depends on the derivative of the potential in the $N$-direction,
not in the $T$-direction, and thus this can be large during slow-roll inflation.
We define the slow-roll parameters as
\begin{align}
	\epsilon \; = \; -\frac{\dot{H}}{H^2} \; = \; \frac{\dot{\phi}_0^2}{2M_P^2H^2} \,,
	\quad
	\eta^a \; = \; -\frac{1}{H\dot{\phi}_0}\mathcal{D}_t \dot{\phi}_0^a \,.
\end{align}
By using the equation of motion of $\phi_0^a$, we can show that
\begin{align}
	\eta^a \; = \; \eta_\parallel T^a + \eta_\perp N^a,
	\quad
	\eta_\parallel \; = \; -\frac{\ddot{\phi}_0}{H\dot{\phi}_0} \, ,
	\quad
	\eta_\perp \; = \; \frac{\dot{\theta}}{H} \; = \; \frac{V_N}{H\dot{\phi}_0} \, .
\end{align}
The slow-roll condition requires $\epsilon, \eta_\parallel \ll 1$, but not that $\eta_\perp$ is small. 
Finally, we derive the time evolution of $N^a$.
We note that
\begin{align}
	\frac{d}{dt}\left(T_a N^a\right) \; = \; 0 \; = \; -\dot{\theta} + T_a \mathcal{D}_t N^a,
	\quad
	N_a \mathcal{D}_t N^a \; = \; 0\,,
\end{align}
and thus we can write down
\begin{align}
	\mathcal{D}_t N^a \; = \; \dot{\theta}T^a + \xi^a\,,
\end{align}
where $\xi^a$ satisfies
\begin{align}
	P^{ab}\xi_b \; = \; \xi^a\,,
	\quad
	P_{ab} \; = \; h_{ab} - T_a T_b - N_a N_b \,.
\end{align}
It is then straightforward to show that
\begin{align}
	P_{ab} \mathcal{D}_t^2 T^b \; = \; -P_{ab}\nabla_\phi V^b \,,
	\quad
	\nabla_\phi \; = \; T^c \nabla_c\,,
\end{align}
and thus we obtain
\begin{align}
	\xi^a \; = \; \frac{1}{H\eta_\perp}P^{ab}\nabla_\phi V_b \,.
\end{align}
In summary, the background equations of motion are governed by
\begin{align}
	&\ddot{\phi}_0 + 3H\dot{\phi}_0 + V_\phi \; = \; 0 \,,
	\quad
	H^2 \; = \; \frac{1}{3M_P^2}\left(\frac{1}{2}\dot{\phi}_0^2 + V\right)\,,
	\quad
	\dot{H} \; = \; -\frac{\dot{\phi}_0^2}{2M_P^2} \,,
	\\
	&T^a \equiv \frac{\dot{\phi}_0^a}{\dot{\phi}_0} \,, 
	\quad
	\mathcal{D}_t T^a \; = \; - \dot{\theta} N^a \; = \; - H\eta_\perp N^a \,,
	\quad
	\mathcal{D}_t N^a \; = \; H\eta_\perp T^a + \frac{1}{H\eta_\perp} P^{ab}\nabla_\phi V_b \,,
	\\
	&\epsilon \; = \; -\frac{\dot{H}}{H^2} \,,
	\quad
	\eta^a \; = \; -\frac{1}{H\dot{\phi}_0}\mathcal{D}_t \dot{\phi}_0^a \; = \; \eta_\parallel T^a + \eta_\perp N^a \,,
	\quad
	\eta_\parallel \; = \; -\frac{\ddot{\phi}_0}{H\dot{\phi}_0} \,,
	\quad
	\eta_\perp \; = \; \frac{V_N}{H\dot{\phi}_0} \, .
\end{align}
The equations are not closed if we have more than two-fields,
as the equation of motion of $N^a$ includes the vector not spanned by $T^a$ nor $N^a$.
However, Higgs-$R^2$ inflation has only two fields, the scalaron and the radial mode of the Higgs
(Goldstone modes, or equivalently, the longitudinal gauge bosons, are treated separately).
In this case, $P_{ab} = 0$, and the above equations are closed.

%%%%%%%%%%%%%%%%%%%%%%%%%%%%%%%%%%%%%%%
\subsection{Perturbation: general discussion}
\label{app:covariant_perturbation}
%%%%%%%%%%%%%%%%%%%%%%%%%%%%%%%%%%%%%%%

Next we discuss the perturbation around the above background in the covariant formalism.
We may use the 3+1 decomposition in the ADM formalism discussed in App.~\ref{app:ADM}.
By using the decomposition of the Ricci scalar, the action is given by
\begin{align}
	S \; = \; \int d^4x \sqrt{\gamma}&\left[
	\frac{M_P^2}{2}NR^{(3)} + \frac{M_P^2}{2N}\left(E_{ij}E^{ij}-E^2\right)
	+ \frac{N}{2}h_{ab}g^{\mu\nu}\partial_\mu \phi^a \partial_\nu \phi^b - NV
	\right] \,,
\end{align}
where the extrinsic curvature is given by
\begin{align}
	E_{ij} \; = \; \frac{1}{2}\left(\dot{\gamma}_{ij} - \nabla_i^{(3)}\beta_j - \nabla_j^{(3)}\beta_i\right) \,,
\end{align}
and the quantities with the superscript ``$(3)$" are constructed from the spatial metric $\gamma_{ij}$.
We take the flat gauge and expand the metric as
\begin{align}
	N \; = \; 1 + \alpha \,,
	\quad
	\beta_i \; = \; a^2 \partial_i \beta\,,
	\quad
	\gamma_{ij} \; = \; a^2 \delta_{ij} \,,
\end{align}
where we ignore the vector and tensor parts.

Since we expand the fields up to the third order, extra care is required when expanding the scalar fields. Our goal is to express everything in terms of geometrical quantities up to the third order, as discussed in~\cite{Gong:2011uw,Elliston:2012ab}.
We may think of $\phi^a$ as connected from $\phi^a_0$ by the geodesics with an appropriate initial velocity $\varphi^a$.
Thus, we expand the fields as
\begin{align}
	\phi^a \; = \; 
	\left[\phi^a_0 + \frac{d\phi^a}{d\lambda} + \frac{1}{2}\frac{d^2\phi^a}{d\lambda^2} 
	+ \frac{1}{6}\frac{d^3\phi^a}{d\lambda^3} + \cdots\right]_{\lambda=0} \, ,
\end{align}
where $\lambda$ is the affine parameter and the geodesic satisfies
\begin{align}
	\mathcal{D}_\lambda \frac{d\phi^a}{d\lambda} \; \equiv \; \frac{d^2\phi^a}{d\lambda^2} 
	+ \Gamma^{a}_{bc}\frac{d\phi^b}{d\lambda}\frac{d\phi^c}{d\lambda} = 0 \, .
\end{align}
We take $\varphi^a$ as the initial velocity, which means
\begin{align}
	\left.\frac{d\phi^a}{d\lambda}\right\vert_{\lambda=0} \; = \; \varphi^a \,.
\end{align}
Then, by using the geodesic equation repeatedly, we obtain\footnote{
	There seems to be a typo in Eq.~(2.4) of~\cite{Elliston:2012ab}.
	This expression agrees with~\cite{Gong:2011uw}.
}
\begin{align}
	\phi^a = \phi^a_0 + \varphi^a - \frac{1}{2}\Gamma^{a}_{bc} \varphi^b \varphi^c + 
	\frac{1}{6}\left(2\Gamma^{a}_{de}\Gamma^{e}_{bc}-\partial_d\Gamma^{a}_{bc}\right)\varphi^b\varphi^c\varphi^d
	+ \cdots \, .
\end{align}
With this definition, $\varphi^a$ lives in the tangent space, allowing everything to be expressed in terms of geometrical quantities. The potential is expanded as
\begin{align}
	V(\phi^a) = \left[
	V + \frac{d V}{d\lambda} + \frac{1}{2}\frac{d^2V}{d\lambda^2} + \frac{1}{6}\frac{d^3V}{d\lambda^3}+\cdots
	\right]_{\lambda=0}.
\end{align}
For instance, the second derivative is computed as
\begin{align}
	\frac{d^2V}{d\lambda^2} = \frac{d}{d\lambda}\left(\frac{d\phi^a}{d\lambda}V_a\right)
	= \left(\frac{d^2\phi^a}{d\lambda^2}V_a + \frac{d\phi^a}{d\lambda}\frac{d\phi^b}{d\lambda}\partial_b V_a\right)
	= \frac{d\phi^a}{d\lambda}\frac{d\phi^b}{d\lambda}\left(\partial_b V_a - \Gamma^{c}_{ab}V_c\right)
	= \frac{d\phi^a}{d\lambda}\frac{d\phi^b}{d\lambda}\nabla_b V_a,
\end{align}
where we used the geodesic equation in the intermediate step. By repeating a similar computation for the third order term,
we can expand the potential as
\begin{align}
	V(\phi) = V + \varphi^a V_a + \frac{1}{2}\varphi^a \varphi^b \nabla_b V_a + \frac{1}{6}\varphi^a \varphi^b \varphi^c
	\nabla_c \nabla_b V_a + \cdots,
\end{align}
where the quantities without the arguments are evaluated by $\phi_0^a$.
The expansion of the kinetic term is more non-trivial. The first order term is given by
\begin{align}
	\frac{d}{d\lambda}\left[\frac{1}{2}h_{ab}\partial_\mu \phi^a \partial^\mu \phi^b\right]
	\; = \; h_{ab}\partial_\mu\left(\frac{d\phi^a}{d\lambda}\right)\partial^\mu \phi^b
	+ \frac{1}{2}h_{ab,c}\frac{d\phi^c}{d\lambda}\partial_\mu \phi^a \partial^\mu \phi^b \,.
\end{align}
We may use that
\begin{align}
	h_{ab,c} \; = \; h_{ad}\Gamma^{d}_{bc} + h_{bd}\Gamma^{d}_{ac} \, .
\end{align}
The above is then simplified as
\begin{align}
	\frac{d}{d\lambda}\left[\frac{1}{2}h_{ab}\partial_\mu \phi^a \partial^\mu \phi^b\right]
	\; = \; h_{ab}\left[
	\partial_\mu\left(\frac{d\phi^a}{d\lambda}\right)
	+ \Gamma^{a}_{cd}(\partial_\mu \phi^c)\frac{d\phi^d}{d\lambda}\right]\partial^\mu \phi^b
	\; \equiv \; h_{ab}\mathcal{D}_\mu\left(\frac{d\phi^a}{d\lambda}\right)\partial^\mu \phi^b \, .
\end{align}
To compute the second and third order terms, it is convenient to note that
\begin{align}
	\left[\mathcal{D}_\lambda, \mathcal{D}_\mu\right]V^a 
	&= {\mathcal{R}^{a}}_{bcd}V^b\frac{d\phi^c}{d\lambda}\partial_\mu \phi^d \,,
\end{align}
for an arbitrary vector $V^a$.
Notice that here $\lambda$ in $\mathcal{D}_\lambda$ 
is the affine parameter and not the spacetime coordinate index.
The second order term is then easily obtained as
\begin{align}
	\frac{d^2}{d\lambda^2}\left[\frac{1}{2}h_{ab}\partial_\mu \phi^a \partial^\mu \phi^b\right]
	&= h_{ab}\left[\mathcal{D}^\mu\left(\frac{d\phi^a}{d\lambda}\right)\mathcal{D}_\mu\left(\frac{d\phi^b}{d\lambda}\right)
	+ \partial^\mu \phi^b \mathcal{D}_\lambda \mathcal{D}_\mu \left(\frac{d\phi^a}{d\lambda}\right)\right]
	\nonumber \\
	&= h_{ab}\mathcal{D}^\mu\left(\frac{d\phi^a}{d\lambda}\right)\mathcal{D}_\mu\left(\frac{d\phi^b}{d\lambda}\right)
	+ \mathcal{R}_{abcd}\partial^\mu \phi^a \partial_\mu \phi^d \frac{d\phi^b}{d\lambda}\frac{d\phi^c}{d\lambda} \,,
\end{align}
where we used the geodesic equation $\mathcal{D}_\lambda  (d\phi^a/d\lambda) = 0$ and the commutator in the last line.
The third order term is also easy to obtain:
\begin{align}
	\frac{d^3}{d\lambda^3}\left[\frac{1}{2}h_{ab}\partial_\mu \phi^a \partial^\mu \phi^b\right]
	&= (\nabla_e \mathcal{R}_{abcd})\partial^\mu \phi^a \partial_\mu \phi^d 
	\frac{d\phi^b}{d\lambda}\frac{d\phi^c}{d\lambda}\frac{d\phi^e}{d\lambda}
	+ 4\mathcal{R}_{abcd}D^\mu\left(\frac{d\phi^a}{d\lambda}\right)\partial_\mu \phi^d 
	\frac{d\phi^b}{d\lambda}\frac{d\phi^c}{d\lambda} \,.
\end{align}
In summary, the kinetic term and the potential are expanded as
\begin{align}
	&\frac{1}{2}h_{ab}(\phi)\partial_\mu \phi^a \partial^\mu \phi^b =
	\frac{1}{2}g^{00}\dot{\phi}_0^2 + \dot{\phi}_0^a g^{0\mu} \mathcal{D}_\mu\varphi_a
	+ \frac{1}{2}\left[h_{ab}g^{\mu\nu}\mathcal{D}_\mu \varphi^a \mathcal{D}_\nu \varphi^b 
	+ g^{00}\dot{\phi}_0^c \dot{\phi}_0^d \mathcal{R}_{acdb}\varphi^a \varphi^b\right]
	\nonumber \\
	&~~~~~~~~~~~~~~~~~~~~~~~~~~~~~~~
	+\frac{1}{6}\left[(\nabla_c \mathcal{R}_{adeb})g^{00}\dot{\phi}_0^d \dot{\phi}_0^e \varphi^a \varphi^b \varphi^c
	+4 \mathcal{R}_{abcd}\dot{\phi}_0^d g^{0\mu}(\mathcal{D}_\mu \varphi^a) \varphi^b \varphi^c\right] + \cdots,
	\\
	&V(\phi) = V + V_a\varphi^a + \frac{1}{2}(\nabla_b V_a) \varphi^a \varphi^b 
	+ \frac{1}{6}(\nabla_c \nabla_b V_a) \varphi^a \varphi^b \varphi^c + \cdots,
\end{align}
where we take $\partial_\mu \phi_0^a = \delta_\mu^0 \dot{\phi}_0^a$.
Note that we have not expanded the metric in the kinetic term yet.

Next, we discuss the constraint equations.
The lapse function $N$ and the shift vector $\beta^i$ do not have the time derivatives,
and thus they are constrained quantities.
To compute the action up to third order, we need to solve the constraints only 
up to first order~\cite{Maldacena:2002vr}.
With this in mind, the quadratic action is given by
\begin{align}
	S_\mathrm{quad}
	\; = \; \int d^4x a^3&\left[\frac{1}{2}h_{ab}\mathcal{D}_t \varphi^a \mathcal{D}_t \varphi^b 
	- \frac{1}{2a^2}h_{ab}\partial_i \varphi^a \partial_i \varphi^b
	-\frac{1}{2}\left(\nabla_b V_a - \dot{\phi}_0^c \dot{\phi}_0^d \mathcal{R}_{acdb}\right)\varphi^a \varphi^b
	\right. \nonumber \\ &\left.
	-2M_P^2 H \left(\alpha - \frac{\dot{\phi}_0}{2M_P^2 H}\varphi\right)\partial_i^2 \beta
	-\left(3M_P^2 H^2 -\frac{\dot{\phi}_0^2}{2}\right)\alpha^2
	-\left(\dot{\phi}_0^a \mathcal{D}_t \varphi_a + V_a \varphi^a\right)\alpha
	\right] \,,
\end{align}
where $\varphi = T^a \varphi_a$, and we used
\begin{align}
	NR^{(3)} + \frac{1}{N}(E_{ij}E^{ij} -E^2) = \frac{1}{N}\left(-6H^2 + 4H\partial_i^2 \beta + 
	(\partial_i \partial_j \beta)^2 - (\partial_i^2\beta)^2\right) \, .
\end{align}
The constraint equation is solved as
\begin{align}
	\alpha \; = \; \frac{H}{\dot{\phi}_0}\epsilon \varphi,
	\quad
	\partial_i^2\beta 
	\; = \; -\frac{1}{2HM_P^2}\left[\left(6M_P^2 H^2 - \dot{\phi}_0^2\right)\alpha + \dot{\phi}_0 T_a \mathcal{D}_t \varphi^a
	+ V_a \varphi^a\right]\, .
\end{align}
The latter is equivalent to
\begin{align}
	\partial_i^2\beta 
	\; = \; -\frac{1}{2HM_P^2}\left[\left(6M_P^2 H^2 - \dot{\phi}_0^2\right)\alpha + \dot{\phi}_0 \dot{\varphi}
	+ V_\phi \varphi 
	+ 2\dot{\phi}_0 H \eta_\perp \chi\right] \, .
\end{align}
%%

%%%%%%%%%%%%%%%%%%%%%%%%%%%%%%%%%%%%%%%
\subsection{Quadratic action}
\label{app:covariant_quadratic}
%%%%%%%%%%%%%%%%%%%%%%%%%%%%%%%%%%%%%%%

After substituting the solution of the constraint equations to the action and performing integration by parts, we arrive at the quadratic action
\begin{align}
	S_\mathrm{quad} &= \int dtd^3x a^3 \left[\frac{1}{2}h_{ab}\mathcal{D}_t \varphi^a \mathcal{D}_t \varphi^b
	- \frac{1}{2a^2}h_{ab}\partial_i \varphi^a \partial_i \varphi^b
	- \frac{1}{2}M_{ab}^2 \varphi^a \varphi^b
	\right] \,,
\end{align}
where the mass term is given by
\begin{align}
	M_{ab}^2 &= \nabla_b V_a - \dot{\phi}_0^2 T^c T^d \mathcal{R}_{acdb} + \frac{2H}{\dot{\phi}_0}\epsilon
	(V_a T_b + V_b T_a) + 2(3-\epsilon)\epsilon H^2 T_a T_b \,.
\end{align}
Note that we have not used the slow-roll approximation to derive it. In the two-field case, we parametrize the fields as
\begin{align}
	\varphi^a \; = \; T^a \varphi + N^a \chi
	= -T^a \frac{\dot{\phi}_0}{H}\zeta + N^a \chi \,,
\end{align}
where $\varphi = -\dot{\phi}_0 \zeta/H$ is the adiabatic mode and $\chi$ is the isocurvature mode. We note that
\begin{align}
	&\dot{\varphi} \; = \; -\frac{\dot{\phi}_0}{H}\left[\dot{\zeta} -\left( (3-\epsilon)H + \frac{V_\phi}{\dot{\phi}_0}\right)\zeta\right] \,,
	\\
	&\frac{1}{\dot{\phi}_0}\frac{dV_\phi}{dt} \; = \; -\dot{\theta}^2 + T^a T^b \nabla_b V_a,
	\quad
	\frac{1}{\dot{\phi}_0}\frac{dV_N}{dt} \; = \; \dot{\theta}\frac{V_\phi}{\dot{\phi}_0} + T^a N^b \nabla_b V_a\,.
\end{align}
After some computation, the action is simplified as
\begin{align}
	S_\mathrm{quad} \; = \; \int d^4x a^3\left[\frac{1}{2}\frac{\dot{\phi}_0^2}{H^2} 
	\left(\dot{\zeta}^2-\frac{1}{a^2}\left(\partial_i \zeta\right)^2\right)
	+ \frac{1}{2}\left(\dot{\chi}^2 - \frac{1}{a^2}\left(\partial_i \chi\right)^2
	-m_\chi^2 \chi^2\right)
	-\frac{2\dot{\theta} \dot{\phi}_0}{H} \dot{\zeta}\chi\right],
\end{align}
where
\begin{align}
	m_\chi^2 \; = \; N^a N^b \nabla_b V_a + \frac{\dot{\phi}_0^2}{2}\mathcal{R} - \dot{\theta}^2,
	\quad
	\mathcal{R} = \mathcal{R}_{ab}h^{ab} \; = \; \mathcal{R}_{acbd}h^{ab}h^{cd} \, .
\end{align}
It is now clear that the turning rate $\dot{\theta}$ controls the mixing between the adiabatic and isocurvature modes.

%%%%%%%%%%%%%%%%%%%%%%%%%%%%%%%%%%%%%%%
\subsection{Cubic action}
\label{app:covariant_cubic}
%%%%%%%%%%%%%%%%%%%%%%%%%%%%%%%%%%%%%%%

Using the results from App.~\ref{app:covariant_perturbation},
the cubic action is straightforward to derive.
The pure gravity part is given by
\begin{align}
	S_\mathrm{cubic}^{(\mathrm{grav})} &= \int d^4x a^3
	\left[\left(3M_P^2 H^2-\frac{\dot{\phi}_0^2}{2}\right) \alpha^3 + 2M_P^2 H \alpha^2 \partial_i^2 \beta
	- \frac{M_P^2}{2}\alpha \left((\partial_i \partial_j \beta)^2 - (\partial_i^2\beta)^2\right)\right] \, .
\end{align}
The gravity-matter mixing part is given by
\begin{align}
	S_\mathrm{cubic}^{(\mathrm{mix})} = \int d^4x a^3
	&\left[(\partial_i \beta)\left(\dot{\phi}_0\alpha \partial_i \varphi 
	- \partial_i \varphi_a \mathcal{D}_t\varphi^a\right)
	+ \alpha^2 \dot{\phi}_0^a \mathcal{D}_t\varphi_a
	\right. \nonumber \\ &\left.
	-\frac{\alpha}{2}\left(\mathcal{D}_t\varphi^a \mathcal{D}_t\varphi_a
	+ \frac{1}{a^2}h_{ab}\partial_i \varphi^a \partial_i \varphi^b
	+ \left(\nabla_b V_a + \dot{\phi}_0^2 T^c T^d \mathcal{R}_{acdb}\right) \varphi^a\varphi^b
	\right)\right] \, .
\end{align}
Finally, the pure matter sector is given by
\begin{align}
	S_\mathrm{cubic}^{(\mathrm{matter})}
	= \int d^4x a^3
	\left[
	\frac{1}{6}(\nabla_c \mathcal{R}_{adeb})\dot{\phi}_0^d \dot{\phi}_0^e \varphi^a \varphi^b \varphi^c
	+ \frac{2}{3}\mathcal{R}_{abcd}\dot{\phi}_0^d (\mathcal{D}_t \varphi^a) \varphi^b \varphi^c
	-\frac{1}{6}(\nabla_c \nabla_b V_a)\varphi^a \varphi^b \varphi^c\right]\,.
\end{align}
They correctly reproduce the results from Refs.~\cite{Gong:2011uw,Elliston:2012ab}.

%%%%%%%%%%%%%%%%%%%%%%%%%%%%%%%%%%%%%%%
\section{Schwinger-Keldysh propagators in de~Sitter spacetime}
\label{app:propagators}
%%%%%%%%%%%%%%%%%%%%%%%%%%%%%%%%%%%%%%%
In this appendix, we review the Schwinger-Keldysh propagators of 
the scalar, fermion, and massive gauge boson during inflation.
We ignore the slow-roll parameters and take the background spacetime as the pure de~Sitter one.
The spacetime dimension is taken to be $d$ in this appendix.
See e.g.~\cite{Chen:2017ryl} and references therein for more details on the Schwinger-Keldysh formalism
in the context of cosmological collider physics.

%%%%%%%%%%%%%%%%%%%%%%%%%%%%%%%%%%%%%%%
\subsection{Preliminary}
%%%%%%%%%%%%%%%%%%%%%%%%%%%%%%%%%%%%%%%
First, we summarize several equations that are repeatedly used in the derivation of the propagators.
We will encounter the mode equation of the form
\begin{align}
	0 \; = \; \left[\frac{d^2}{d\tau^2}+k^2 -\frac{\nu^2-1/4}{\tau^2} \right]v_k \,,
	\label{eq:Bessel}
\end{align}
where $\tau$ is the conformal time and $v_k$ is the mode function with $k$ 
the size of its momentum.
By defining $z = -k\tau$ and $\tilde{v}_k = v_k/\sqrt{z}$, 
this is recast as Bessel's differential equation
\begin{align}
	z^2 \frac{d^2 \tilde{v}_k}{dz^2} + z\frac{d\tilde{v}_k}{dz} + \left(z^2 - \nu^2\right)\tilde{v}_k \; = \; 0.
\end{align}
A general solution is given by a linear combination of the Hankel functions as
\begin{align}
	v_k \; = \; z^{1/2}\left(c_1 H_{\nu}^{(1)}(z) + c_2H_{\nu}^{(2)}(z)\right) \, .
\end{align}
The Hankel function of the first (second) kind corresponds to the positive (negative) frequency mode 
in the asymptotic past $z \to \infty$. Consequently, the Bunch-Davies vacuum condition eliminates the latter mode.

The Hankel function satisfies several relations useful for our purpose. First, the Hankel functions of the first and second kind are related to each other 
by the complex conjugates as
\begin{align}
	\left(H_\nu^{(1)}(z)\right)^* \; = \; H_{\nu^*}^{(2)}(z) \,,
\end{align}
where we assume that the argument $z$ is real.
They also satisfy
\begin{align}
	\left(\frac{d}{dz}+\frac{\nu}{z}\right)H_\nu^{(i)}(z) \; = \; H_{\nu-1}^{(i)}(z) \, ,
	\quad
	H_{-\nu}^{(1)} \; = \; e^{i\pi \nu} H_{\nu}^{(1)} \,,
	\quad
	H_{-\nu}^{(2)} \; = \; e^{-i\pi \nu} H_{\nu}^{(2)} \,,
	\label{eq:Hankel_rel1}
\end{align}
and the latter two equations indicate that
\begin{align}
	\left(e^{i\pi\nu/2}H_\nu^{(1)}(z)\right)^*
	\; = \; e^{-i\pi\nu/2}H_{\nu}^{(2)}(z) \, ,
	\label{eq:Hankel_rel2}
\end{align}
if $\nu$ is either real or pure imaginary.
The Wronskian of the Hankel functions is given by
\begin{align}
	H_{\nu+1}^{(1)}(z) H_{\nu}^{(2)}(z) - H_{\nu}^{(1)}(z)H_{\nu+1}^{(2)}(z) \; = \; -\frac{4i}{\pi z}.
	\label{eq:Hankel_Wronskian}
\end{align}

When we compute the propagators in the coordinate space, 
we need to integrate the product of the Hankel functions.
The result is the hypergeometric function
\begin{align}
	\frac{\pi}{4}\int \frac{d^{d-1}k}{(2\pi)^{d-1}}e^{i\vec{k}\cdot(\vec{x}_1 - \vec{x}_2)} H_{\nu}^{(1)}(z_1) 
	H_{\nu}^{(2)}(z_2)
	\; = \; \frac{I_\nu(Z_{12})}{(4\pi)^{d/2}(\tau_1\tau_2)^{\frac{d-1}{2}}} \, ,
	\label{eq:Hankel_integral}
\end{align}
where $z_1 = -k\tau_1$, $z_2 = -k\tau_2$,
\begin{align}
	{I}_{\nu}(Z)
	&\equiv \frac{\Gamma\left(\frac{d-1}{2}+\nu\right)\Gamma\left(\frac{d-1}{2}-\nu\right)}
	{\Gamma(d/2)}
	{}_2F_1\left(\frac{d-1}{2}+\nu, \frac{d-1}{2}-\nu; \frac{d}{2};\frac{1+Z}{2}\right) \,,
\end{align}
and
\begin{align}
	Z_{ij} \; = \; 1-\frac{x_{ij}^2 - (\tau_i - \tau_j-i0)^2}{2\tau_i \tau_j} \, ,
\end{align}
with $x_{ij} = \vert \vec{x}_i - \vec{x}_j\vert$.
Here we keep the $i\epsilon$ prescription 
required to make the integral convergent explicit.
We may note that $I_\nu$ satisfies the differential equation
\begin{align}
	0 &= (1-Z^2)I_\nu''(Z) - dZ I_\nu'(Z) - \left(d-2+\frac{m^2}{H^2}\right)I_\nu(Z) \, .
	\label{eq:Ideriv}
\end{align}
We also define
\begin{align}
	Z_{++} &=
	1-\frac{x_{12}^2 - (\vert \tau_1 - \tau_2\vert - i0)^2}{2\tau_1 \tau_2} \, ,
	\quad
	Z_{--} = 1-\frac{x_{12}^2 - (\vert \tau_1 - \tau_2\vert + i0)^2}{2\tau_1 \tau_2} \,,
	\\
	Z_{+-} &= 1-\frac{x_{12}^2 - ( \tau_1 - \tau_2 + i0)^2}{2\tau_1 \tau_2} \,,
	\quad
	~~Z_{-+} = 1-\frac{x_{12}^2 - ( \tau_1 - \tau_2 - i0)^2}{2\tau_1 \tau_2}\,,
\end{align}
where the subscripts indicate different contours in the Schwinger-Keldysh formalism. 
Finally, the following relations turn out to be useful to derive the gauge boson propagator:
\begin{align}
	&\partial_{\tau_1}Z_{12} = \frac{1}{\tau_2}-\frac{Z_{12}}{\tau_1},
	\quad
	\partial_{\tau_2}Z_{12} = \frac{1}{\tau_1}-\frac{Z_{12}}{\tau_2},
	\quad
	\partial_{\tau_1}\partial_{\tau_2}Z_{12} 
	= -\left(\frac{1}{\tau_1^2} + \frac{1}{\tau_2^2}\right)+\frac{Z_{12}}{\tau_1\tau_2},
	\label{eq:Zderiv1}
	\\
	&\partial_{1i}Z_{12} = -\partial_{2i}Z_{12} = -\frac{(x_1-x_2)_i}{\tau_1\tau_2},
	\quad
	\partial_{1i}\partial_{2j}Z_{12} = \frac{\delta_{ij}}{\tau_1\tau_2},
	\quad
	\frac{x_{12}^2}{\tau_1^2\tau_2^2} = \frac{1}{\tau_1^2}+\frac{1}{\tau_2^2} - \frac{2Z_{12}}{\tau_1\tau_2}.
	\label{eq:Zderiv2}
\end{align}
It follows that
\begin{align}
	&(\partial_{1\alpha}Z_{12})(\partial_1^\alpha Z_{12}) = \frac{Z_{12}^2-1}{\tau_1^2},
	\quad
	(\partial_{2\alpha}Z_{12})(\partial_2^\alpha Z_{12}) = \frac{Z_{12}^2-1}{\tau_2^2},
	\\
	&(\partial_{1\alpha}\partial_{2\beta}Z_{12})(\partial_{1}^{\alpha}\partial_2^\beta Z_{12})
	= \frac{Z_{12}^2+3}{\tau_1^2\tau_2^2},
	\quad
	(\partial_{1\alpha}Z_{12})(\partial_{2\beta}Z_{12})\partial_1^\alpha\partial_2^\beta Z_{12}
	= \frac{Z_{12}(Z_{12}^2-1)}{\tau_1^2\tau_2^2} \,,
\end{align}
where the contractions are taken with respect to $\eta_{\alpha\beta}$.

%%%%%%%%%%%%%%%%%%%%%%%%%%%%%%%%%%%%%%%
\subsection{Scalar}
\label{app:scalar}
%%%%%%%%%%%%%%%%%%%%%%%%%%%%%%%%%%%%%%%

We now review the derivation of the scalar field propagators in de~Sitter spacetime.
We consider a massive real scalar field $\chi$ with a non-minimal coupling $\xi$ 
(just for completeness) as
\begin{align}
	S_\chi \; = \; \int d^dx \sqrt{-g}\left[\frac{1}{2}g^{\mu\nu}\partial_\mu \chi \partial_\nu \chi - \frac{m^2}{2}\chi^2
	+ \frac{\xi}{2}R \chi^2\right] \, .
\end{align}
Here $\chi$ can be either adiabatic or isocurvature modes.

%%%%%%%%%%
\subsubsection*{Mode equation}
%%%%%%%%%%

We may quantize the field as
\begin{align}
	\hat{\chi}(\tau, \vec{x}) \; = \; a^{-\frac{d-2}{2}}\int\frac{d^3k}{(2\pi)^3}
	e^{i\vec{k}\cdot\vec{x}} \left[\chi(\vec{k}, \tau) \hat{a}_{\vec{k}} + 
	\chi^*(-\vec{k}, \tau) \hat{a}_{-\vec{k}}^\dagger\right] \,.
\end{align}
The mode function satisfies
\begin{align}
	\left[\frac{d^2}{d\tau^2} + k^2 + \left(\frac{m^2}{H^2}-\frac{d(d-2)}{4}-d(d-1)\xi\right)\frac{1}{\tau^2}\right]\chi \; = \; 0\,,
\end{align}
and the creation-annihilation operator satisfies
\begin{align}
	\left[\hat{a}_{\vec{k}}, \hat{a}^\dagger_{\vec{k}'}\right]
	&= (2\pi)^{d-1}\delta^{(d-1)}(\vec{k}-\vec{k}') \,.
\end{align}
This is Bessel differential equation~\eqref{eq:Bessel}, and hence the solution is given by
\begin{align}
	\chi \; = \; \sqrt{\frac{-\pi\tau}{4}}e^{i\pi\nu/2} H_\nu^{(1)}(-k\tau)\,,
	\quad
	\nu \; = \; \sqrt{\frac{(d-1)^2}{4}+d(d-1)\xi - \frac{m^2}{H^2}}\, .
\end{align}
where we assume the Bunch-Davies vacuum condition.

%%%%%%%%%%
\subsubsection*{Propagators}
%%%%%%%%%%
For our purpose, it is convenient to define the scalar propagators in the momentum space.
We define the scalar two-point function as
\begin{align}
	\vev{\hat{\chi}(\tau_1,\vec{x}_1) \hat{\chi}(\tau_2, \vec{x}_2)}
	&= \int\frac{d^{d-1}k}{(2\pi)^{d-1}}e^{i\vec{k}\cdot(\vec{x}_1-\vec{x}_2)}
	G(k;\tau_1, \tau_2),
	\quad
	G(k;\tau_1, \tau_2)
	= \frac{\pi\sqrt{\tau_1\tau_2}}{4(a_1a_2)^{\frac{d-2}{2}}}H_\nu^{(1)}(-k\tau_1)H_\nu^{(2)}(-k\tau_2),
\end{align}
where we used Eq.~\eqref{eq:Hankel_rel2}.
The corresponding expression in the coordinate space is given by
\begin{align}
	\vev{\hat{\chi}(\tau_1,\vec{x}_1) \hat{\chi}(\tau_2, \vec{x}_2)}
	&= \frac{H^{d-2}}{(4\pi)^{d/2}}I_\nu(Z_{12})\,,
\end{align}
where we used Eq.~\eqref{eq:Hankel_integral}.
In the Schwinger-Keldysh formalism, there are four distinct propagators, each corresponding to the selection of fields on different contours. These are given by
\begin{align}
	G_{++}(k;\tau_1,\tau_2) &= G(k;\tau_1,\tau_2)\theta(\tau_1-\tau_2) + G(k;\tau_2,\tau_1)\theta(\tau_2-\tau_1) \,,
	\\
	G_{--}(k;\tau_1,\tau_2) &= G(k;\tau_2,\tau_1)\theta(\tau_1-\tau_2) + G(k;\tau_1,\tau_2)\theta(\tau_2-\tau_1)
	= G_{++}^*(k;\tau_1,\tau_2)\,,
	\\
	G_{-+}(k;\tau_1,\tau_2) &= G(k;\tau_1,\tau_2)\,,
	\quad
	G_{+-}(k;\tau_1,\tau_2) = G(k;\tau_2,\tau_1)
	= G_{-+}^*(k;\tau_1,\tau_2)\,,
\end{align}
where the subscript ``$\pm$" indicates the time-ordered and anti-time-ordered contours.
We may use $G$ for a general massive scalar field (and hence the isocurvature mode), and $\Delta$
specific to a massless scalar field with $\xi=0$ (including the adiabatic mode),
corresponding to $\nu = (d-1)/2$.
In particular, in $d=4$, we obtain
\begin{align}
	\Delta(k;\tau_1,\tau_2) &= \frac{H^2}{2k^3}(1+ik\tau_1)(1-ik\tau_2) e^{-ik(\tau_1-\tau_2)} \,.
\end{align}
It is useful to note that
\begin{align}
	\partial_\tau \Delta(k;\tau,0) \; = \; \frac{H^2\tau}{2k}e^{-ik\tau} \,,
	\quad
	\partial_\tau \Delta(k;0,\tau) \; = \; \frac{H^2\tau}{2k}e^{ik\tau} \, .
\end{align}
%%

%%%%%%%%%%%%%%%%%%%%%%%%%%%%%%%%%%%%%%%
\subsection{Fermion}
\label{app:fermion}
%%%%%%%%%%%%%%%%%%%%%%%%%%%%%%%%%%%%%%%

Next, we review the derivation of the massive Dirac field propagator in de~Sitter spacetime~\cite{Candelas:1975du,Allen:1986qj,Miao:2006pn,Koksma:2009tc}.
The action is given by
\begin{align}
	S_\psi \; = \; \int d^d x \sqrt{-g}\,\bar{\psi}\left(i\slashed{\nabla}-m\right)\psi \,.
\end{align}
In de~Sitter spacetime, this reduces to
\begin{align}
	S_\psi \; = \; \int d\tau d^{d-1}x\,\bar{\tilde{\psi}}\left[i\gamma^b \partial_b - a m\right] \tilde{\psi}\,,
\end{align}
where we used 
\begin{align}
	i\slashed{\nabla} \; = \; a^{-\frac{d+1}{2}}i\gamma^b \partial_b \,a^{\frac{d-1}{2}} \,,
\end{align}
and redefined the fermion as $\tilde{\psi} = a^{(d-1)/2}\psi$, where $b$ runs the local Lorentz indices.
This indicates that the massless fermion is conformal in an arbitrary spacetime dimension. The Dirac equation follows as
\begin{align}
	\left[i\gamma^b \partial_b - am\right] \tilde{\psi} \; = \; 0 \, .
\end{align}
%%

%%%%%%%%%%
\subsubsection*{Mode equation}
%%%%%%%%%%

By going to the Fourier space
\begin{align}
	\psi \; = \; a^{-\frac{d-1}{2}}\int \frac{d^{d-1}k}{(2\pi)^{d-1}}e^{i\vec{k}\cdot \vec{x}} \tilde{\psi}(\vec{k},\tau) \,,
\end{align}
we obtain the mode equation 
\begin{align}
	\left[i\gamma^0 \frac{d}{d\tau} - \vec{k}\cdot\vec{\gamma} - a m\right]\tilde{\psi} \; = \; 0 \,.
	\label{eq:Dirac_mom_bfrquad}
\end{align}
To solve this, we act the operator ``$i\gamma^0 d/d\tau - \vec{k}\cdot\vec{\gamma} + a m$", and obtain
\begin{align}
	\left[\frac{d^2}{d\tau^2} + k^2 + i m \frac{da}{d\tau} \gamma^0 + a^2 m^2\right]\tilde{\psi} \; = \; 0 \, .
\end{align}
We decompose the modes as
\begin{align}
	\tilde{\psi} \; = \; \sum_{\lambda, h = \pm} \tilde{\psi}_{\lambda h} \,\xi_\lambda \otimes \chi_h \, ,
\end{align}
where we define the spinors in the subspaces as
\begin{align}
	\gamma^0 \xi_\lambda \; = \; \lambda \xi_\lambda,
	\quad
	\vec{\sigma}\cdot\vec{k}\,\chi_h = hk \chi_h \,,
	\quad
	\sum_\lambda \xi_\lambda \xi^\dagger_\lambda
	\; = \; 
	\sum_h \chi_h \chi^\dagger_h = \mathbbm{1}_{2\times 2} \, ,
\end{align}
with $h$ denoting the helicity and $\lambda$ denoting the eigenvalue of $\gamma^0$.
The mode equation is then given by
\begin{align}
	\left[\partial_\tau^2 + k^2 + \frac{1}{\tau^2}\left(\frac{m^2}{H^2}+ \lambda \frac{im}{H}\right)\right]\tilde{\psi}_{\lambda h} \; = \; 0 \,,
\end{align}
which is of the form~\eqref{eq:Bessel}, and thus we obtain
\begin{align}
	\tilde{\psi}_{\lambda h} &= \sqrt{\frac{\pi z}{4}}
	\left[e^{i \pi \nu_\lambda/2}
	H_{\nu_\lambda}^{(1)}(z) \times b_{\lambda h}
	+ 
	e^{-i \pi \nu_\lambda/2}
	H_{\nu_\lambda}^{(2)}(z) \times d_{\lambda -h}^*\right] \, ,
\end{align}
where $z = -k\tau$.
Here we keep both the positive and negative energy solutions, corresponding to the particle and anti-particle, and define
\begin{align}
	\nu_\lambda \; = \; \frac{1}{2}-\lambda\frac{im}{H} \,.
\end{align}
We now plug this expression back into Eq.~\eqref{eq:Dirac_mom_bfrquad} and obtain\footnote{
	Here we implicitly fix the convention for the relative sign between $\xi_+$ and $\xi_-$,
	such that it can be expressed as $\xi_\pm^\dagger = (1, \pm1)/\sqrt{2}$ in the Weyl representation.
}
\begin{align}
	&\left(i\frac{d}{d\tau} + \frac{m}{H\tau}\right)\tilde{\psi}_{+h} + h k \tilde{\psi}_{-h} = 0,
	\quad
	\left(i\frac{d}{d\tau} - \frac{m}{H\tau}\right)\tilde{\psi}_{-h} + h k \tilde{\psi}_{+h} = 0 \,.
	\label{eq:Dirac_original_eq}
\end{align}
With the relations~\eqref{eq:Hankel_rel1}, this is solved as
\begin{align}
	b_{-h} \; = \; -h b_{+h} \,,
	\quad
	d_{-h} \; = \; -h d_{+h} \,.
\end{align}
After quantization, we obtain
\begin{align}
	\hat{\psi}(\vec{x},\tau)
	&= a^{-(d-1)/2}
	\int \frac{d^{d-1}k}{(2\pi)^{d-1}}
	\sqrt{\frac{-\pi k \tau}{4}} e^{i\vec{k}\cdot\vec{x}}
	\sum_{h=\pm} \chi_h 
	\nonumber \\
	&\otimes 
	\left[\left(e^{i\pi\nu_+/2}H_{\nu_+}^{(1)} \xi_+ - he^{i\pi \nu_-/2} H_{\nu_-}^{(1)}\xi_-\right)
	\hat{b}_{\vec{k},h}
	%\right. \nonumber \\ &\left.
	+ \left(e^{-i\pi\nu_+/2}H_{\nu_+}^{(2)} \xi_+ + he^{-i\pi \nu_-/2} H_{\nu_-}^{(2)}\xi_-\right)
	\hat{d}^\dagger_{-\vec{k},-h}
	\right] \,.
\end{align}
The quantization condition is
\begin{align}
	a^{d-1}\left\{\hat{\psi}(\vec{x},\tau), \hat{\psi}^\dagger(\vec{y},\tau)\right\}
	=  \delta^{(d-1)}(\vec{x}-\vec{y})\times \mathbbm{1} \,,
\end{align}
which results in
\begin{align}
	\left\{\hat{b}_{\vec{k},h}, \hat{b}_{\vec{k}',h'}^\dagger\right\}
	&= \left\{\hat{d}_{\vec{k},h}, \hat{d}_{\vec{k}',h'}^\dagger\right\}
	= (2\pi)^{d-1}\delta^{(d-1)}(\vec{k}-\vec{k}') \delta_{hh'} \,.
\end{align}
Here we used
\begin{align}
	e^{i\pi(\nu_+ - \nu_-)/2} H_{\nu_+}^{(1)} H_{\nu_-}^{(2)}
	+ e^{-i\pi(\nu_+-\nu_-)/2} H_{\nu_-}^{(1)}H_{\nu_+}^{(2)}
	= \frac{4}{\pi z} \,,
\end{align}
which follows from Eqs.~\eqref{eq:Hankel_rel1} and~\eqref{eq:Hankel_Wronskian}.

%%%%%%%%%%
\subsubsection*{Propagators}
%%%%%%%%%%

Next, we compute the propagators.
We define them as
\begin{align}
	\left[S_{1}(\vec{x}_1 - \vec{x}_2; \tau_1,\tau_2)\right]_{ab}
	&= \langle {\psi}_{a}(\tau_1, \vec{x}_1) \bar{\psi}_{b}(\tau_2,\vec{x}_2)\rangle \,,
	\quad
	\left[S_{2}(\vec{x}_1 - \vec{x}_2; \tau_1,\tau_2)\right]_{ab}
	= -\langle \bar{\psi}_{b}(\tau_2,\vec{x}_2) {\psi}_{a}(\tau_1, \vec{x}_1) \rangle \,.
\end{align}
To evaluate these expressions, we note that\footnote{
	We assume the convention for the relative sign between $\xi_\pm$
	that is consistent with Eq.~\eqref{eq:Dirac_original_eq}.
	With this and the previous choices,
	the propagators do not depend on the sign convention.
}
\begin{align}
	\sum_{h}\chi_h \chi_h^\dagger \otimes \xi_\pm \xi^\dagger_\pm = \frac{1\pm\gamma^0}{2},
	\quad
	\sum_h h \chi_h \chi_h^\dagger \otimes \xi_\mp \xi^\dagger_\pm = 
	\pm \frac{\vec{k}\cdot\vec{\gamma}}{k}\frac{1\pm\gamma^0}{2},
\end{align}
which one can check, for example, by using a specific representation of the Dirac matrices. We then obtain
\begin{align}
	S_1(\vec{x}_1-\vec{x}_2; \tau_1, \tau_2)
	=&~a_1^{-\frac{d-1}{2}}a_2^{-\frac{d-1}{2}}\frac{\pi \sqrt{\tau_1\tau_2}}{4}
	\int\frac{d^{d-1}k}{(2\pi)^{d-1}} e^{i\vec{k}\cdot(\vec{x}_1-\vec{x}_2)}
	\nonumber \\
	\times& \left[k \left(e^{i\pi\frac{\nu_+-\nu_-}{2}} H_{\nu_+}^{(1)}(z_1)H_{\nu_-}^{(2)}(z_2)
	\frac{1+\gamma^0}{2}
	- e^{-i\pi\frac{\nu_+-\nu_-}{2}}H_{\nu_-}^{(1)}(z_1)H_{\nu_+}^{(2)}(z_2) \frac{1-\gamma^0}{2}\right)
	\right. \nonumber \\ &~\left.
	-\vec{k}\cdot\vec{\gamma}
	\left(
	H_{\nu_-}^{(1)}(z_1)H_{\nu_-}^{(2)}(z_2) \frac{1+\gamma^0}{2}
	+ H_{\nu_+}^{(1)}(z_1)H_{\nu_+}^{(2)}(z_2) \frac{1-\gamma^0}{2}
	\right)
	\right]\,,
\end{align}
and
\begin{align}
	S_2(\vec{x}_1-\vec{x}_2; \tau_1, \tau_2)
	=&~a_1^{-\frac{d-1}{2}}a_2^{-\frac{d-1}{2}}\frac{\pi \sqrt{\tau_1\tau_2}}{4}
	\int\frac{d^{d-1}k}{(2\pi)^{d-1}} e^{i\vec{k}\cdot(\vec{x}_1-\vec{x}_2)}
	\nonumber \\
	\times 
	&\left[
	k\left(-e^{-i\pi\frac{\nu_+-\nu_-}{2}}H_{\nu_+}^{(2)}(z_1)H_{\nu_-}^{(1)}(z_2)\frac{1+\gamma^0}{2}
	+ e^{i\pi\frac{\nu_+-\nu_-}{2}}H_{\nu_-}^{(2)}(z_1)H_{\nu_+}^{(1)}(z_2)\frac{1-\gamma^0}{2}\right)
	\right. \nonumber \\ &~\left.
	- \vec{k}\cdot\vec{\gamma}
	\left(
	H_{\nu_-}^{(2)}(z_1)H_{\nu_-}^{(1)}(z_2)\frac{1+\gamma^0}{2}
	+ H_{\nu_+}^{(2)}(z_1)H_{\nu_+}^{(1)}(z_2)\frac{1-\gamma^0}{2}
	\right)
	\right] \,,
\end{align}
where $a_i = a(\tau_i)$ and $z_i = -k\tau_i$.
Eq.~\eqref{eq:Hankel_rel1} tells us that
\begin{align}
	k e^{i\pi\frac{\nu_\lambda-\nu_{-\lambda}}{2}}H_{\nu_\lambda}^{(1)}(z)
	= i\left(\frac{d}{d\tau}+\frac{\nu_{-\lambda}}{\tau}\right)H_{\nu_{-\lambda}}^{(1)}(z),
	\quad
	k e^{-i\pi\frac{\nu_\lambda-\nu_{-\lambda}}{2}}H_{\nu_\lambda}^{(2)}(z)
	= -i\left(\frac{d}{d\tau}+\frac{\nu_{-\lambda}}{\tau}\right)H_{\nu_{-\lambda}}^{(2)}(z),
\end{align}
and this allows us to extract the derivatives as
\begin{align}
	S_1(\vec{x}_1-\vec{x}_2; \tau_1, \tau_2)
	&= a_1^{-\frac{d-1}{2}}a_2^{-\frac{d-1}{2}}\left[i\gamma^b \partial_{b} + a m\right]_{1}
	\nonumber \\
	&\times \int\frac{d^{d-1}k}{(2\pi)^{d-1}}e^{i\vec{k}\cdot(\vec{x}_1-\vec{x}_2)}
	\frac{\pi\sqrt{\tau_1\tau_2}}{4}\left[
	H_{\nu_-}^{(1)}(z_1)H_{\nu_-}^{(2)}(z_2) \frac{1+\gamma^0}{2}
	+ H_{\nu_+}^{(1)}(z_1)H_{\nu_+}^{(2)}(z_2) \frac{1-\gamma^0}{2}
	\right],
	\\
	S_2(\vec{x}_1-\vec{x}_2; \tau_1, \tau_2)
	&= a_1^{-\frac{d-1}{2}}a_2^{-\frac{d-1}{2}}\left[i\gamma^b \partial_{b} + a m\right]_{1}
	\nonumber \\
	&\times \int\frac{d^{d-1}k}{(2\pi)^{d-1}}e^{i\vec{k}\cdot(\vec{x}_1-\vec{x}_2)}
	\frac{\pi\sqrt{\tau_1\tau_2}}{4}\left[
	H_{\nu_-}^{(2)}(z_1)H_{\nu_-}^{(1)}(z_2) \frac{1+\gamma^0}{2}
	+ H_{\nu_+}^{(2)}(z_1)H_{\nu_+}^{(1)}(z_2) \frac{1-\gamma^0}{2}
	\right],
\end{align}
where the subscript ``1" indicates that $a$ is evaluated at $\tau=\tau_1$ 
and the derivatives are acting on $\tau_1, \vec{x}_1$.
Now the momentum integral is of the form~\eqref{eq:Hankel_integral},
and by rewriting the ordinary derivative to the covariant derivative,
we obtain
\begin{align}
	S_1(\vec{x}_{1}-\vec{x}_2;\tau_1, \tau_2)
	&= \frac{H^{d-2}}{(4\pi)^{d/2}}\left[a\left(i\slashed{\nabla}+m\right)\right]_{1}
	\frac{1}{\sqrt{a_1 a_2}}\left[\frac{1+\gamma^0}{2} I_{\nu_-}(Z_{12}) + \frac{1-\gamma^0}{2}I_{\nu_+}(Z_{12})\right],
	\\
	S_2(\vec{x}_{1}-\vec{x}_2;\tau_1, \tau_2)
	&= \frac{H^{d-2}}{(4\pi)^{d/2}}\left[a\left(i\slashed{\nabla}+m\right)\right]_{1}
	\frac{1}{\sqrt{a_1 a_2}}\left[\frac{1+\gamma^0}{2} I_{\nu_-}(Z_{21}) + \frac{1-\gamma^0}{2}I_{\nu_+}(Z_{21})\right].
\end{align}
From this, we obtain the four propagators in the Schwinger-Keldysh formalism as
\begin{align}
	S_{\lambda \lambda'}(\vec{x}_{1}-\vec{x}_2;\tau_1, \tau_2)
	&= \frac{H^{d-2}}{(4\pi)^{d/2}}\left[a\left(i\slashed{\nabla}+m\right)\right]_{1}
	\frac{1}{\sqrt{a_1 a_2}}\left[\frac{1+\gamma^0}{2} I_{\nu_-}(Z_{\lambda \lambda'}) 
	+ \frac{1-\gamma^0}{2}I_{\nu_+}(Z_{\lambda \lambda'})\right] \,,
\end{align}
where $\lambda, \lambda' = \pm$ correspond to the different contours in the Schwinger-Keldysh formalism.
The time-ordered propagator correctly reproduces the results in~\cite{Koksma:2009tc,Chen:2016hrz}.
Note that the propagators differ only in the sign of the $i\epsilon$ prescription in the embedding distance,
as in the flat spacetime case. 
This indicates that, in the late-time expansion, i.e. $\tau_1, \tau_2 \to 0$ while keeping $x_{12}$ finite,
these differences disappear and all the propagators give the same result.

%%%%%%%%%%%%%%%%%%%%%%%%%%%%%%%%%%%%%%%
\subsection{Massive gauge boson}
\label{app:gauge_boson}
%%%%%%%%%%%%%%%%%%%%%%%%%%%%%%%%%%%%%%%

Finally, we review the derivation of the massive gauge boson propagator 
in de~Sitter spacetime~\cite{Allen:1985wd,Frob:2013qsa}.
We consider the action in the $R_\xi$-gauge, given by
\begin{align}
	S_A \; = \; \int d^dx \sqrt{-g}\left[-\frac{1}{4}F^{\mu\nu}F_{\mu\nu} + \frac{m^2}{2}A_\mu A^\mu
	- \frac{1}{2\xi_g}\left(\nabla^\mu A_\mu\right)^2\right] \,,
\end{align}
where $\xi_g$ is the gauge fixing parameter.
Here we do not write down the kinetic mixing part and the pure Goldstone part in the gauge fixing.
The former cancels with the mixing from the kinetic term of the Higgs,
while the latter (together with the Goldstone modes themselves) can be ignored 
in the unitary gauge $\xi_g \to \infty$, which we take in this paper.
Below we derive the propagator in the standard canonical quantization method,
following Ref.~\cite{Frob:2013qsa}.

%%%%%%%%%%
\subsubsection*{Mode equation}
%%%%%%%%%%

In de~Sitter spacetime, the action is given by
\begin{align}
	S_A \; = \; \int d\tau d^{d-1}x&\left[
	\frac{a^{d-4}}{2}\left(\partial_\tau A_i - \partial_i A_0\right)^2
	- \frac{a^{d-4}}{2}\left((\partial_i A_j)^2 - (\partial_i A_i)^2\right)
	\right. \nonumber \\ &\left.
	+ \frac{m^2}{2}a^{d-2}\left(A_0^2 - A_i^2\right)
	- \frac{a^{d-4}}{2\xi_g}\left(\partial_\tau A_0 + (d-2)\frac{da/d\tau}{a}A_0 - \partial_i A_i\right)^2
	\right] \,.
\end{align}
Note that $A_0$ has the kinetic term since we keep $\xi_g$ finite at this moment.
It is convenient to define the conjugate momenta to solve the mode equations. 
They are given by
\begin{align}
	\pi^0 \; = \; -\frac{a^{d-4}}{\xi_g}\left[\partial_\tau A_0 + (d-2)\frac{da/d\tau}{a}A_0 - \partial_i A_i\right] \, ,
	\quad
	\pi^i \; = \; a^{d-4}\left(\partial_\tau A_i - \partial_i A_0\right)\,,
\end{align}
and with them the equation of motion is given by
\begin{align}
	0 &= 
	\partial_\tau \pi^i - a^{d-4}(\nabla^2 A_i - \partial_i \partial_j A_j) + m^2 a^{d-2}A_i - \partial_i \pi^0 \, ,
	\\
	0 &=
	\partial_\tau \pi^0 - (d-2)\frac{da/d\tau}{a}\pi^0 - m^2 a^{d-2}A_0 - \partial_i \pi^i \,,
\end{align}
where $\nabla^2 = \partial_i^2$.
To disentangle the equations, we may define the transverse and longitudinal modes as
\begin{align}
	A_i \; = \; A_{Ti} + \partial_i A_L,
	\quad
	\pi^i \; = \; \pi_T^i + \partial^i \pi_L = \pi_T^i - \partial_i \pi_L \,.
\end{align}
The transverse part is conveniently written in terms of $A_{Ti}$ as
\begin{align}
	0 &= 
	\left[\partial_\tau^2 + (d-4)\frac{da/d\tau}{a}\partial_\tau - \nabla^2 + m^2 a^2\right]A_{Ti},
	\quad
	\pi^i_T = a^{d-4}\partial_\tau A_{Ti}.
\end{align}
The longitudinal and temporal modes are mixed in terms of $A_L$ and $A_0$, but are
decoupled in terms of $\pi_L$ and $\pi^0$ as
\begin{align}
	0 &=
	\left[\partial_\tau^2 - (d-2)\frac{da/d\tau}{a}\partial_\tau - \nabla^2 + m^2 a^2\right] \pi_L,
	\\
	0 &= 
	\left[\partial_\tau^2 - (d-2)\frac{da/d\tau}{a}\partial_\tau - \nabla^2 + \xi_g m^2 a^2
	-(d-2)\frac{d}{d\tau}\left(\frac{da/d\tau}{a}\right)
	\right] \pi^0.
\end{align}
After solving these equations, $A_L$ and $A_0$ are given by
\begin{align}
	A_L \; = \; \frac{a^{2-d}}{m^2}\left(\partial_\tau \pi_L + \pi^0\right) \,,
	\quad
	A_0 \; = \; \frac{a^{2-d}}{m^2}\left[\partial_\tau \pi^0 - (d-2)\frac{da/d\tau}{a}\pi^0 + \nabla^2 \pi_L\right] \,.
\end{align}
We quantize the fields as
\begin{align}
	&\hat{A}_{Ti}(\vec{x},t) = a^{-\frac{d-4}{2}}\int\frac{d^{d-1}k}{(2\pi)^{d-1}}
	e^{i\vec{k}\cdot\vec{x}} 
	\sum_{\lambda=\pm}\left[\epsilon_i^{(\lambda)}(\vec{k}) A_T(\vec{k},\tau) 
	\hat{a}_{\vec{k},\lambda} + 
	{\epsilon_i^{(\lambda)}}^*(-\vec{k}) A_T^*(-\vec{k},\tau) \hat{a}^\dagger_{-\vec{k},\lambda} \right] \,,
	\\
	&\hat{\pi}_L(\vec{x},t) = a^{\frac{d-2}{2}}\int\frac{d^{d-1}k}{(2\pi)^{d-1}} e^{i\vec{k}\cdot\vec{x}}
	\left[\pi_L(\vec{k},\tau)\hat{a}_{\vec{k},L} + {\pi_L^*(-\vec{k},\tau)} \hat{a}_{-\vec{k},L}^\dagger \right] \,,
	\\
	&\hat{\pi}^0(\vec{x},t) = a^{\frac{d-2}{2}}\int\frac{d^{d-1}k}{(2\pi)^{d-1}} e^{i\vec{k}\cdot\vec{x}}
	\left[\pi^0(\vec{k},\tau)\hat{a}_{\vec{k},0} + {\pi^0}^*(-\vec{k},\tau) \hat{a}_{-\vec{k},0}^\dagger \right] \,,
\end{align}
where the polarization vector and creation-annihilation operators satisfy
\begin{align}
	\sum_{\lambda=\pm}\epsilon_i^{(\lambda)}{\epsilon_j^{(\lambda)}}^* \; = \; \delta_{ij}-\frac{k_i k_j}{k^2} \,,
	\quad
	\left[\hat{a}_{\vec{k},\lambda}, \hat{a}_{\vec{k}',\lambda'}^\dagger\right]
	&= -(2\pi)^{d-1}\eta_{\lambda\lambda'}\delta^{(d-1)}(\vec{k}-\vec{k}') \,,
\end{align}
where we take $\eta_{++} = \eta_{--} = \eta_{LL} = -\eta_{00} = -1$ 
with all the other components vanishing.
The temporal component has the negative metric, but it does not cause an issue
since this degree of freedom is killed by the gauge redundancy (or more precisely the BRST symmetry) 
and is unphysical. The mode equations are given by
\begin{align}
	0 &= \left[\frac{d^2}{d\tau^2} + k^2 -\frac{\nu^2-1/4}{\tau^2}\right]A_T,
	\quad
	0 = \left[\frac{d^2}{d\tau^2} + k^2 -\frac{\nu^2-1/4}{\tau^2}\right]\pi_L,
	\quad
	0 =  \left[\frac{d^2}{d\tau^2} + k^2 -\frac{\nu_\xi^2-1/4}{\tau^2}\right]\pi^0,
\end{align}
where
\begin{align}
	\nu \; = \; \sqrt{\frac{(d-3)^2}{4} - \frac{m^2}{H^2}} \,,
	\quad
	\nu_\xi \; =\;  \sqrt{\frac{(d-1)^2}{4} - \frac{\xi_gm^2}{H^2}} \,.
\end{align}
Note that $\pi^0$ becomes infinitely heavy in the unitarity gauge $\xi_g \to \infty$.
These are again Bessel differential equations~\eqref{eq:Bessel} 
and the solutions can be written in terms of the Hankel functions.
The Bunch-Davies vacuum condition together with the canonical commutation relation
\begin{align}
	\left[\hat{A}_\mu(\vec{x},\tau), \hat{\pi}^\nu(\vec{y},\tau)\right] \; = \; i\delta_{\mu}^\nu \delta^{(d-1)}(\vec{x}-\vec{y}) \,,
\end{align}
fixes the mode functions as
\begin{align}
	A_T \; = \; \sqrt{\frac{-\pi\tau}{4}}e^{i\pi\nu/2} H_\nu^{(1)}(-k\tau) \, ,
	\quad
	\pi_L \; = \; \frac{m}{k}\sqrt{\frac{-\pi\tau}{4}}e^{i\pi\nu/2} H_\nu^{(1)}(-k\tau) \,,
	\quad
	\pi^0 \; = \; m\sqrt{\frac{-\pi\tau}{4}}e^{i\pi\nu_\xi/2} H_{\nu_\xi}^{(1)}(-k\tau) \,.
\end{align}
%%

%%%%%%%%%%
\subsubsection*{Propagators}
%%%%%%%%%%
We are now ready to compute the propagators.
The transverse part is easy to compute as
\begin{align}
	\langle A_{Ti}(\vec{x}_1,\tau_1) A_{Tj}(\vec{x}_2,\tau_2)\rangle
	&= \frac{H^{d-2}a_1 a_2}{(4\pi)^{d/2}}\left(\delta_{ij}-\frac{\partial_{1i}\partial_{1j}}{\nabla_1^2}\right)
	{I}_\nu(Z_{12}) \,,
\end{align}
where we used Eqs.~\eqref{eq:Hankel_rel2} and~\eqref{eq:Hankel_integral},
and the subscripts ``1" of the derivatives indicate that they act on $\vec{x}_1$.
To derive the longitudinal and temporal parts, we note that
\begin{align}
	\langle \pi_L(\vec{x}_1,\tau_1) \pi_L(\vec{x}_2,\tau_2)\rangle
	&= -\frac{(Ha_1a_2)^{d-2}}{(4\pi)^{d/2}}\frac{m^2}{\nabla_1^2}
	I_\nu(Z_{12}),
	~~
	\langle \pi^0(\vec{x}_1,\tau_1) \pi^0(\vec{x}_2,\tau_2)\rangle
	= -\frac{m^2 (Ha_1a_2)^{d-2}}{(4\pi)^{d/2}}
	I_{\nu_\xi}(Z_{12}),
\end{align}
where we again used Eqs.~\eqref{eq:Hankel_rel2} and~\eqref{eq:Hankel_integral}.
From this we obtain
\begin{align}
	&\langle \partial_{1i}A_L(\vec{x}_1,\tau_1) \partial_{2j}A_L(\vec{x}_2,\tau_2)\rangle
	= -\frac{H^{d-2}}{m^2(4\pi)^{d/2}}
	\left[\frac{1}{(a_1a_2)^{d-2}}\frac{\partial_{1i}\partial_{2j}}{\nabla_1^2}
	\partial_{\tau_1}\partial_{\tau_2}\left((a_1a_2)^{d-2}I_\nu(Z_{12})\right)
	+ \partial_{1i}\partial_{2j}I_{\nu_\xi}(Z_{12})
	\right],
	\\
	&\langle A_0(\vec{x}_1,\tau_1) A_0(\vec{x}_2,\tau_2)\rangle
	= -\frac{H^{d-2}}{m^2(4\pi)^{d/2}}\left[\nabla_1^2 I_\nu(Z_{12}) 
	+ \partial_{\tau_1}\partial_{\tau_2}I_{\nu_\xi}(Z_{12})\right],
\end{align}
and
\begin{align}
	&\langle A_0(\vec{x}_1,\tau_1) \partial_{2i}A_L(\vec{x}_2,\tau_2)\rangle
	= -\frac{H^{d-2}}{m^2(4\pi)^{d/2}}\left[\frac{1}{a_2^{d-2}}\partial_{\tau_2}\partial_{2i} \left(a^{d-2}_2 I_\nu(Z_{12}) \right)
	+ \partial_{\tau_1}\partial_{2i}I_{\nu_\xi}(Z_{12})\right] \,,
	\\
	&\langle \partial_{1i}A_{L}(\vec{x}_1,\tau_1) A_0(\vec{x}_2,\tau_2)\rangle
	= -\frac{H^{d-2}}{m^2(4\pi)^{d/2}}\left[\frac{1}{a_1^{d-2}}\partial_{\tau_1}\partial_{1i} \left(a^{d-2}_1 I_\nu(Z_{12}) \right)
	+ \partial_{\tau_2}\partial_{1i}I_{\nu_\xi}(Z_{12})\right] \,.
\end{align}

Next, we make these expressions more concise.
With the relations~\eqref{eq:Zderiv1} and~\eqref{eq:Zderiv2}, it is relatively easy to show that
\begin{align}
	\langle A_\alpha(\vec{x}_1,\tau_1) A_0(\vec{x}_2,\tau_2)\rangle
	&= \frac{H^{d-2}}{m^2(4\pi)^{d/2}}\left[K_{\alpha 0}(Z_{12}) - \partial_{1\alpha}\partial_{\tau_2} I_{\nu_\xi}(Z_{12})\right] \,,
	\\
	\langle A_0(\vec{x}_1,\tau_1) A_\alpha(\vec{x}_2,\tau_2)\rangle
	&= \frac{H^{d-2}}{m^2(4\pi)^{d/2}}\left[K_{0 \alpha}(Z_{12}) - \partial_{\tau_1}\partial_{2\alpha} I_{\nu_\xi}(Z_{12})\right]\,,
\end{align}
where the bitensor function $K_{\alpha\beta}$ is defined as
\begin{align}
	K_{\alpha\beta} (Z_{12}) 
	&= \left[Z_{12}I_\nu'' + (d-1)I_\nu'\right] (\partial_{1\alpha}Z_{12})(\partial_{2\beta}Z_{12})
	+ \left[(1-Z_{12}^2)I_\nu'' - (d-1)Z_{12}I_\nu'\right] \partial_{1\alpha}\partial_{2\beta}Z_{12} \,,
\end{align}
with the prime denoting the derivatives with respect to $Z_{12}$.
The pure spatial component is more complicated.
The part that includes $\nu_\xi$ is trivial, and thus we focus on the other part.
Our goal is to show that
\begin{align}
	\tilde{K}_{ij}(Z_{12}) = \left[\frac{m^2}{H^2 \tau_1\tau_2}\left(\delta_{ij}+\frac{\partial_{1i}\partial_{2j}}{\nabla_1^2}\right)
	-\frac{1}{(a_1a_2)^{d-2}}\frac{\partial_{1i}\partial_{2j}}{\nabla_1^2}
	\partial_{\tau_1}\partial_{\tau_2} (a_1a_2)^{d-2}\right]I_\nu \,,
\end{align}
is equivalent to
\begin{align}
	K_{ij}(Z_{12}) \; = \; \frac{\delta_{ij}}{\tau_1\tau_2}
	\left[(1-Z_{12}^2)I_\nu'' - (d-1)Z_{12}I_\nu'\right]
	- \frac{(x_{12})_{i}(x_{12})_j}{\tau_1^2\tau_2^2}
	\left[Z_{12}I_\nu'' + (d-1)I_\nu'\right] \,,
\end{align}
where $\vec{x}_{12} = \vec{x}_1 - \vec{x}_2$.
For this purpose, we act $\nabla_1^2$ on both expressions and check if they agree.\footnote{
	This is fine up to the functions that have vanishing Laplacian,
	but these functions do not allow the Fourier transform and are not important.
} 
By acting $\nabla_1^2$, we obtain
\begin{align}
	\nabla_1^2\left(K_{ij} - \tilde{K}_{ij}\right)
	&= \frac{\delta_{ij}}{\tau_1\tau_2}\nabla_1^2 \left[
	(1-Z_{12}^2)I_\nu'' - (d-1)Z_{12}I_\nu' - \frac{m^2}{H^2}I_\nu\right]
	- \nabla_1^2
	\left[ \frac{(x_{12})_i (x_{12})_j}{\tau_1^2\tau_2^2}
	\left(Z_{12}I_\nu'' + (d-1)I_\nu'\right)\right]
	\nonumber \\
	&+\left[-\frac{m^2}{H^2\tau_1\tau_2} + (\tau_1\tau_2)^{d-2}
	\partial_{\tau_1}\partial_{\tau_2}\frac{1}{(\tau_1\tau_2)^{d-2}}\right]
	\left[-\frac{(x_{12})_i (x_{12})_j}{\tau_1^2\tau_2^2} I_\nu'' + \frac{\delta_{ij}}{\tau_1\tau_2}I_\nu'\right] \,.
\end{align}
There are two structures, $\delta_{ij}$ and $(x_{12})_i(x_{12})_j$.
The former is given by
\begin{align}
	\left.\nabla_1^2\left(K_{ij} - \tilde{K}_{ij}\right)\right\vert_{\delta_{ij}}
	&= \frac{\nabla_1^2}{\tau_1\tau_2} \left[
	(1-Z_{12}^2)I_\nu'' - (d-1)Z_{12}I_\nu' - \frac{m^2}{H^2}I_\nu
	\right]
	-\frac{2}{\tau_1^2\tau_2^2}\left(Z_{12}I_\nu'' + (d-1)I_\nu'\right)
	\nonumber \\
	&+ 
	\frac{1}{\tau_1\tau_2}\left[-\frac{m^2}{H^2\tau_1\tau_2} 
	+ (\tau_1\tau_2)^{d-1}\partial_{\tau_1}\partial_{\tau_2}\frac{1}{(\tau_1\tau_2)^{d-1}}\right]
	I_\nu' \,.
\end{align}
After some computation, we obtain
\begin{align}
	\left.\nabla_1^2\left(K_{ij} - \tilde{K}_{ij}\right)\right\vert_{\delta_{ij}}
	&= \frac{1}{\tau_1\tau_2}\left[\frac{1}{\tau_1\tau_2}\frac{d}{dZ_{12}} + \nabla_1^2\right]
	\left[(1-Z_{12}^2)I_\nu'' - dI_\nu' - \left(d-2+\frac{m^2}{H^2}\right)I_\nu\right]
	= 0 \,,
\end{align}
where we used  Eq.~\eqref{eq:Ideriv} in the last equality.
The latter is given by
\begin{align}
	\left.\nabla_1^2\left(K_{ij} - \tilde{K}_{ij}\right)\right\vert_{(x_{12})_i (x_{12})_j}
	= -\frac{1}{\tau_1^2\tau_2^2}
	&\left[
	\left(\nabla_1^2 - \frac{4}{\tau_1\tau_2}\frac{d}{dZ_{12}}\right)\left(Z_{12}I_\nu'' + (d-1)I_\nu'\right)
	\right. \nonumber \\ &\left.
	+ \left(-\frac{m^2}{H^2\tau_1\tau_2} 
	+ (\tau_1\tau_2)^{d}\partial_{\tau_1}\partial_{\tau_2}\frac{1}{(\tau_1\tau_2)^{d}}
	\right) I_\nu''
	\right] \,.
\end{align}
After several steps, we obtain
\begin{align}
	\left.\nabla_1^2\left(K_{ij} - \tilde{K}_{ij}\right)\right\vert_{(x_{12})_i (x_{12})_j}
	&= -\frac{1}{\tau_1^3\tau_2^3}\frac{d^2}{d^2Z_{12}}
	\left[(1-Z_{12}^2)I_\nu'' - dI_\nu' - \left(d-2+\frac{m^2}{H^2}\right)I_\nu\right]
	= 0 \,,
\end{align}
where we again used Eq.~\eqref{eq:Ideriv}.
Therefore we conclude that $\tilde{K}_{ij} = K_{ij}$, and we find
\begin{align}
	\langle A_\alpha(\vec{x}_1, \tau_1) A_\beta(\vec{x}_2,\tau_2)\rangle
	&= \frac{H^{d-2}}{m^2(4\pi)^{d/2}}\left[K_{\alpha\beta}(Z_{12}) - \partial_{1\alpha}\partial_{2\beta}I_{\nu_\xi}(Z_{12})\right] \,.
\end{align}
In the unitarity gauge $\xi_g\to\infty$, $\pi^0$ becomes infinitely massive, and we can
simply ignore $I_{\nu_\xi}$.
Furthermore, different propagators correspond to merely different choices of $Z$ 
in the argument of $K_{\alpha\beta}$.
Therefore, we obtain the Schwinger-Keldysh propagator of the gauge bosons
\begin{align}
	G_{\lambda\lambda';\alpha\beta}(\vec{x}_1-\vec{x}_2;\tau_1, \tau_2)
	&= \frac{H^{d-2}}{m^2(4\pi)^{d/2}}K_{\alpha\beta}(Z_{\lambda\lambda'}) \, ,
\end{align}
with $\lambda, \lambda' = \pm$. One can show its equivalence to 
the expressions in~\cite{Allen:1985wd,Chen:2016hrz}, as demonstrated in~\cite{Frob:2013qsa}.

%%%%%%%%%%%%%%%%%%%%%%%%%%%%%%%%%%%%%%%
\small
\bibliographystyle{utphys}
\bibliography{ref}
%%%%%%%%%%%%%%%%%%%%%%%%%%%%%%%%%%%%%%%
  
%%%%%%%%%%%%%%%%%%%%%%%%%%%%%%%%%%%%%%%
\end{document}